# Generalizability of Functional Forms for Interatomic Potential Models Discovered by Symbolic Regression

Alberto Hernandez and Tim Mueller

**ABSTRACT**

In recent years there has been great progress in the use of machine learning algorithms to develop interatomic potential models. Machine-learned potential models are typically orders of magnitude faster than density functional theory but also orders of magnitude slower than physics-derived models such as the embedded atom method. In our previous work, we used symbolic regression to develop fast, accurate and transferrable interatomic potential models for copper with novel functional forms that resemble those of the embedded atom method. To determine the extent to which the success of these forms was specific to copper, here we explore the generalizability of these models to other face-centered cubic transition metals and analyze their out-of-sample performance on several material properties. We found that these forms work particularly well on elements that are chemically similar to copper. When compared to optimized Sutton-Chen models, which have similar complexity, the functional forms discovered using symbolic regression perform better across all elements considered except gold where they have a similar performance. They perform similarly to a moderately more complex embedded atom form on properties on which they were trained, and they are more accurate on average on other properties. We attribute this improved generalized accuracy to the relative simplicity of the models discovered using symbolic regression. The genetic programming models are found to outperform other models from the literature about 50% of the time in a variety of property predictions, with about 1/10$^{th}$ the model complexity on average. We discuss the implications of these results to the broader application of symbolic regression to the development of new potentials and highlight how models discovered for one element can be used to seed new searches for different elements.

## I. INTRODUCTION

Researchers across several fields apply molecular dynamics and Monte Carlo simulations to advance the scientific understanding, discovery, and design of materials and molecules. Using these methods, the thermodynamic and kinetic properties of a material can be computed with knowledge of the potential energy surface. *Ab initio* methods such density functional theory [1] (DFT), which has demonstrated good predictive accuracy [2-4] across many chemistries and configurations of atoms, can be used to compute the potential energy surface, but the computational cost and non-linear scaling of these methods severely limits the time scale and number of atoms that can be practically modeled. Surrogate models, such as cluster expansions [5] and interatomic potential models (or force fields) [6-15], are normally orders of magnitude faster than *ab initio* methods and usually scale linearly with respect to system size. The improved speed and scaling of surrogate models enable atomistic simulations that inform the design of materials at larger time and length scales.

Different types of interatomic potentials are commonly used for materials modeling. Classical (or empirical) interatomic potential models [6, 16-21], like the embedded atom method [16] (EAM), are usually derived from physical principles. This physical basis often makes it practical to apply these models to atomic configurations that are significantly different from those in their training set [22-24]. However classical models generally have limited accuracy due to their fixed functional forms.

Researchers have recently made great progress in the development of accurate interatomic potential models via supervised machine learning [5, 9, 10, 25-45], but these models are typically 2-3 orders of magnitude slower than classical interatomic potentials [34]. In addition, the complexity of many machine-learned interatomic potentials may lead to poor transferability to atomic configurations unlike the ones they were trained on. There has recently been progress in addressing this issue, but it remains an important challenge in the field [46-49].

Symbolic regression, in which machine learning is used to find simple expressions for unknown functions, has attracted increasing interest due to the speed, simplicity, and interpretability of the resulting models [10, 36, 50-59]. Our research group previously demonstrated the use of symbolic regression to develop accurate many-body interatomic potential models for copper that have simplicity and speed comparable to, and in several cases better than, classical interatomic potential models [36]. These models demonstrated low prediction errors for material properties on which they were not trained, suggesting that their functional forms encode underlying physics. The models were generated using genetic programming to explore a hypothesis space [52] of simple models designed to have physically meaningful terms. Our algorithm has been implemented in the open-source software package Potential Optimization by Evolutionary Techniques (POET) [60].

Here, we extend our analysis of this approach to the face-centered cubic (fcc) transition metals from groups 9, 10 and 11 on the periodic table: Cu, Ag, Au, Ni, Pd, Pt, Rh, and Ir. We demonstrate that the functional forms discovered for copper in our previous work (i.e., GP1, GP2, and GP3) [36] generalize well to elemental systems that are chemically similar to Cu, like Ag, Ni, and Pd, and that POET can find new, accurate functional forms for these elemental systems. Across these elements, POET models tend to have much lower errors than potentials based on the Sutton Chen [61] model, which has similar complexity. Compared to a moderately more complex classical EAM-type model trained for each of these elements, the POET models are found to have similar performance when predicting forces, energies, and stresses (on which the models were trained) but generalize better to properties on which the models were not trained, which is consistent with the expectation that simpler models have lower generalization errors.

## II. METHODS

### A. Developing the interatomic potential models.

In our previous work [36], we used POET to develop new interatomic potential models for Cu which we labeled GP1, GP2, and GP3. The corresponding functional forms are shown in equations (1), (2) and (3) respectively.

$$E_i = \sum_j (r^{x_0 - x_1 r} - x_2^{\,r}) f(r) + x_3 / \sum_j (x_4^{\,r} f(r)) \tag{1}$$

$$E_i = x_0 \sum_j r^{x_1 - x_2 r} f(r) + \left( x_3 - \sum_j \left( x_4 + x_5 r^{x_6 - x_7 r} \right) f(r) \right) / \sum_j f(r) \tag{2}$$

$$E_i = x_0 \sum_j r^{x_1 - x_2 r} f(r) + \left( x_3 - x_4 \sum_j r^{x_5 - x_6 r} f(r) \right) / \sum_j f(r) \quad (3)$$

Here $E_i$ is the energy associated with atom i, $\sum$ is the summation over neighbors of atom i, r is the distance between atom *i* and atom *j*, $x_i$ are the potential parameters, and $f(r)$ is a smoothing function [62]. These models all have a pair-wise repulsive term and a separate many-body attraction term that is a function of a sum over pairwise interactions. Although this form resembles that of a typical embedded-atom type potential, in which the many-body term is associated with metallic bonding, it was discovered by POET directly from the energy, force, and stress data. In all three potentials the attractive term is proportional to the inverse of the "density", a pairwise sum over nearest neighbors. In GP2 and GP3 the density is simply a sum over the smoothing function, making one component of the attractive term effectively a weighted average of pairwise interactions. Thus as the weight of one interaction increases, the weights of the other interactions decrease accordingly, resembling the behavior of bond order potentials. None of the three forms directly accounts for bond angles, as these are not included in the input features considered by POET.

In this work we optimized the parameters of these functional forms for Ag, Au, Cu, Ir, Ni, Pd, Pt, and Rh. The parameters were optimized to minimize a weighted average of the normalized mean squared errors on energies, forces and stresses with respect to DFT training data:

$$fitness = 0.5MSE_{energy} + 0.4MSE_{force} + 0.1MSE_{stress} \quad (4)$$

We note that by this measure of fitness, lower values imply better models. The weights were chosen to be consistent with our previous work [36], which we found gave good results for copper. The normalized energies were calculated by subtracting the minimum model-calculated energy and dividing by the standard deviation, and the forces and stresses were standardized by subtracting the mean and dividing by the standard deviation.

*Table 1. List of metrics used in the training and validation datasets. Where MAE is Mean Absolute Error, MAPE is the Mean Absolute Percent Error, and APE is the Absolute Percent Error. The low-index surface energies were not used to validate GP3 because these were included in the training data for this model*

| Metric name | Used in training | Used in validation |
|---|---|---|
| Fitness * | yes | yes |
| MAE of energies | no *** | yes |
| MAE of forces | no *** | yes |
| MAE of stresses | no *** | yes |
| MAPE of C11, C12, and C44 | no | yes |
| MAPE of 7 high-symmetry phonon frequencies ** | no | yes |
| APE of vacancy formation energy | no | yes |
| APE of vacancy migration energy | no | yes |
| APE of dumbbell formation energy | no | yes |
| MAPE of 13 low-index surface energies | only for GP3 | yes except for GP3 |
| APE of intrinsic stacking fault energy | no | yes |
| APE of unstable stacking fault energy | no | yes |
| APE of hcp formation energy | no | yes |
| APE of bcc formation energy | no | yes |
| APE of fcc lattice parameter | no | yes |
| APE of bcc lattice parameter | no | yes |

* Fitness from equation (4)

** $v_L(X)$, $v_T(X)$, $v_L(L)$, $v_T(L)$, $v_L(K)$, $v_{T1}(K)$, and $v_{T2}(K)$

*** The mean absolute error (MAE) is not used in training, but the mean squared error (MSE) is used to calculate the fitness used for training

To optimize the parameters we used the Covariance Matrix Adaptation Evolution Strategy (CMA-ES) [63] restarting with increasing population size (IPOP-CMA-ES) [64], and the conjugate gradient optimizer [65]. The population sizes used in IPOP-CMA-ES were 10, 30, 50, 70, 90 and 110, and after each IPOP-CMA-ES run, we optimized the parameters with the conjugate gradient optimizer. We executed 30 runs of IPOP-CMA-ES and conjugate gradient for each model and each element, for a total of 720 runs. We then took the model with the best training fitness for each element and functional form. The models obtained in this way are labeled GP1-c, GP2-c, and GP3-c (Table 2). As this optimization method is much more extensive (for the parameter space) than what we used when discovering GP1, GP2 and GP3 for Cu, we also re-optimized the parameters for Cu using the same method for consistency.

*Table 2. Acronyms of the interatomic potential models discussed in this work.*

| Acronym | Description |
|---|---|
| GP1-c | Optimized the parameters in GP1 [36] with IPOP-CMAES and conjugate gradient |
| GP2-c | Optimized the parameters in GP2 [36] with IPOP-CMAES and conjugate gradient |
| GP3-c | Optimized the parameters in GP3 [36] with IPOP-CMAES and conjugate gradient |
| SC4-c | Optimized the parameters in Sutton Chen with IPOP-CMAES and conjugate gradient, but maintained the square root |
| SC5-c | Optimized the parameters in Sutton Chen with IPOP-CMAES and conjugate gradient, allowing the power of 0.5 to change |
| LB-c | Optimized the parameters in [66] with IPOP-CMAES and conjugate gradient |
| GPn | New functional forms discovered in this work using POET, for Cu, Ag, Au, Ni, Pd, Pt, Rh and Ir. |

To compare the performance of our models against a simple and physics-derived interatomic potential model, we optimized the parameters of the Sutton Chen [61] EAM model with IPOP-CMA-ES and conjugate gradient starting from known parameters for each fcc element (see Supplementary Information Table 10 for the initial parameters). Equation (5) shows the Sutton Chen EAM model, where $x_i$ are the parameters.

$$E_i = \sum_j x_0 r^{x1} f(r) - \left( \sum_j x_2 r^{x_3} f(r) \right)^{x_4} \tag{5}$$

In the original Sutton Chen EAM model, $x_4$ is 0.5. For each element we ran 30 optimizations of Sutton Chen models keeping this value fixed. We refer to the resulting models, in which only four parameters were optimized, as SC4-c. We also ran 30 optimizations for each element in which we allowed $x_4$ to change, and we refer to the resulting models as SC5-c (Table 2). As the Sutton Chen model is in most cases slightly less complex than the models found by symbolic regression, for comparison purposes we also added a recently-published EAM functional form that is more complex than the GP models [66] (Equations (6), (7), (8), and (9)).

$$E_i = \frac{1}{2}\sum_j D_e((1-e^{-a(r-r_e)})^2 - 1)\Psi\left(r\right) + F(n_i) \tag{6}$$

$$n_i = \sum_j r^{-\beta}(1 + a_1\cos(\alpha r + \varphi))\Psi\left(r\right) \tag{7}$$

$$F(n) = F_0(1-\gamma\ln(n))n^{\gamma} + F_1 n \tag{8}$$

$$\Psi\left(r\right) = \frac{((r-r_c)/h)^4}{1+((r-r_c)/h)^4} \tag{9}$$

Starting from the original parameters described in [66], we optimized the potential using IPOP-CMA-ES and conjugate gradient for each fcc element, for a total of 240 runs. We label the resulting models LB-c.

We also used POET to discover new functional forms for each of the transition metal fcc elements in groups 9, 10, and 11 (Table 3). We ran the search using 5 different initializations: seeding with GP1, GP2, or GP3 from our previous work [36], seeding with a Sutton Chen model, or starting from randomly generated functions using the grow or full methods [67] with equal probability. For each of these initializations, we considered two smoothing functions: one from reference [62], which we used in our previous work [36], and another shown in equation (10). We added this smoothing function because its second derivative is zero at the inner and outer cutoff distances. The outer cutoff radii were set to include the 3rd-nearest neighbor in the DFT-relaxed fcc structure, and the inner cutoff radii were set to 1 Å less than the outer cutoff radii. The cutoff radii we used are provided in the Supplementary Information SI-Table 2.

$$f_2(r) = 6x^5 - 15x^4 + 10x^3, \quad x = \left(r_{out} - r\right)/\left(r_{out} - r_{in}\right) \tag{10}$$

We ran 30 optimizations with each of these starting configurations for each element, for a total of 300 runs for each of the 8 elements. For each model we determined the complexity, fitness, and computational cost. Complexity was measured as the number of nodes in the tree-graph representation of the potential model, the fitness was calculated using equation (4), and the computational cost was measured as the number of summations over neighbors. For each element, we selected a single model by constructing a set of 2-dimensional convex hulls of the models discovered by POET with respect to complexity and training fitness, with one convex hull for each computational cost. For the models appearing on the set of convex hulls for each element, we created a 2-dimensional convex hull with respect to complexity (x-axis) and the average of the normalized fitness and the normalized mean absolute percent error (MAPE) on elastic constants (y-axis). Plots of these models can be found in the Supplementary Information Figure 1 to 8 and an example for Ni is shown on Figure 1. The normalization of the fitness was done by taking the natural logarithm of the training fitness value, and then doing a min-max normalization. For the MAPE on elastic constants, we directly performed a min-max normalization (as shown on the Supplementary Information Equation 1.1). We selected the model at the elbow of this 2D hull, defined as the last model on the hull before the slope increased above -0.005. We label this model for each element as “GPn”.

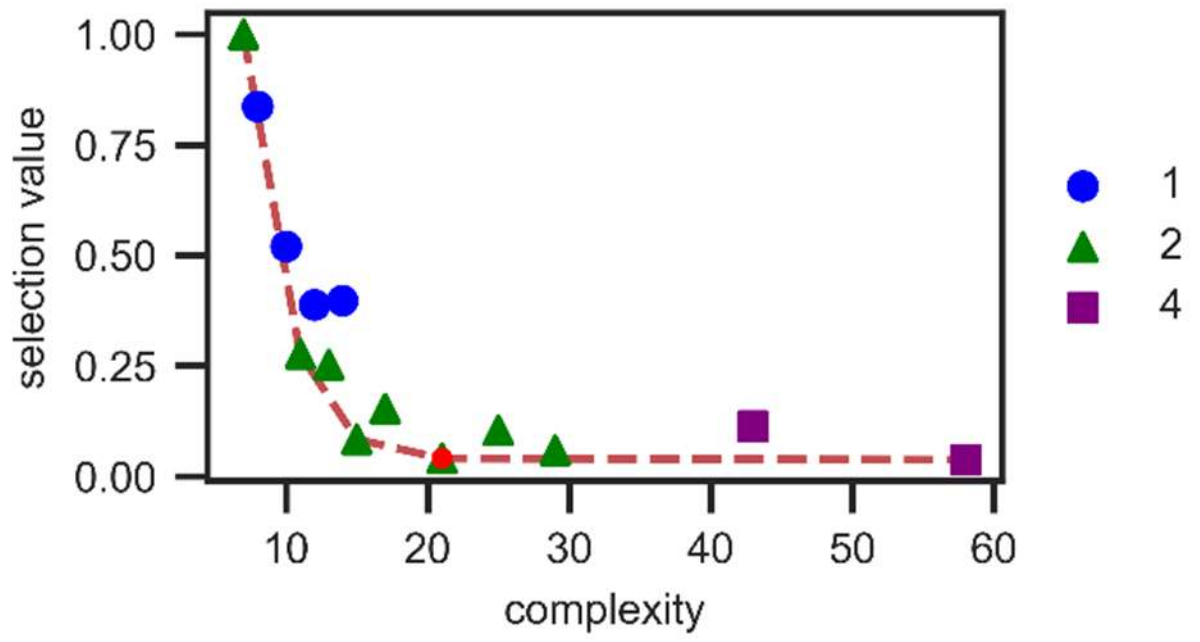

*Figure 1. Convex hull used to select a single model from the 300 runs of POET for Ni. The selection value is the average of the normalized fitness and the normalized mean absolute percent error (MAPE) on elastic constants. The legend indicates the speed (i.e., number of summations over neighbors) of each interatomic potential model.*

### B. Density functional theory data generation.

The DFT data were computed using the Vienna Ab initio Simulation Package [68] (VASP) with the Perdew-Burke-Ernzerhof [69] (PBE) generalized gradient approximation (GGA) exchange correlation functional. The following projector augmented wave method [70] (PAW) pseudopotentials were used: Cu_pv, Ag, Au, Ni_pv, Pd, Pt, Rh_pv, and Ir. Efficient *k*-point grids were obtained from the *k*-point grid server with MINDISTANCE = 50 Å [71], where MINDISTANCE is the minimum distance between lattice points in the real space lattice that corresponds to the k-point lattice. A plane-wave cutoff energy of 750 eV and ADDGRID = TRUE in VASP were used. The DFT point defect energies were computed by linear extrapolation (i.e., linear extrapolation of the values at 2×2×2 and 3×3×3 supercells with respect to the inverse of the supercell size [72]). The phonon dispersion curves were computed on a 3×3×3 supercell. The radial distribution function molecular dynamics simulations were performed in the NVT ensemble at the experimental liquid density in a 3×3×3 supercell. To calculate the radial distribution function, DFT molecular dynamics was performed with a plane-wave cutoff energy of at least 400 eV, the electronic self-consistency convergence was $10^{-5}$ eV, and only the *k*-point at Γ was used.

The DFT training data for copper was the same as was used in reference [36]. For the other elements we used a similar approach. We ran 3 types of DFT molecular dynamics simulations: NVT at 300K on the fcc phase, NVT at a temperature between the melting temperature and the boiling temperature using the liquid density, and NPT at 300K starting from the fcc phase (Supplementary Information Table 9). From each molecular dynamics simulation we collected 50 snapshots (one snapshot every 100 steps with a time step of 1 fs). We took the first 25 snapshots from each simulation as the training set and the last 25 snapshots from each simulation as a part of the validation set. Thus for each element, the training set had 75 structures, 75 energies, 7200 components of force, and 450 components of the virial stress.

### C. Computing properties with interatomic potential model

The data used to validate the interatomic potential models developed in this work were computed with LAMMPS [73]. The supercell sizes were the same as the ones used for DFT calculations. Instructions and files required to use models in LAMMPS are provided on the Supplementary Information.

## III. RESULTS AND DISCUSSION

### A. Assessing the transferability of functional forms developed with POET for Cu to other elemental systems.

Using our measure of fitness (equation (4)) on the validation data as a metric, the interatomic potential models developed by optimizing the parameters of the functional forms of GP1, GP2, and GP3 outperform the Sutton Chen models by an order of magnitude across the elements considered except Au, where SC5-c has a similar validation error (Figure 2). This order of magnitude improvement comes with only a slight increase in complexity; the number of nodes in GP1-c, GP2-c, GP3-c, and Sutton Chen are 19, 26, 24, and 15, respectively (Supplementary Information Figure 9) and the models from genetic programming have 2 or 3 summations over neighbors, compared to 2 for Sutton Chen. The SC5-c model for Au is particularly good compared to Sutton Chen models for other elements; it has a similar fitness as the GP1-c, GP2-c, and GP3-c models, which have reasonable validation mean absolute errors: 6 meV/atom, around 90 meV/Å, and around 0.3 GPa. On average, the performance of the model LB-c [66] using our measure of fitness (equation (4)) is similar to the performance of the GPn models and better than the performance of GP1-c, GP2-c and GP3-c. LB-c has a complexity of 85 (SI-Figure 9) which potentially allows it to reproduce the validation energies, forces, and stresses better than the simpler Sutton Chen model.

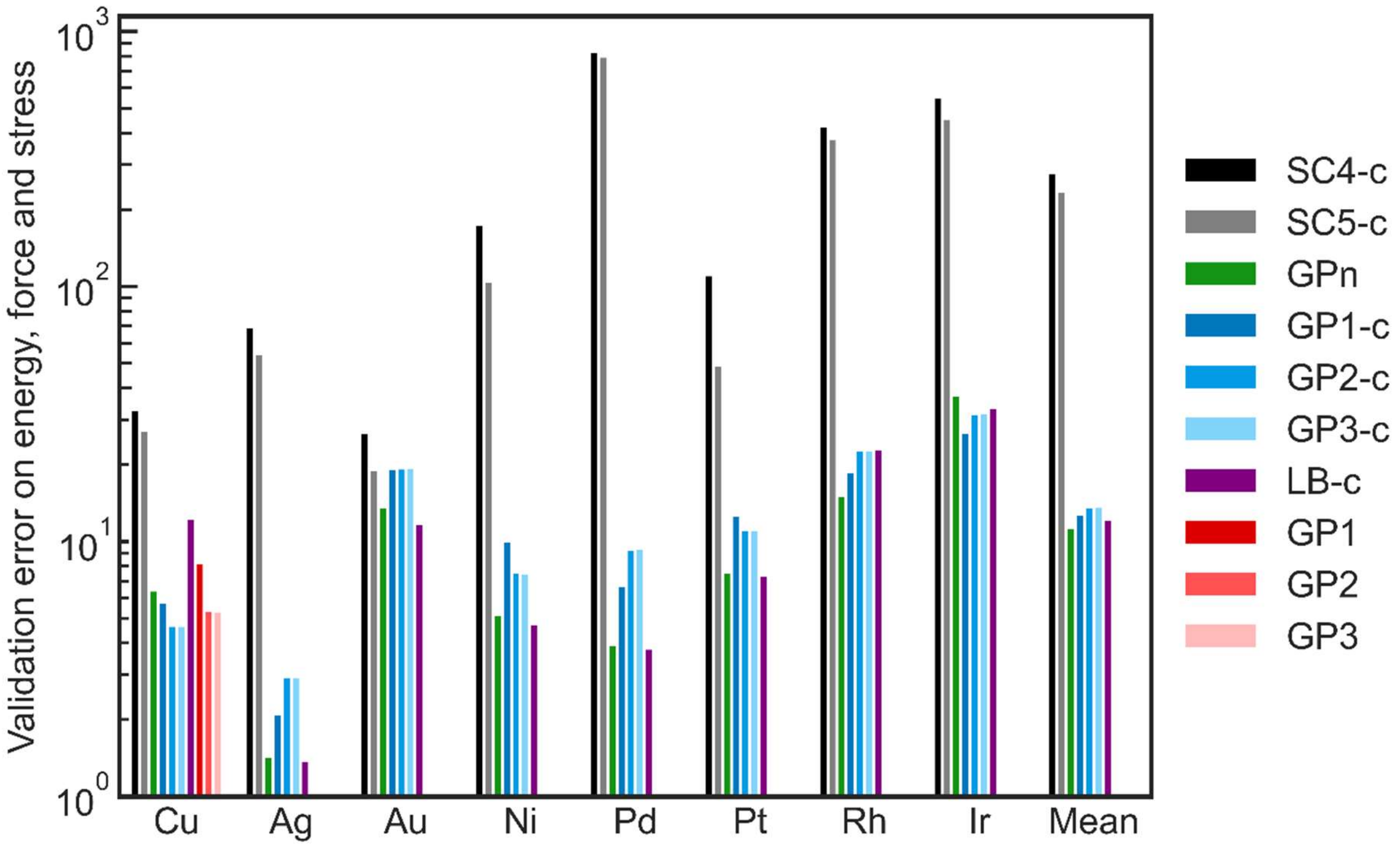

*Figure 2. Error on the validation set of energies, forces and stresses. This error metric is the same as that used to calculate the fitness in equation* (4) *multiplied by 1000. The models are ordered in approximately increasing complexity. The GPn correspond to new functional forms developed with POET. The models SC4-c, SC5-c, GP1-c, GP2-c, GP3-c, and LB-c were developed by optimizing the parameters of the corresponding functional forms using IPOP CMA-ES and the conjugate gradient optimizer.*

For copper, GP1-c, GP2-c and GP3-c have slightly lower validation errors on the energies, forces and stresses than the original models GP1, GP2, and GP3 discovered by POET in our previous work [36] (Figure 2). The parameters for these models were discovered using multiple runs of IPOP-CMA-ES and conjugate gradient, which is much more extensive than what is done in a typical search by POET, suggesting that models with improved fitness could be discovered by POET with more time or tighter convergence criteria.

To better assess the performance of the models derived from POET, we evaluated the models using a wider set of validation metrics shown in Table 1 (excluding the fitness). Detailed validation results on these properties can be found in the Supplementary Information Figure 36 to 57). To summarize the predictive performance of the models, all validation errors were linearly scaled to values between 0 and 1, where 0 is the best performing model and 1 is the worst-performing model on each metric. The normalized metrics were averaged over all properties, including validation errors on energies, forces, and stresses. The performance of the models GP1-c, GP2-c, and GP3-c (Figure 3) indicates that the functional forms of GP1, GP2, and GP3 transfer well to elements that are close to Cu in the periodic table, such as Ag, Ni, and Pd. This finding suggests that GP1, GP2, and GP3 encode information about the physics of the Cu system that can be applied to other systems that are chemically similar to Cu. For all elements other than Au and Cu, the models discovered using genetic programming were significantly more accurate for the validation properties than the Sutton-Chen derived models (Figure 3), consistent with their validation performance on energies, forces, and virial stresses (Figure 2).

The ability to use a single functional form across elements could be used to create parameterized potentials for alloys, where the values of the parameters depend on the element. Our results suggest that this should be feasible for chemically similar elements, but it would be more challenging to find a common functional form for chemically dissimilar elements. An alternative approach to constructing an alloy potential would be to insert branches for each element to the tree-graph representing the potential model. However this latter approach would result in much more complex potentials and scale poorly with the number of elements in the system.

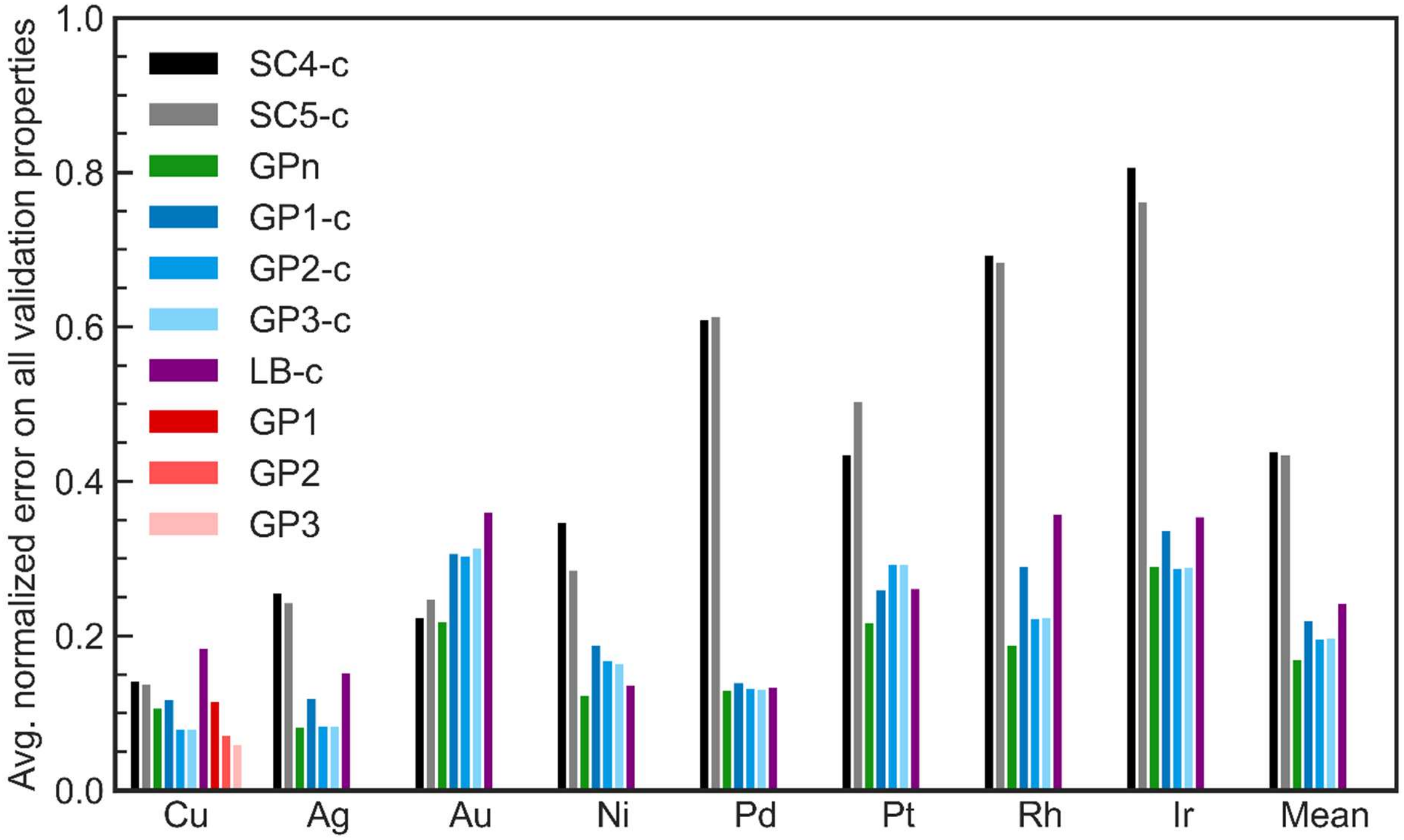

*Figure 3. Average of normalized errors across validation properties. The validation metrics considered on this plot are listed in Table 1 (excluding the fitness) . The normalization was done using min-max scaling: (x-min(x))/(max(x)-min(x)).*

The GP models are more transferrable to properties different to the ones they were trained on than Sutton Chen and LB-c models. The errors of GP models on all validation properties are on average less than the errors of LB-c models and Sutton Chen models (Figure 3). To test whether this result was skewed by elements like Rh and Ir which have relatively large errors across all models, we also calculated the geometric mean of the ratio of the average normalized error of LB-c models and GPn models from Figure 3. The geometric mean is 1.42, which is comparable to the arithmetic mean of 1.46. This supports the observation that the GPn models have smaller errors than LB-c models. Relative to LB-c, GP models performed better on the properties on which they were not trained than on energies, forces, and stresses. This improved generalizability is likely attributable to the relative simplicity of the GP models.

On average, the models with parameters discovered by POET for copper in our previous work [36] have lower validation errors for the wider set of properties than the models that have more extensively optimized parameters. This suggests that the more extensive search for optimal parameter values may have overfit the models. This difference is most notable for GP1, for which the parameters $x_1$ and $x_3$ in Equation (1) decreased by 50% and 80%, respectively (Supplementary Information Table 3). Even though the average absolute percent change in the parameters for GP2 is 53% (Supplementary Information Table 4), it has practically the same average normalized errors on all the validation properties as GP2-c (Figure 3). For energies, forces, and stresses, the MAEs are within 1 meV/atom, 3 meV/Å, and 0.04 GPa of each other respectively. Some of the difference between GP3 and GP3-c may be attributed to the fact that the training data for GP3 included 13 low-index surfaces on its training data, but these surfaces were not used to train any of the other models, including GP3-c. Consequently, the validation properties of GP3 excluded the low index surfaces. It is possible that the inclusion of low index surfaces for developing GP3 gave it access to a wider range of atomic environments useful for predicting validation properties such as stacking fault energies, for which GP3 has an error 8 mJ/m$^2$ lower than GP3-c.

### B. New functional forms identified in this work using POET.

In addition to the parameter optimization of known functional forms described in the previous section, we have also used POET to search for models with new functional forms for the eight fcc elements, which we refer to as "GPn" models (Table 3). One of the advantages to the symbolic regression approach is that searches for new models can be "seeded" with models that are known to perform well and/or have foundations in physical principles. For the generation of the GPn models we performed five searches for each element. In four of the searches POET was seeded with GP1, GP2, GP3 [36], or Sutton Chen, and in the fifth no seeding was used.

*Table 3. POET GPn models. The smoothing function* $f_2(r)$ *is as defined in equation* (10) and the smoothing function $f(r)$ *in [62]*

| Element | Seed | Complexity | POET GPn model |
|---|---|---|---|
| Cu | GP1 | 15 | $64.155\sum\left(r^{-4.461}-0.195^{r}\right)f(r)+\frac{29.764}{\sum f(r)}$ |
| Ag | SC | 19 | $\sum\left(\frac{424.997}{r^{7.322}}-0.011\right)f(r)+8.173\times 0.702^{\sum r^{1-r}f(r)}$ |
| Au | SC | 15 | $\sum\left(\frac{8167.212}{r^{10.860}}-0.011\right)f(r)+\frac{0.054}{\sum r^{-5.377}f(r)}$ |
| Ni | SC | 21 | $\sum\left(\frac{43.845}{r^{3.698}}-0.148\right)f_2(r)-62.560\left(\sum re^{-1.885r}f_2(r)\right)^{0.870}$ |
| Pd | SC | 17 | $42.613\sum\left(r^{-2.136r}f(r)\right)-42.613\left(\sum r^{-4.790}f(r)\right)^{0.071}$ |
| Pt | GP2 | 25 | $\sum\left(10.892r^{5.037-3.650r}-0.037\right)f_2(r)+12.720\times 0.215^{\sum 3.649r^{-r}f_2(r)}$ |
| Rh | No seed | 21 | $\sum\left(-89.551\times 0.264^{r}-0.317+\frac{98.908}{r^{3.580}}\right)f(r)+\frac{0.083}{\sum 0.136^{r}f(r)}$ |
| Ir | SC | 13 | $\sum\left(28.226r^{-1.942^{r}}f(r)\right)+\frac{78.243}{\sum f(r)}$ |

The GPn models have the lowest errors across all validation properties on average, and the mean error of GPn models on the validation energies, forces and stresses are only slightly lower than other models. This suggests that many of the models (except Sutton Chen models) fit the data in a similar way, but the transferability of GPn models to properties on which they were not trained is better. Because the GPn models have similar complexity and a similar number of summations over neighbors as the original GP1, GP2, and GP3 models, they also share the extremely high computation speed of these models [36].

The GPn models developed with POET perform better on elements close to Cu in the periodic table (Figure 4) than in elements far from Cu; a similar trend is observed for GP1-c, GP2-c, and GP3-c (Supplementary Information Figure 58 to 60). The trends in SC4-c and SC5-c models are similar, where their errors are comparable on Cu and elements on its group on the periodic table (Supplementary Information Figure 61 and 62), and the worst relative performances are for Ir, which is furthest from Cu. Unlike the GPn models, SC4-c and SC5-c do not perform well on Pd (Supplementary Information Figure 61 and 62), even though Cu is close to Pd on the periodic table. The GPn functional forms have a better accuracy on validation data than SC4-c and SC5-c for all the elements considered, but the performance of GPn models on Au, Pt, Rh and Ir is significantly worse than on the other elements, suggesting that the hypothesis space (i.e. the space of all functions considered by POET) needs to be expanded to develop suitably accurate models for these elements, e.g. by adding bond-angle terms [20, 21]. The addition of new features describing the local atomic environment should also help POET discover interatomic potential models for systems where covalent bonding is important. LB-c, which is more complex than GPn models, also does not perform well on these elements, suggesting that its functional form has a similar limitation as the GPn models.

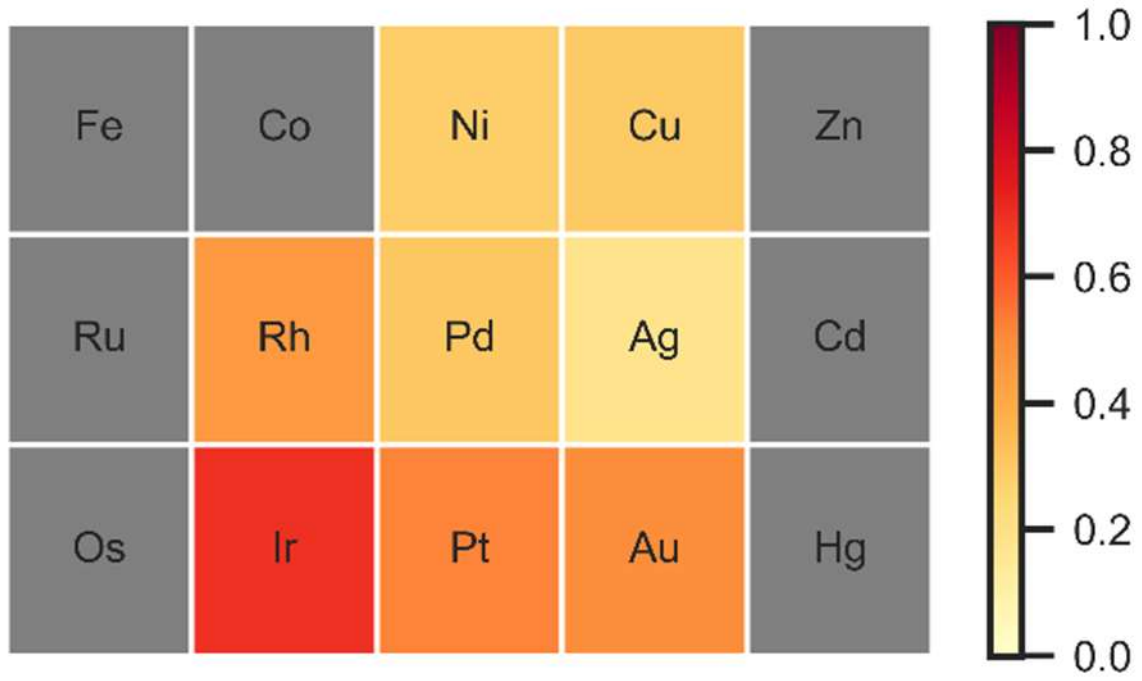

*Figure 4. Average of normalized errors across validation properties for GPn models. The validation metrics considered on this plot are in Table 1 (excluding the fitness). The normalization was done using min-max scaling (x-min(x))/(max(x)-min(x))*

The Generalized Gradient Approximation (GGA) [69] has been reported to have problems in calculating the vacancy formation energy of some fcc metals [74]. Interestingly, the vacancy formation energies predicted by the GPn models are closer to the experimental values than the DFT values are. For the vacancy formation energies computed with DFT and GPn, the average difference with the closest experimental value for GPn is 63 meV compared to 214 meV for DFT (Figure 5). The vacancy formation energies predicted by GPn for Cu, Ni, Pd, and Pt are within the range of experimentally-reported energies, and the errors between the predictions of GPn and the closest experimental value for Ag, Au and Rh are less than 100 meV. For Ir, the GPn model is 281 meV from the only reported experimental value, and DFT is 307 meV from the experimental value. O'Brien et al. observed a similar trend for the vacancy formation energies of Pt and Au calculated using many-body interatomic potentials [75]. They reported a general good agreement between the vacancy formation energies predicted by EAM potentials and experimental values when fitting the models with DFT data excluding atomic configurations with vacancies. Although it is not clear why the models discovered by POET calculate more accurate vacancy formation energies than DFT, it may be due to the fact that the error in DFT-calculated vacancy formation energies is believed to come from errors in the way DFT treats surfaces [74, 76]. Because neither surfaces nor vacancies were included in the data used to train the GPn models, they did not learn this error.

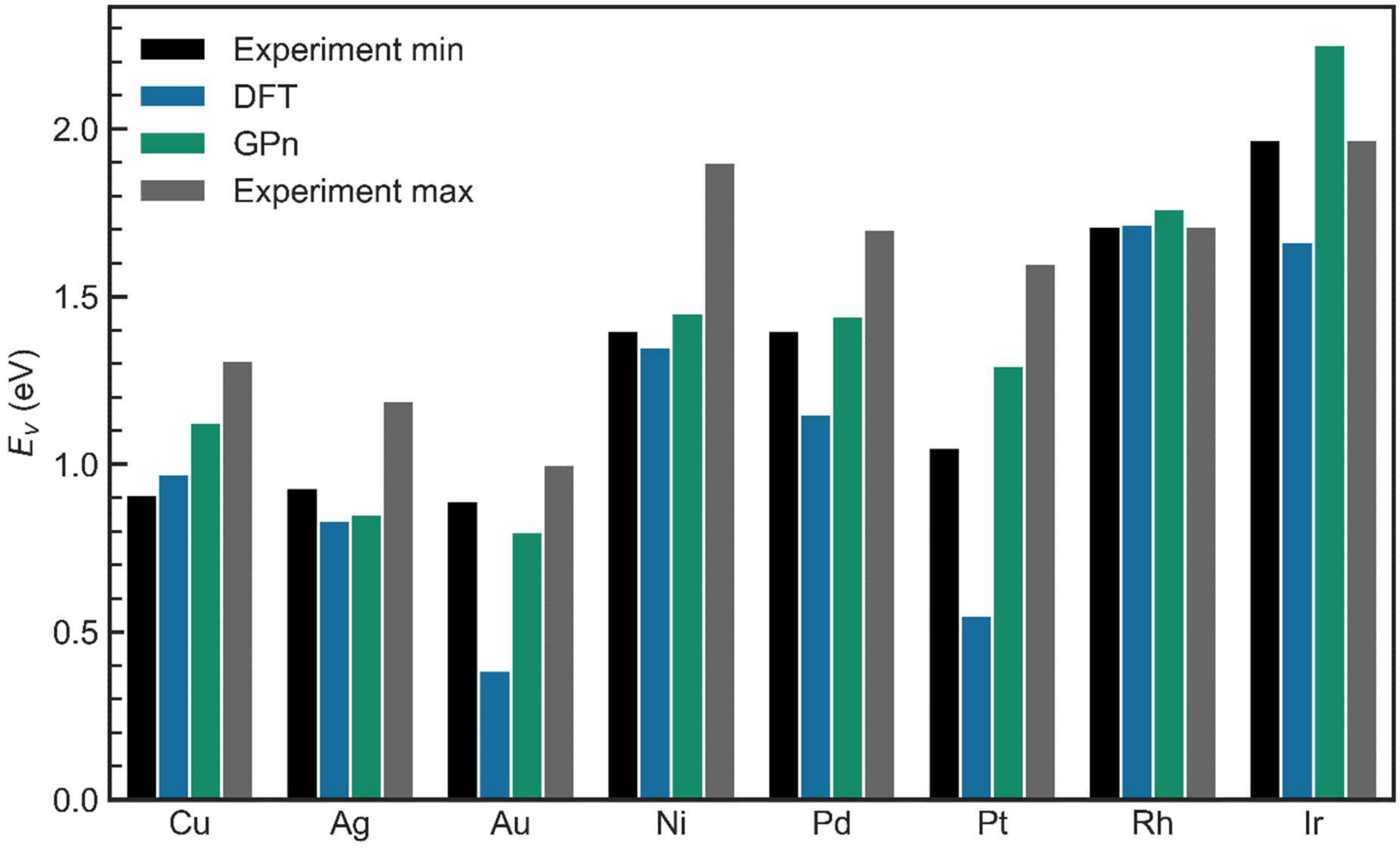

*Figure 5. Vacancy formation energy ($E_v$) predicted by GPn models compared to DFT-calculated $E_v$, maximum experimental $E_v$, and minimum experimental $E_v$. The DFT results were generated using the PBE exchange correlation functional. References: Cu [77, 78], Ag [77, 79], Au [80], Ni [81, 82], Pd [83, 84], Pt [85, 86], Rh [87], Ir [87]*

For four of the eight elements (Ni, Pd, Au, and Ir), the “GPn” model came from a run seeded with the Sutton Chen model, Equation (5) and Table 3, demonstrating the benefits of starting from a physically-derived functional form. The GPn forms for Ni and Pd inherited an embedding function that resembles the embedding function of the Sutton Chen model where the density is exponentiated to a fraction. The pair interaction term of Ni retained a similar form as the original Sutton Chen seed with an additional constant, but the pair term for Pd is different. While the functional form of the density of Pd remained the same as the Sutton Chen seed, the density of Ni was updated by POET using the natural exponential function. The embedding function of the GPn models for Au and Ir, which were seeded using Sutton Chen, have an embedding form like the one of GP1, which is a constant divided by the density. The pair interaction term of Au retained a similar form as the original Sutton Chen seed, and POET added a constant. These observations suggest that POET successfully extracted information from the Sutton Chen seed and combined it with new functional forms to find better models.

The POET run that found the GPn model for Cu was seeded with GP3, and there are similarities in the embedding functions of the two models. The only other element for which the GPn model was seeded by one of the previously-discovered models for copper was Pt, which was seeded with GP1. The GPn models for both Cu and Pt have terms in the form of $r^{a-br}$, which was present in the functions used to seed them. This term is not present in other GPn potentials, suggesting that although this term works well for copper, there are superior alternatives for most other fcc elements. POET did not use a seed to find the GPn models for Ag and Rh. Interestingly, the embedding function for these elements is like the one of GP1, which may be preferred by the algorithm because it has a small number of nodes. Five of the eight potentials have embedding functions which are proportional to the inverse of the density, suggesting that this form is a particularly effective, simple representation of the underlying physical interactions. It is not yet clear how the different functional forms reflect the underlying physics for each element, but such analysis may be possible e.g. by comparing with electronic structure calculations [88].

### C. Assessing the tradeoff between accuracy and complexity: validating against literature EAM-type models.

We compared the performance of the interatomic potential models developed with genetic programming to the EAM-type models in the NIST Interatomic Potentials Repository [8] for Cu, Ag, Au, Ni, Pd, Pt, Rh and Ir and the EAM models from reference [89]. We identified only 155 instances in which validation data was reported for one of the 21 validation properties for the models from the literature, and 37 of those reports were for copper (Supplementary Information Table 11). To assess the accuracy and complexity tradeoff of the EAM-type interatomic potential models developed in this work, we generated a Pareto frontier with respect to validation error and complexity for each element and for each of the 21 validation properties, giving a total of 168 Pareto frontiers. An example of one of these frontiers is given in Figure 6a. The models on the Pareto set are optimal in the sense that for each model on the Pareto frontier, there is no model that is both less complex and has lower validation error. Thus the models on the Pareto frontier are those that are particularly likely to have good transferability.

The models derived from Sutton Chen, SC4-c and SC5-c, show up on 27 and 91 of the 168 Pareto frontiers, respectively These models are the simplest for six of the eight elements, and the simplest model is always on the Pareto frontier. For the remaining two elements (Au and Ir), the complexity of the GPn models is less than or equal to the complexity of Sutton Chen. Au is the element in which Sutton Chen performs the best (Figure 3), but GPn with equal complexity still outperforms Sutton Chen (i.e., SC4-c and SC5-c) on about half of the validation properties (10/21). For Ir, GPn is less complex than Sutton Chen (13 vs 15 nodes) and more accurate on 19 out of 21 properties.

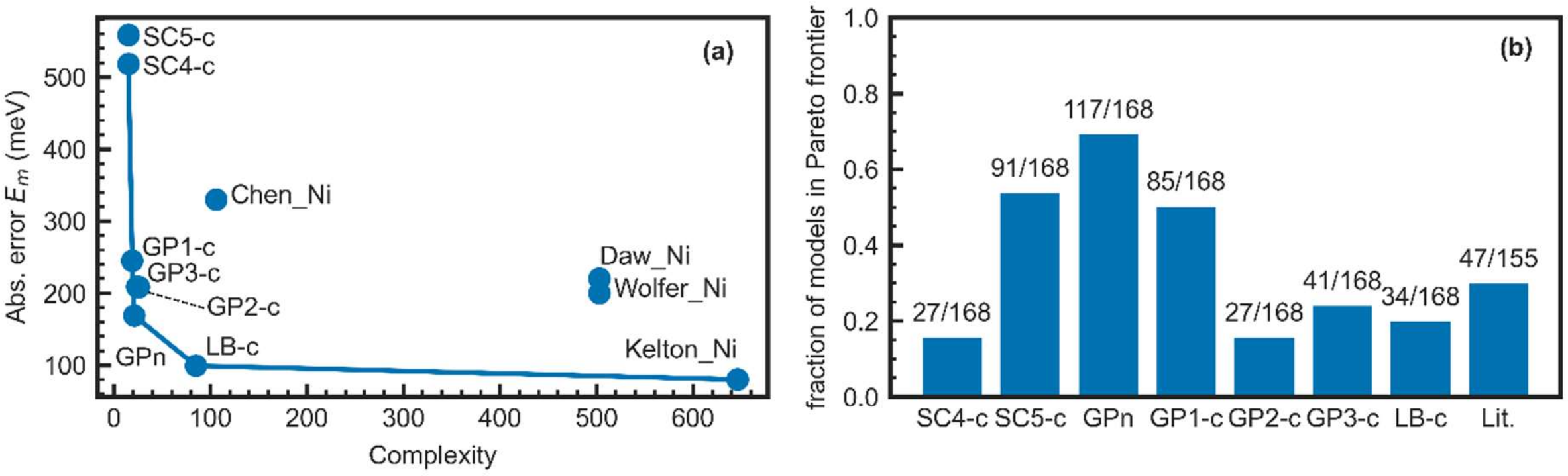

*Figure 6. (a) Pareto frontier of EAM-type interatomic potential models for Ni considering the absolute error on the vacancy migration energy and the number of nodes (complexity). The models SC4-c, GP1-c, GPn and Kelton_Ni belong to the frontier. The references are: Chen_Ni [90], Daw_Ni [91], Wolfer_Ni [86], and Kelton_Ni [92]. (b) Number of times that an EAM-type model belongs to the Pareto frontier divided by the number of times that the model has validation values available across the elements and properties. The metrics considered are the ones shown in Table 1 (excluding the fitness).*

Overall, the new forms discovered by POET, the GPn models, are the most likely to show up on the Pareto frontiers. This is in part due to their simplicity, which averages only 18 nodes. They reduce the percentage of times that SC4-c, SC5-c, GP1-c, GP3-c, GP2-c, LB-c, and literature models belong to the Pareto frontiers by 7, 19, 23, 12, 15, 10 and 1 percent, respectively. On average, the types of models developed using genetic programming (i.e., GP1-c, GP2-c, GP3-c, and GPn) have lower error on the validation properties than 50% of the EAM-type literature models (Supplemental Information Table 12), and they are on average more than 10 times simpler. For comparison, each of the types of Sutton Chen models (i.e., SC4-c and SC5-c) are more accurate than about 35% of EAM-type literature models, and LB-c models are more accurate than 48% of the EAM-type literature models. Models derived from POET (around 20 nodes) are simpler than LB-c (85 nodes) and they still are slightly more accurate than LB-c. GPn, GP1-c, GP2-c, and GP3-c models have a lower error than LB-c models on 93, 88, 90, and 89 out of 168 properties respectively.

## IV. CONCLUSIONS

The functional forms of interatomic potential models developed for Cu with POET have good transferability to Ag, Ni, and Pd, which are close to Cu on the periodic table, but their performance is not as good on other elements further from Cu. We observed a similar trend for new GPn functional forms developed with POET in this work. The good performance on elemental systems like Cu may suggest that the functional forms of GP1, GP2, and GP3 encoded physical information that allows them to transfer well to these elemental systems. Even though POET was able to identify functional forms that outperform the Sutton Chen functions across Cu, Ag, Au, Ni, Pd, Pt, Rh, and Ir, the accuracy on Au, Pt, Rh, and Ir is not high. The accuracy for these elements could potentially be improved by expanding the hypothesis space, e.g. by including bond-angle terms and/or longer-range interactions. The results could also potentially improve by identifying more complex functional forms. The focus on expanding the hypothesis space should be preferred because it would allow maintaining simple functional forms.

When compared to a moderately more complex embedded atom functional form, the models developed by POET had similar validation errors on forces, energies, and stresses (properties on which they were trained) but lower average error on other properties. We attribute the lower generalization errors to the relative simplicity of the POET models. Finally, we showed that the models developed with POET show a good balance of accuracy and complexity when compared to other models, frequently appearing on the Pareto frontiers with respect to accuracy and complexity and outperforming much more complex models about 50% of the time.

## ACKNOWLEDGEMENTS

We acknowledge financial support from the Office of Naval Research, grant number ONR-N00014-15-1-2681. This work was done using high performance computing resources from the Maryland Advanced Research Computing Center (MARCC) and the Cray XC40/50 (Onyx) from the Army Engineer Research and Development Center (ERDC). We would also like to acknowledge Chenyang Li, Adarsh Balasubramanian, Fenglin Yuan, Yunzhe Wang, Chuhong Wang and Pandu Wisesa for helpful discussions.

**SUPPLEMENTAL MATERIAL**

*SI-Table 1. POET GPn models with full precision*

| Element | Seed | POET GPn model |
|---|---|---|
| Cu | GP1 | $64.15515940524648\Sigma\left(\left(\frac{1}{r^{4.461239524986285}}-0.1954983463474226^{r}\right)f(r)\right)+\frac{29.76431804908527}{\Sigma(f(r))}$ |
| Ag | SC | $\Sigma\left(\left(\frac{424.9965825900254}{r^{7.322698914792303}}-0.01102677991908652\right)f(r)\right)+8.172810046531140\times0.7017644070210753^{\Sigma\left(r^{1-r}f(r)\right)}$ |
| Au | SC | $\Sigma\left(\left(\frac{8167.21232353391}{r^{10.8596729498766}}-0.01085427302578209\right)f(r)\right)+\frac{0.05350843770358775}{\Sigma\left(r^{-5.377144047422479}f(r)\right)}$ |
| Ni | SC | $\Sigma\left(\left(\frac{43.844638787564}{r^{3.69765898135618}}-0.1481588631749128\right)f_2(r)\right)-62.55959167587809\left(\Sigma\left(re^{-1.885228718370605r}f_2(r)\right)\right)^{0.870093496934648}$ |
| Pd | SC | $42.61343703107455\Sigma\left(r^{-2.135723519763371r}f(r)\right)-42.61343703107455\left(\Sigma\left(r^{-4.790271382259407}f(r)\right)\right)^{0.070787893131414}$ |
| Pt | GP2 | $\Sigma\left(\left(10.89243794141817r^{5.036647157219199-3.649317739301988r}-0.0373360678671021\right)f_2(r)\right)+12.720343347841920\times0.2146539628086967^{\Sigma\left(3.649317739301988r^{-r}f_2(r)\right)}$ |
| Rh | No seed | $\Sigma\left(\left(-89.55102711997160\times0.2640482761663984^{r}-0.3169171347295917+\frac{98.9083181938882}{r^{3.580044561312461}}\right)f(r)\right)+\frac{0.0826182489861164}{\Sigma\left(0.1359924613515034^{r}f(r)\right)}$ |
| Ir | SC | $\Sigma\left(28.22630935153837r^{-1.941760188276659^{r}}f(r)\right)+\frac{78.24327611794345}{\Sigma(f(r))}$ |

The number of nodes of the models are 15, 19, 24, and 26 nodes for Sutton Chen, GP1-c, GP3-c, and GP2-c, respectively. The average number of nodes of GPn models is 18, and the average for literature models is 348.

The selection value shown on the following convex hulls is:

$$s_i = \frac{1}{2}\left( \frac{\ln fitness_i - \min \ln fitness}{\max \ln fitness - \min \ln fitness} + \frac{MAPE_{EC,i} - \min MAPE_{EC}}{\max MAPE_{EC} - \min MAPE_{EC}} \right) \quad (1.1)$$

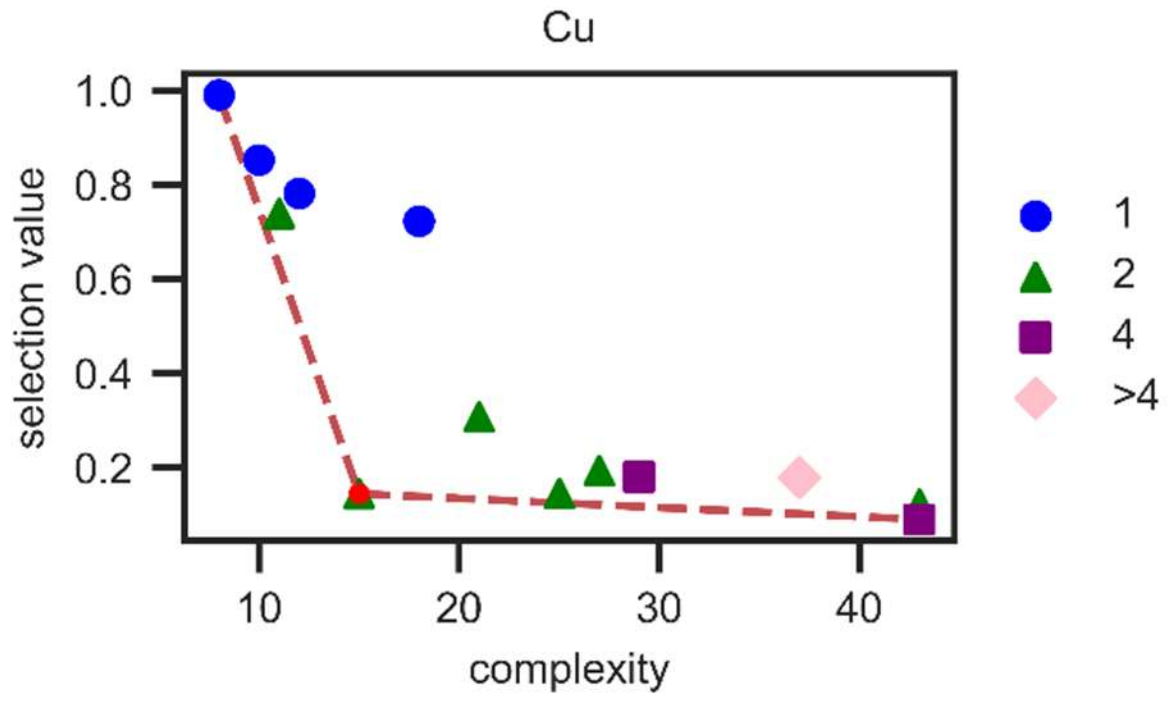

*SI-Figure 1. Convex hull of "selection value" (see methods) and complexity (i.e., number of nodes) for selecting the model GPn for Cu. The convex hull is shown as the red dashed line, and GPn is shown as a red dot. The legend indicates the speed (i.e., number of summations over neighbors) of each interatomic potential model.*

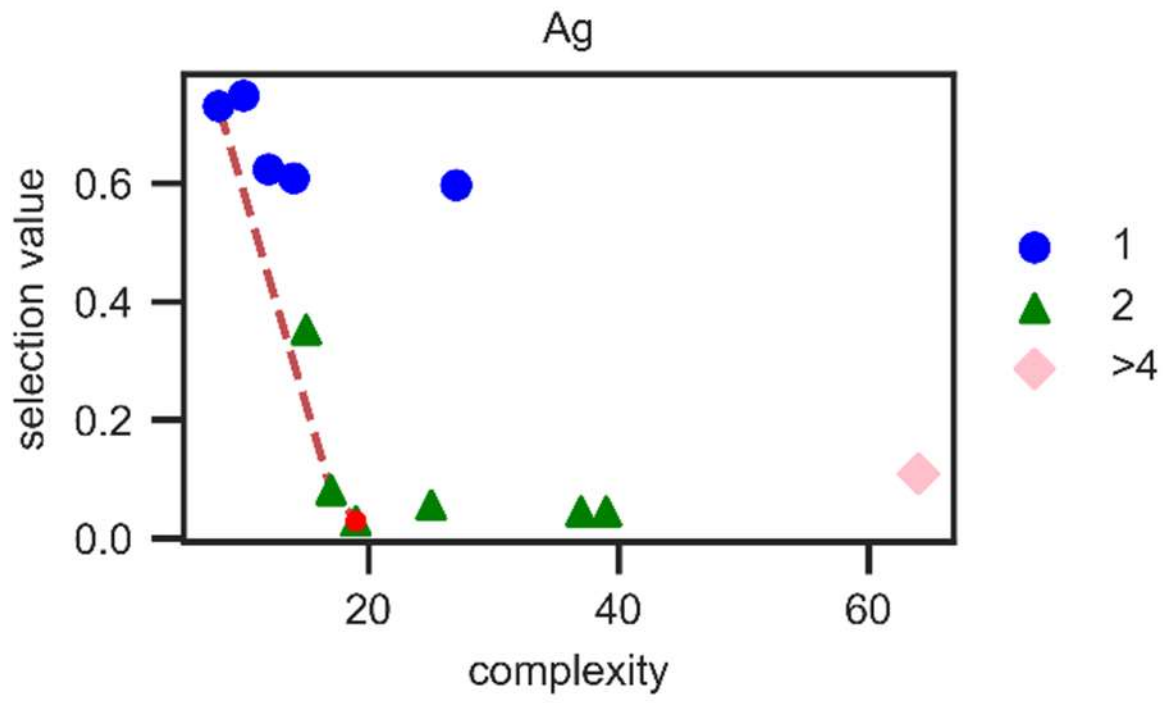

*SI-Figure 2. Convex hull of "selection value" (see methods) and complexity (i.e., number of nodes) for selecting the model GPn for Ag. The convex hull is shown as the red dashed line, and GPn is shown as a red dot. The legend indicates the speed (i.e., number of summations over neighbors) of each interatomic potential model.*

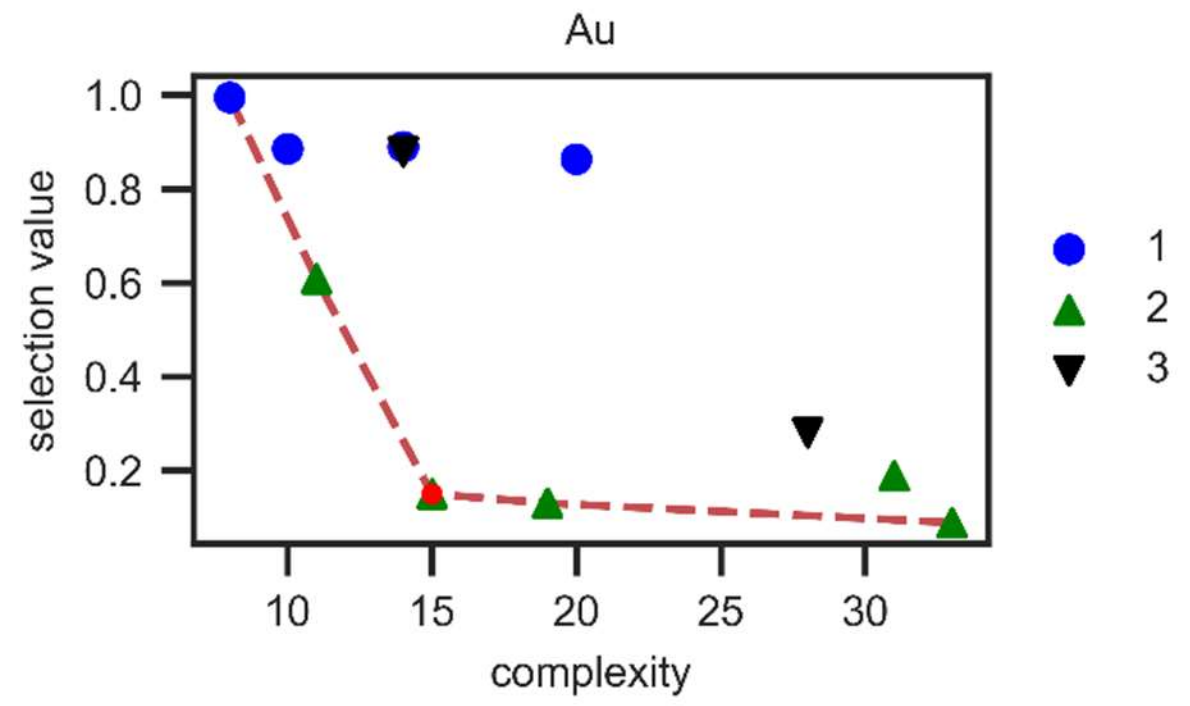

*SI-Figure 3. Convex hull of "selection value" (see methods) and complexity (i.e., number of nodes) for selecting the model GPn for Au. The convex hull is shown as the red dashed line, and GPn is shown as a red dot. The legend indicates the speed (i.e., number of summations over neighbors) of each interatomic potential model.*

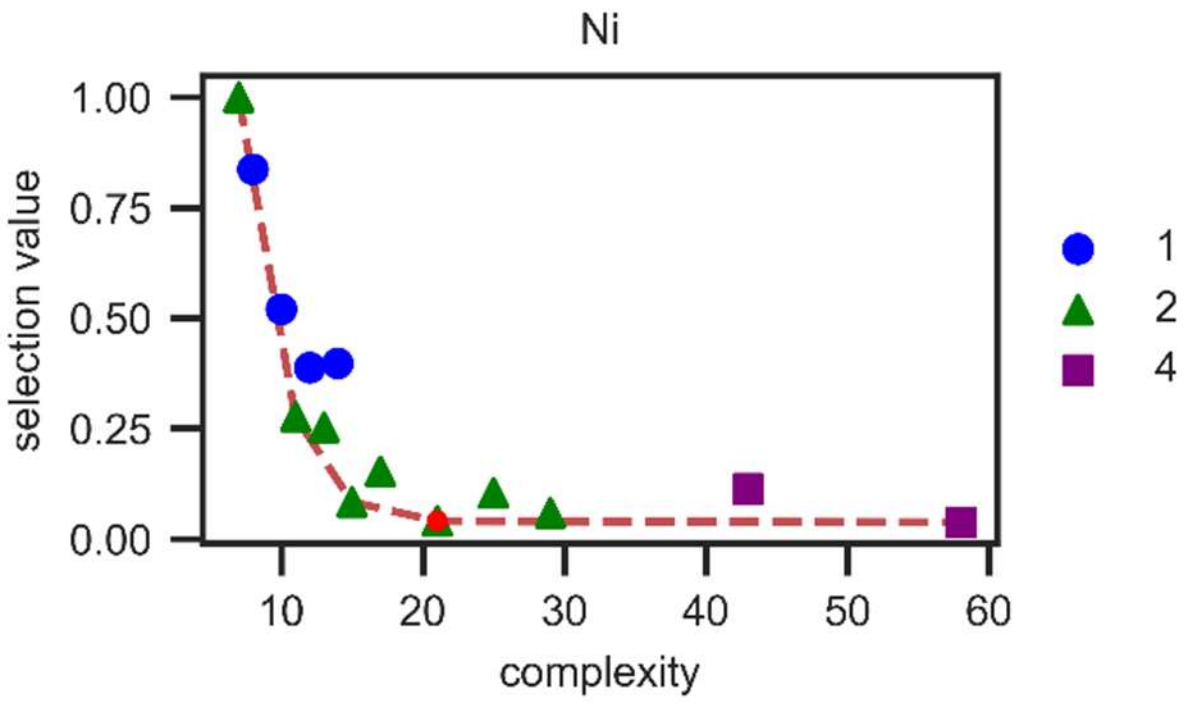

*SI-Figure 4. Convex hull of "selection value" (see methods) and complexity (i.e., number of nodes) for selecting the model GPn for Ni. The convex hull is shown as the red dashed line, and GPn is shown as a red dot. The legend indicates the speed (i.e., number of summations over neighbors) of each interatomic potential model.*

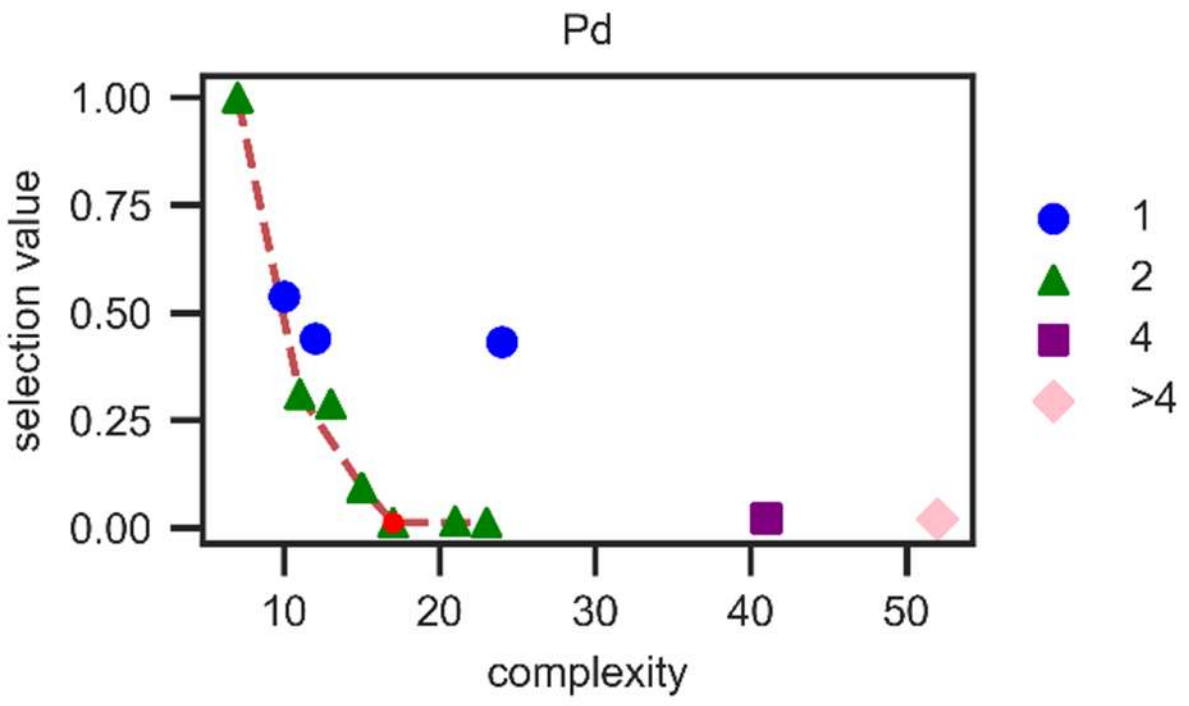

*SI-Figure 5. Convex hull of "selection value" (see methods) and complexity (i.e., number of nodes) for selecting the model GPn for Pd. The convex hull is shown as the red dashed line, and GPn is shown as a red dot. The legend indicates the speed (i.e., number of summations over neighbors) of each interatomic potential model.*

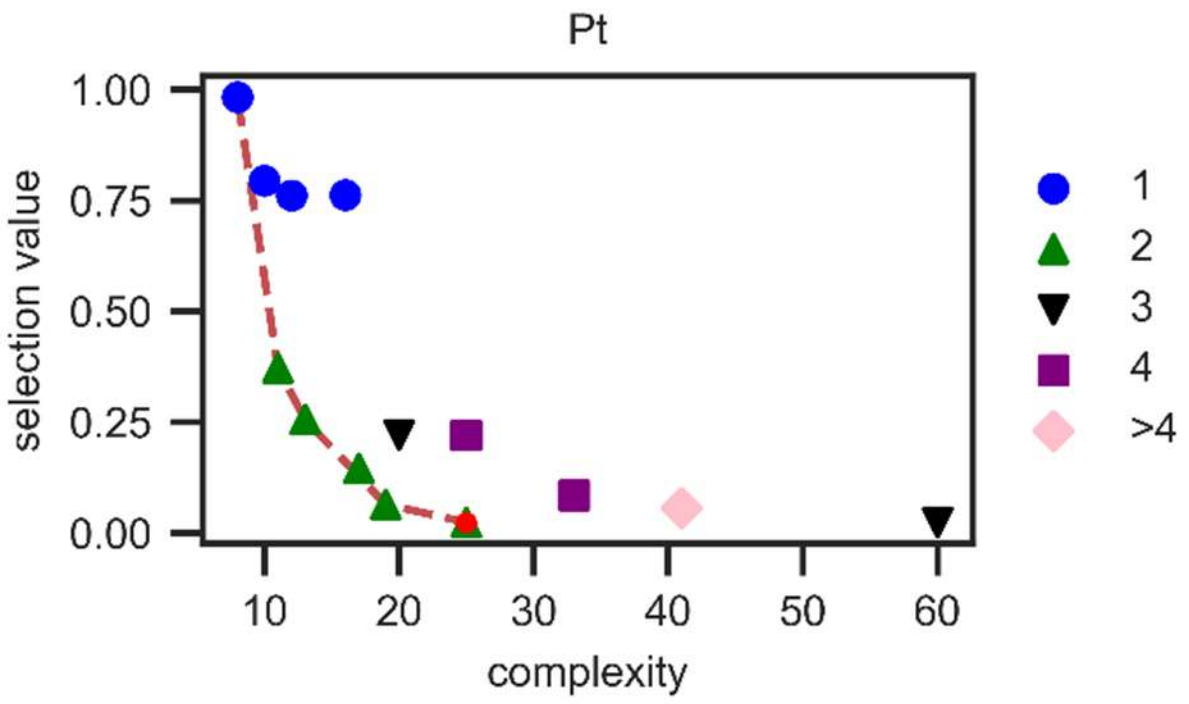

*SI-Figure 6. Convex hull of "selection value" (see methods) and complexity (i.e., number of nodes) for selecting the model GPn for Pt. The convex hull is shown as the red dashed line, and GPn is shown as a red dot. The legend indicates the speed (i.e., number of summations over neighbors) of each interatomic potential model.*

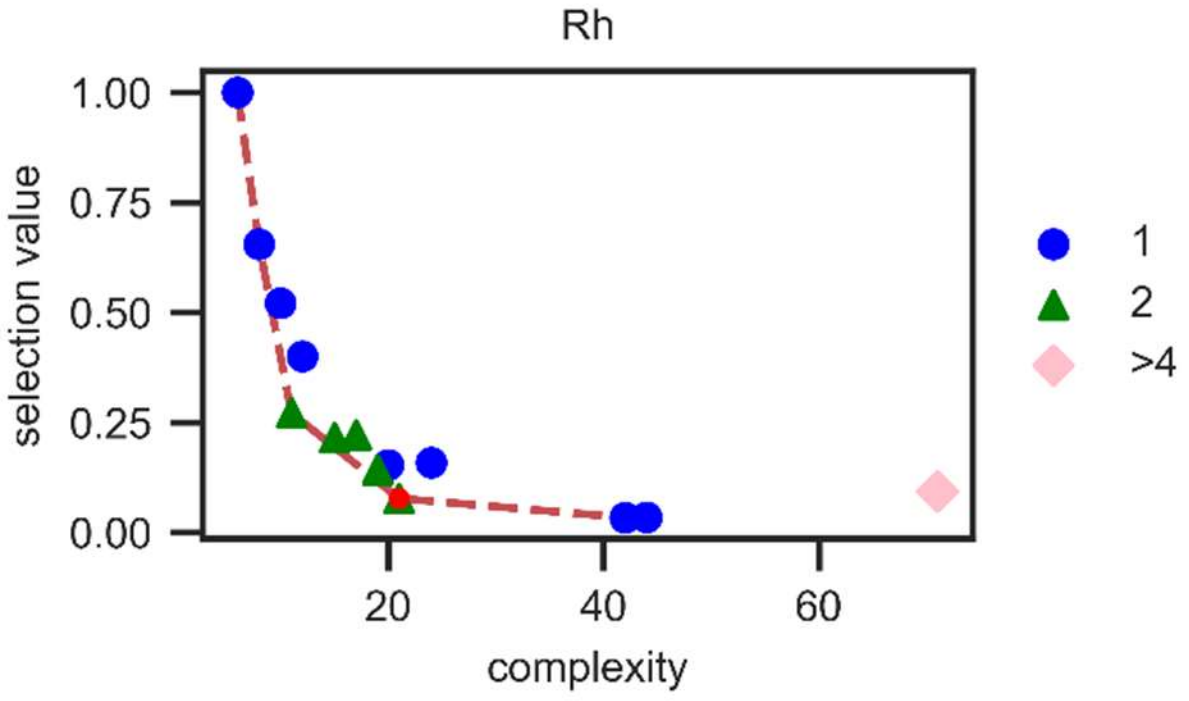

*SI-Figure 7. Convex hull of "selection value" (see methods) and complexity (i.e., number of nodes) for selecting the model GPn for Rh. The convex hull is shown as the red dashed line, and GPn is shown as a red dot. The legend indicates the speed (i.e., number of summations over neighbors) of each interatomic potential model.*

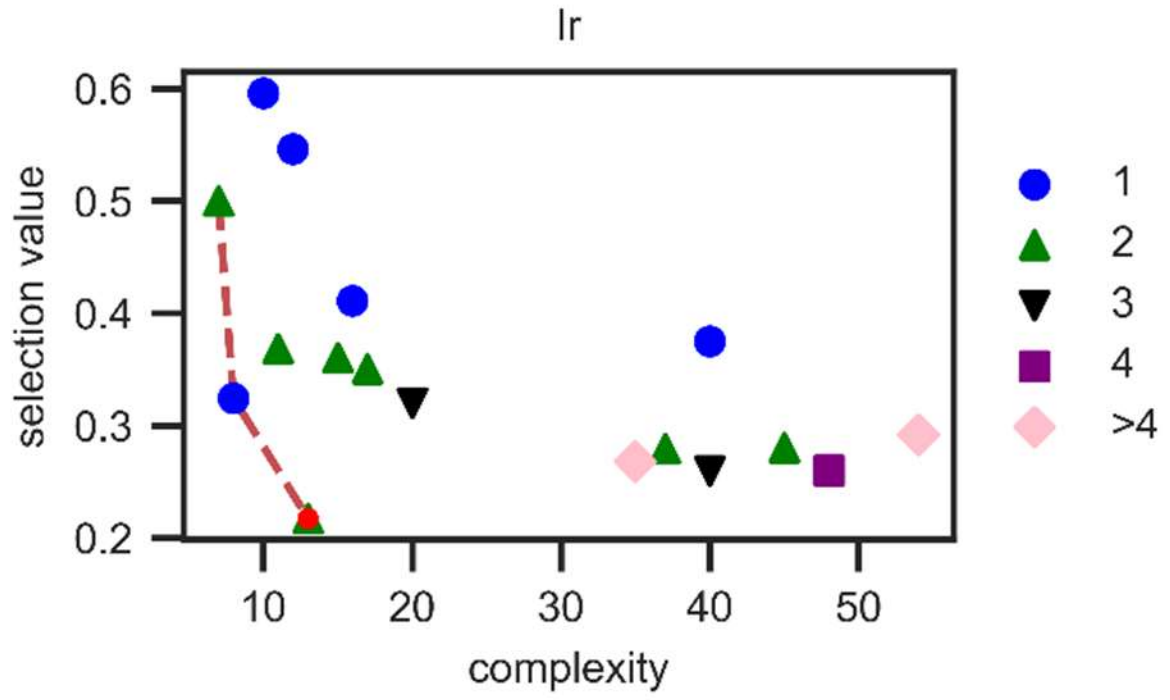

*SI-Figure 8. Convex hull of "selection value" (see methods) and complexity (i.e., number of nodes) for selecting the model GPn for Ir. The convex hull is shown as the red dashed line, and GPn is shown as a red dot. The legend indicates the speed (i.e., number of summations over neighbors) of each interatomic potential model.*

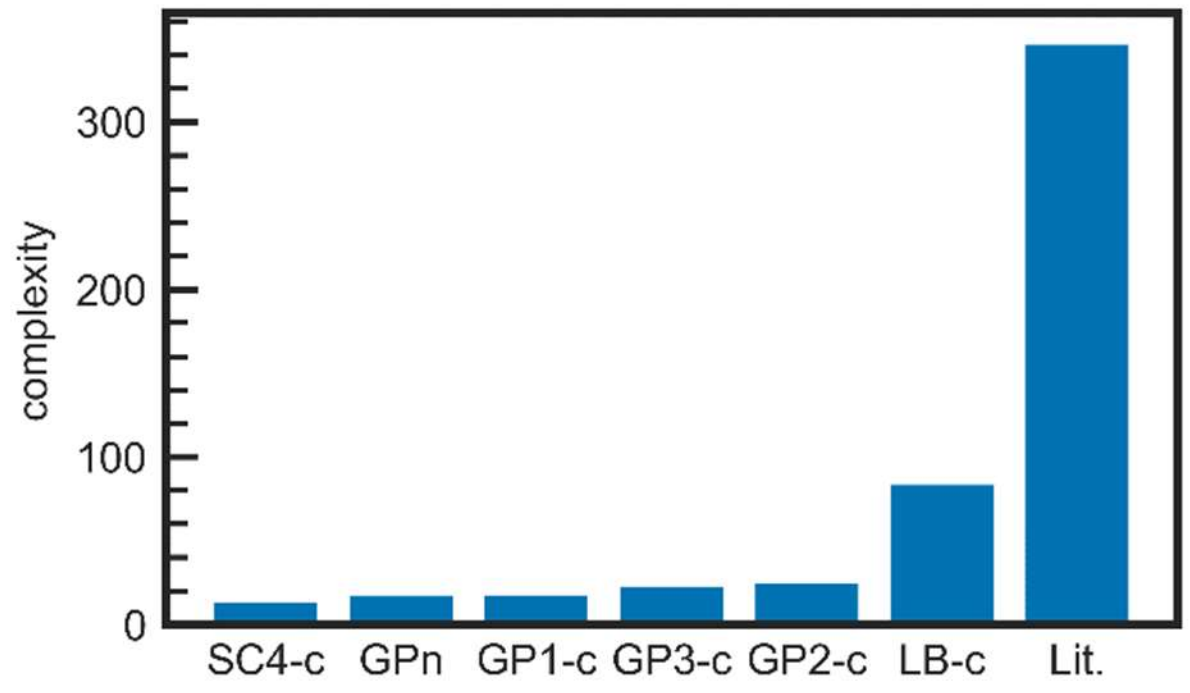

*SI-Figure 9. Number of nodes of each of the model types. The number of nodes of SC4-c and SC5-c are the same. The number of nodes of the literature and GPn models are averages.*

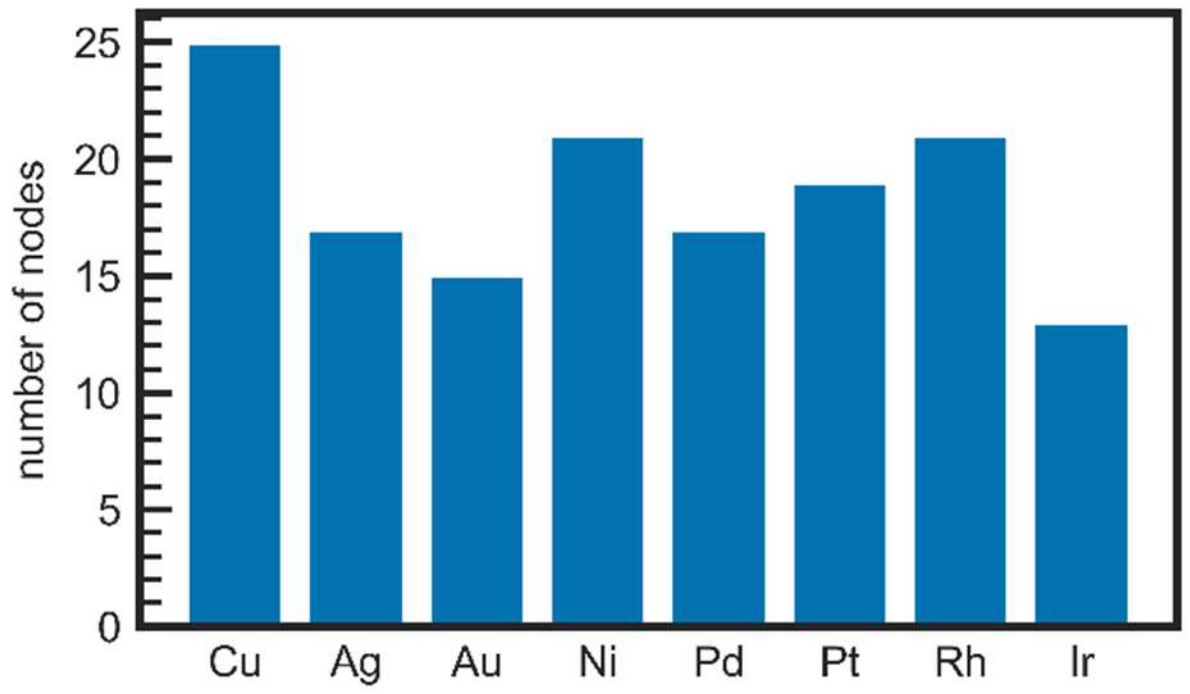

*SI-Figure 10. Complexity of GPn models. For reference, the complexity of the Sutton Chen models is 15.*

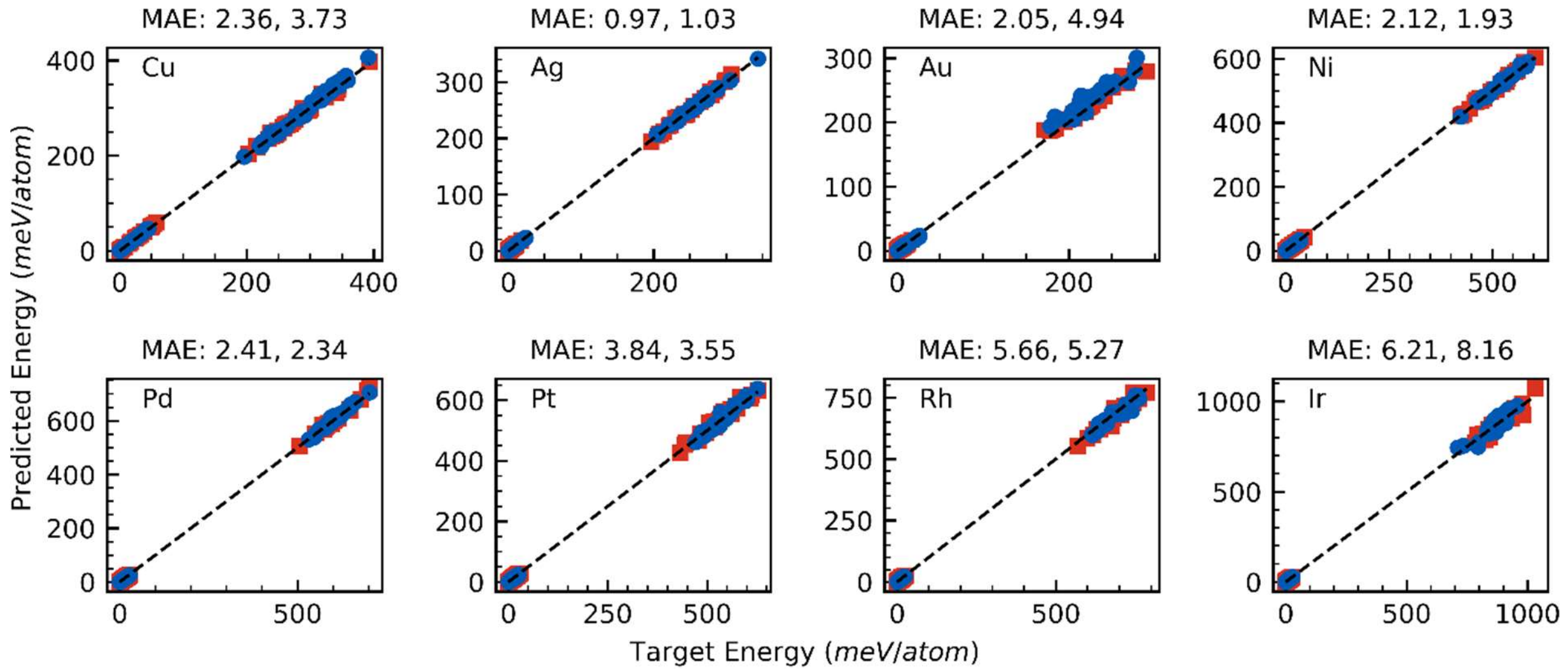

*SI-Figure 11. Parity plots of models discovered with POET (GPn) showing performance on energies. The values on the title of each subplot are the MAE on training and validation, respectively. Validation values are blue circles, and training values are red squares.*

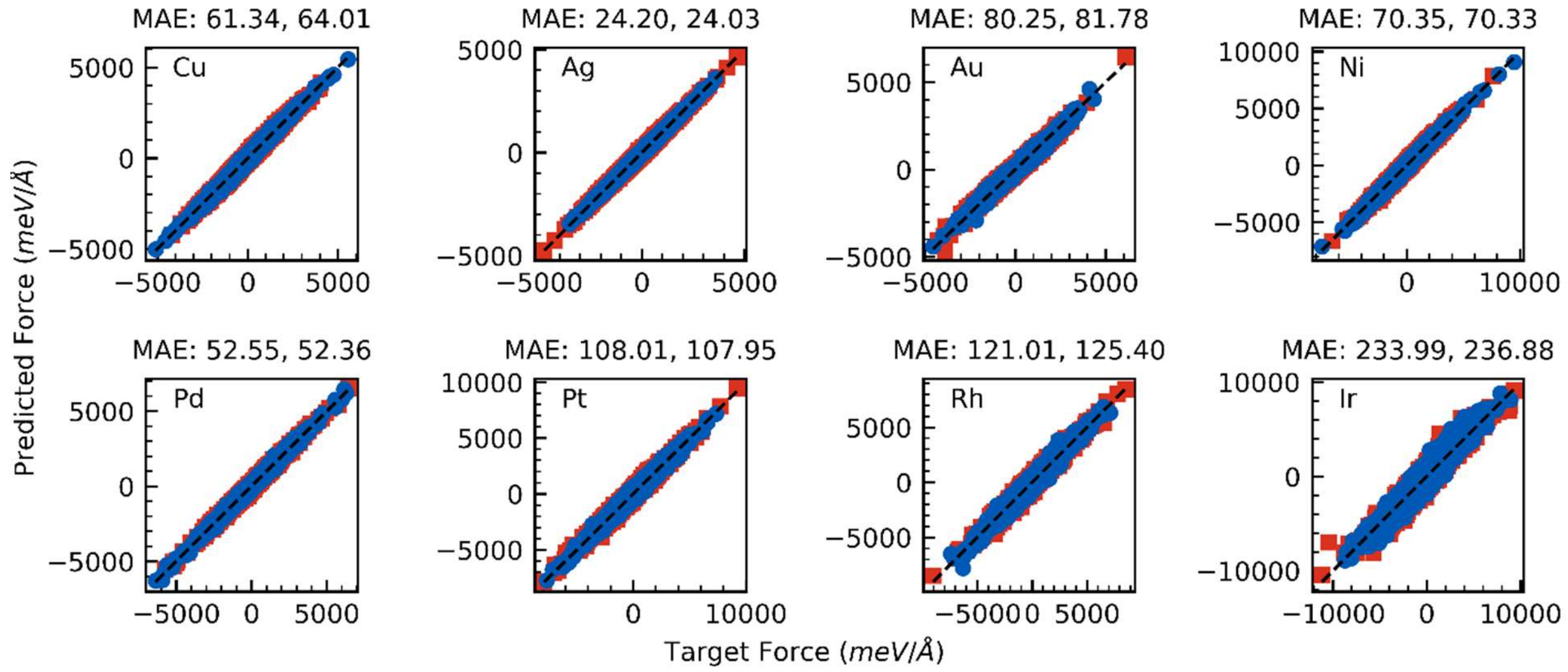

*SI-Figure 12. Parity plots of models discovered with POET (GPn) showing performance on forces. The values on the title of each subplot are the MAE on training and validation, respectively. Validation values are blue circles, and training values are red squares.*

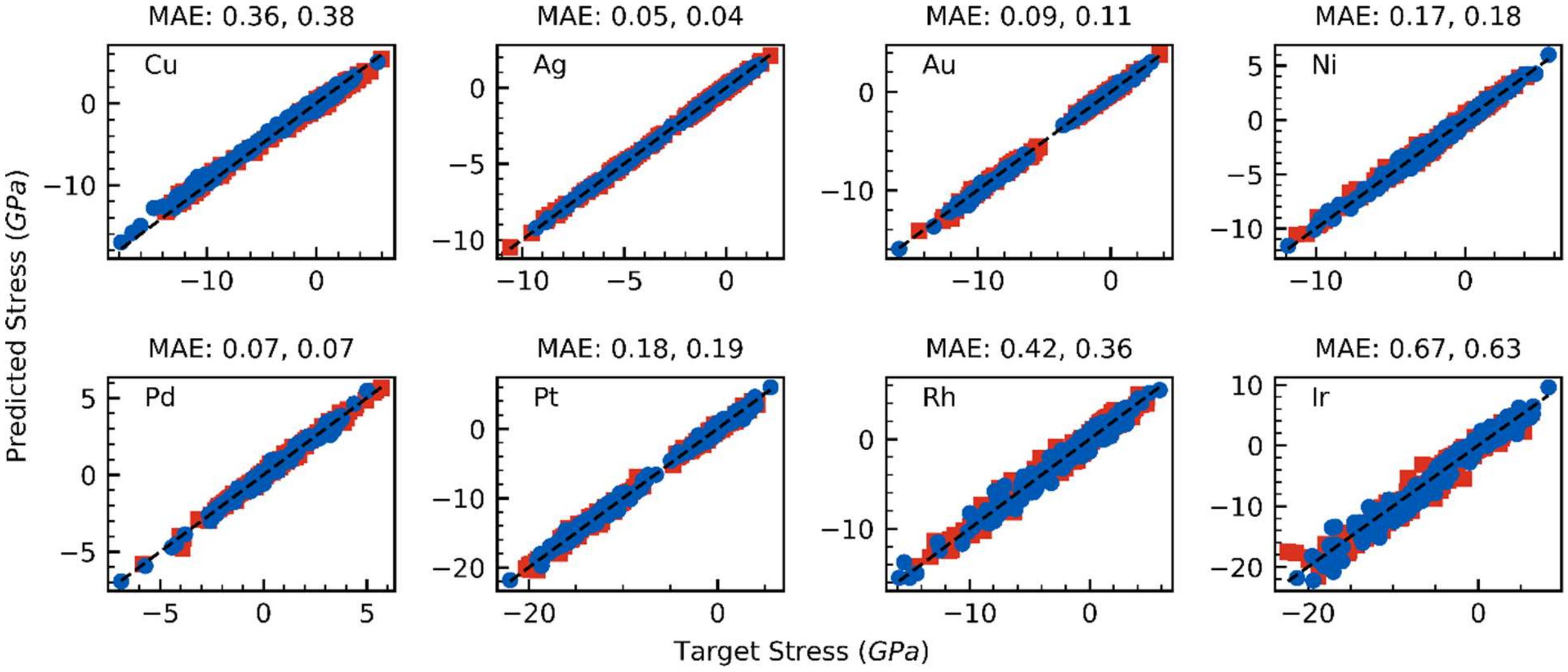

*SI-Figure 13. Parity plots of models discovered with POET (GPn) showing performance on stresses. The values on the title of each subplot are the MAE on training and validation, respectively. Validation values are blue circles, and training values are red squares.*

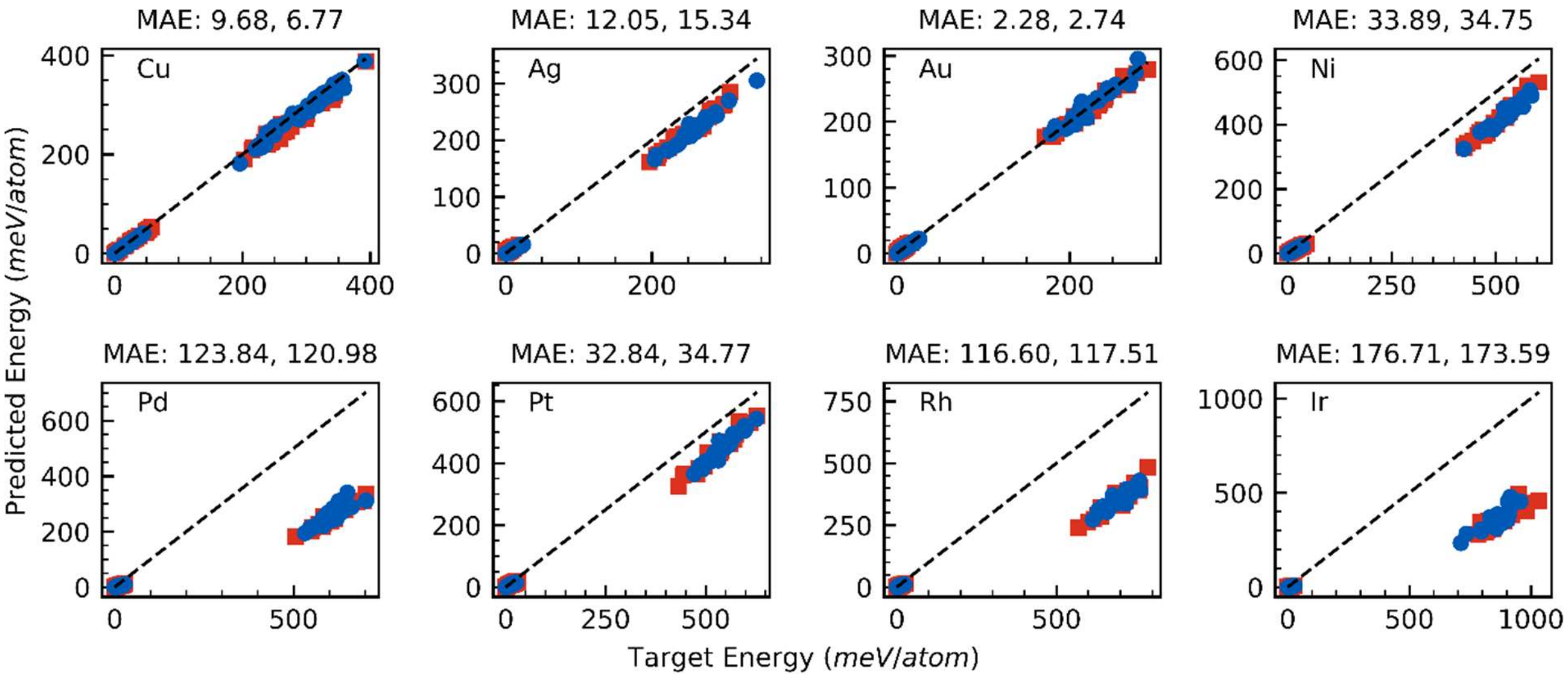

*SI-Figure 14. Parity plots of models discovered with SC-4 showing performance on energies. The values on the title of each subplot are the MAE on training and validation, respectively. Validation values are blue circles, and training values are red squares.*

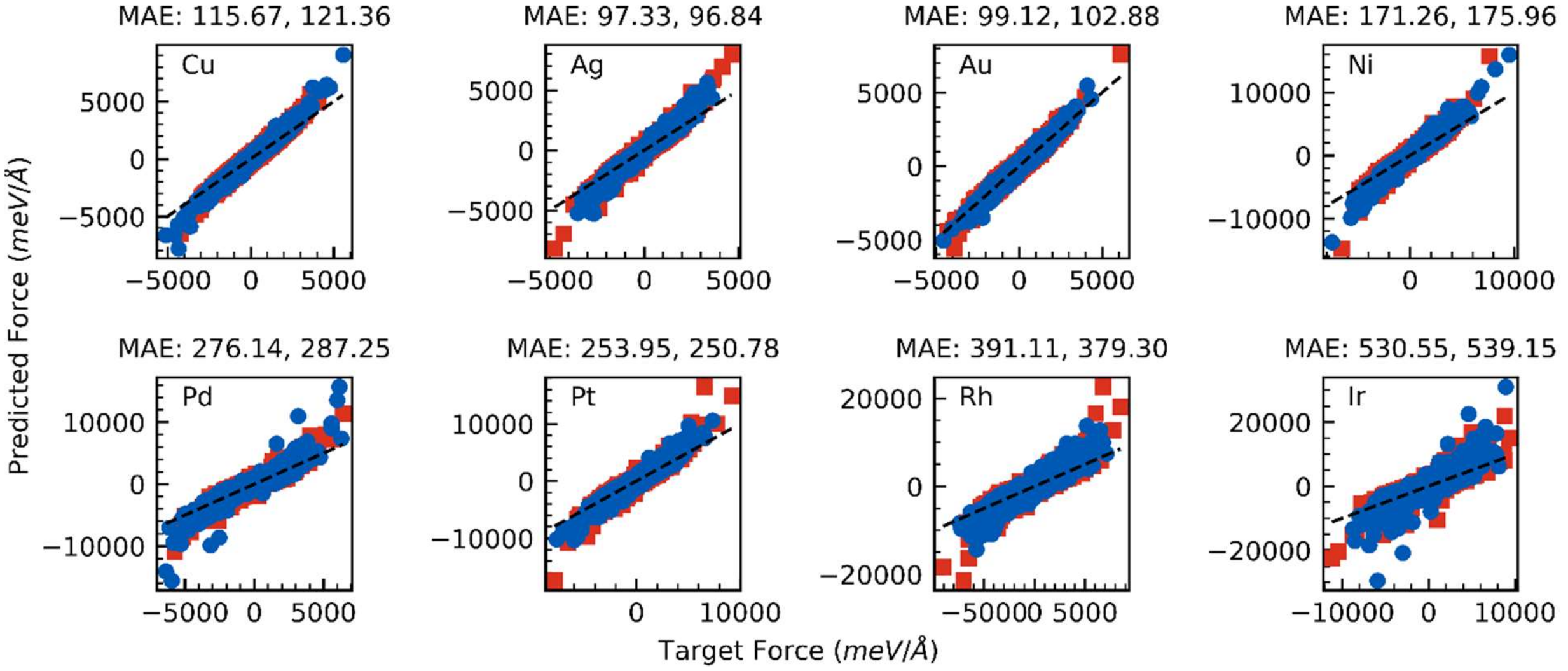

*SI-Figure 15. Parity plots of models discovered with SC-4 showing performance on forces. The values on the title of each subplot are the MAE on training and validation, respectively. Validation values are blue circles, and training values are red squares.*

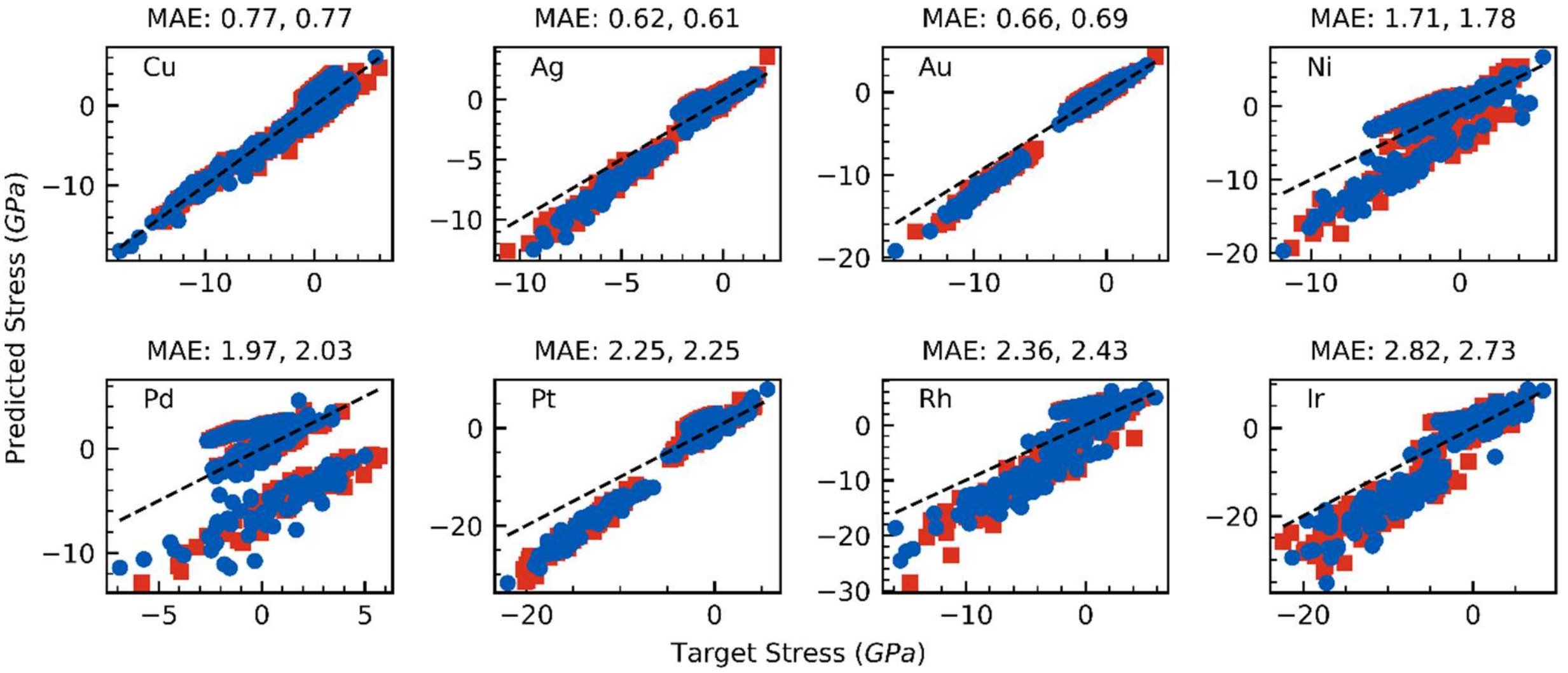

*SI-Figure 16. Parity plots of models discovered with SC-4 showing performance on stresses. The values on the title of each subplot are the MAE on training and validation, respectively. Validation values are blue circles, and training values are red squares.*

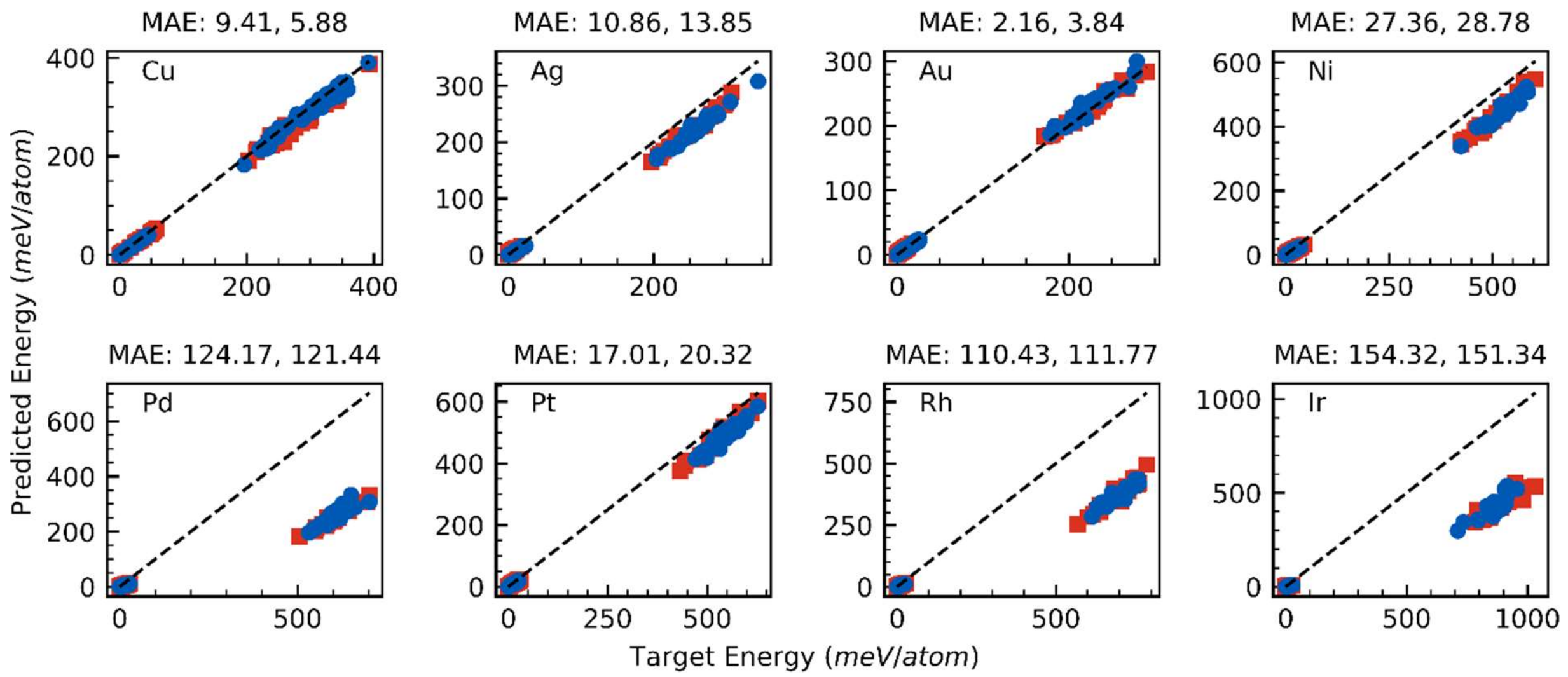

*SI-Figure 17. Parity plots of models discovered with SC-5 showing performance on energies. The values on the title of each subplot are the MAE on training and validation, respectively. Validation values are blue circles, and training values are red squares.*

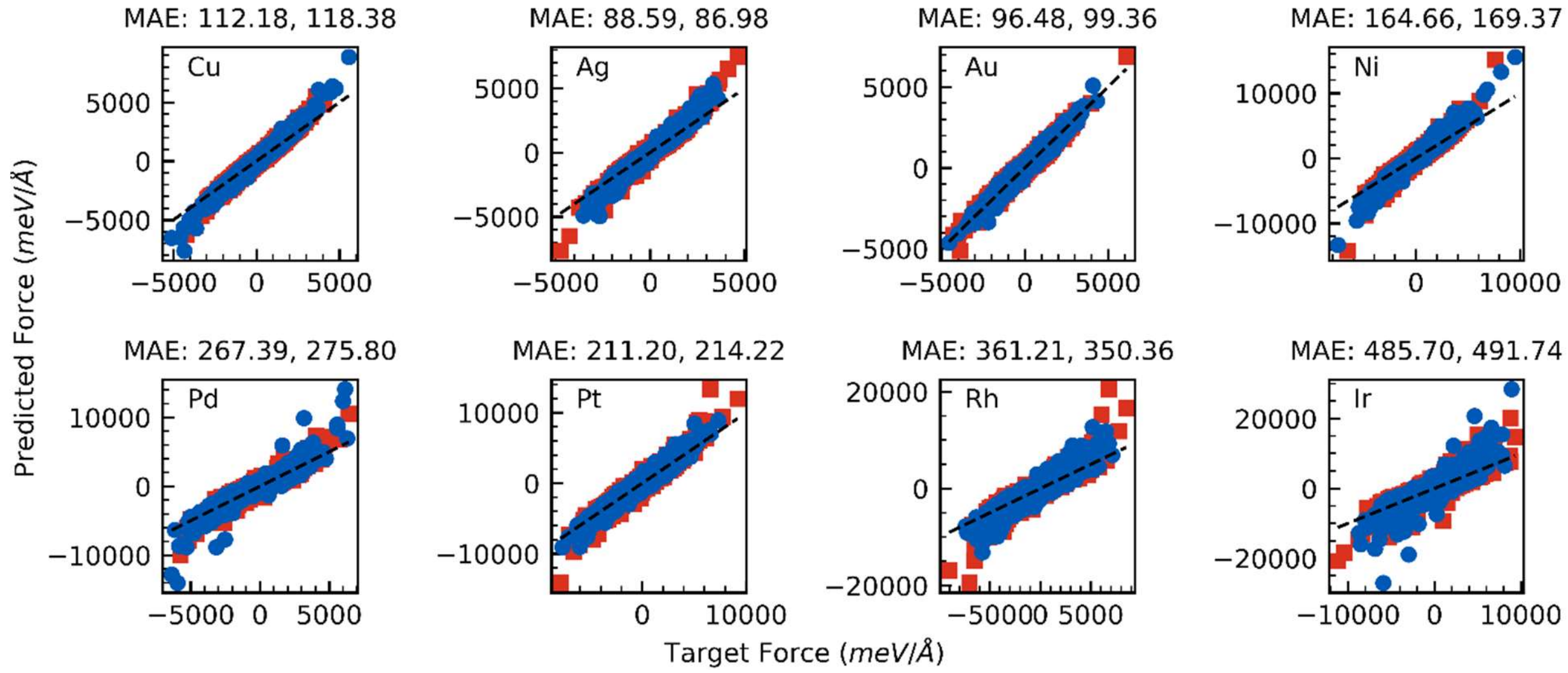

*SI-Figure 18. Parity plots of models discovered with SC-5 showing performance on forces. The values on the title of each subplot are the MAE on training and validation, respectively. Validation values are blue circles, and training values are red squares.*

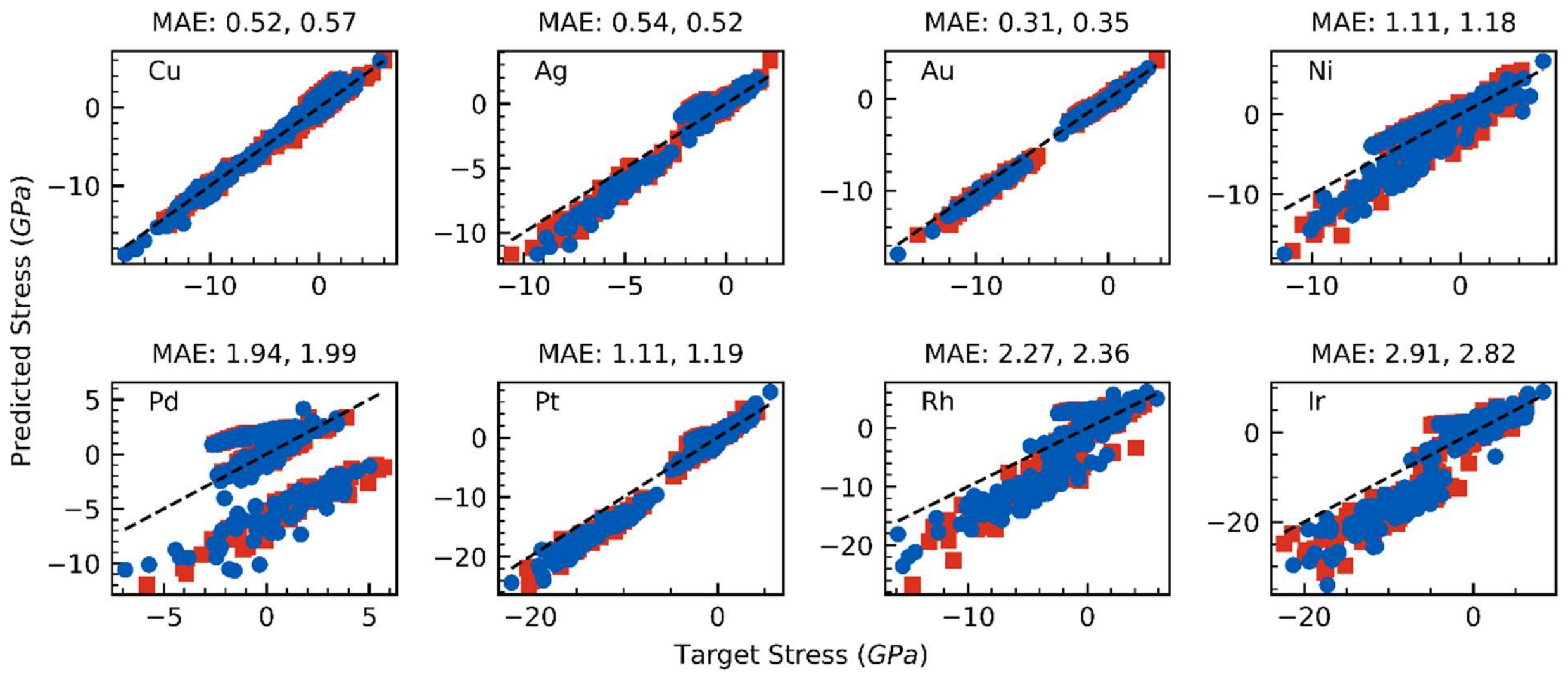

*SI-Figure 19. Parity plots of models discovered with SC-5 showing performance on stresses. The values on the title of each subplot are the MAE on training and validation, respectively. Validation values are blue circles, and training values are red squares.*

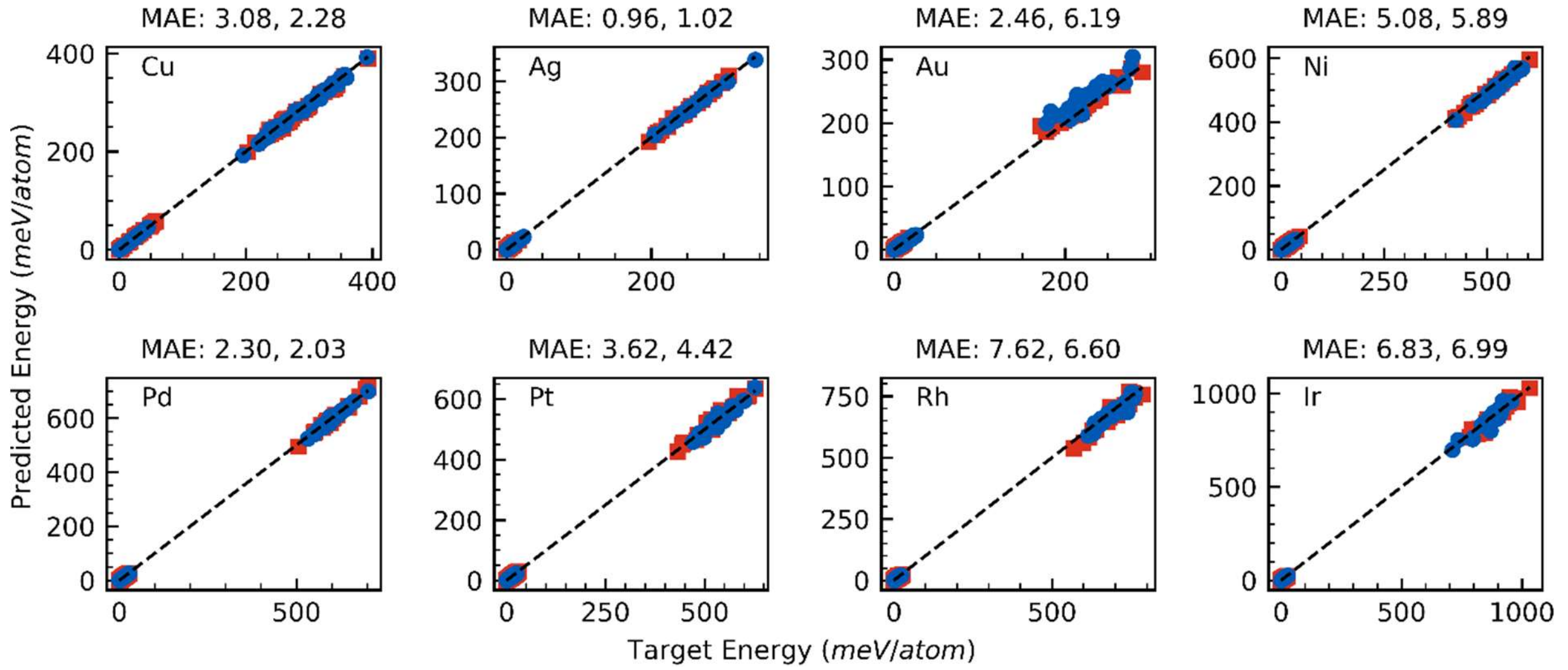

*SI-Figure 20. Parity plots of models discovered with GP1-c showing performance on energies. The values on the title of each subplot are the MAE on training and validation, respectively. Validation values are blue circles, and training values are red squares.*

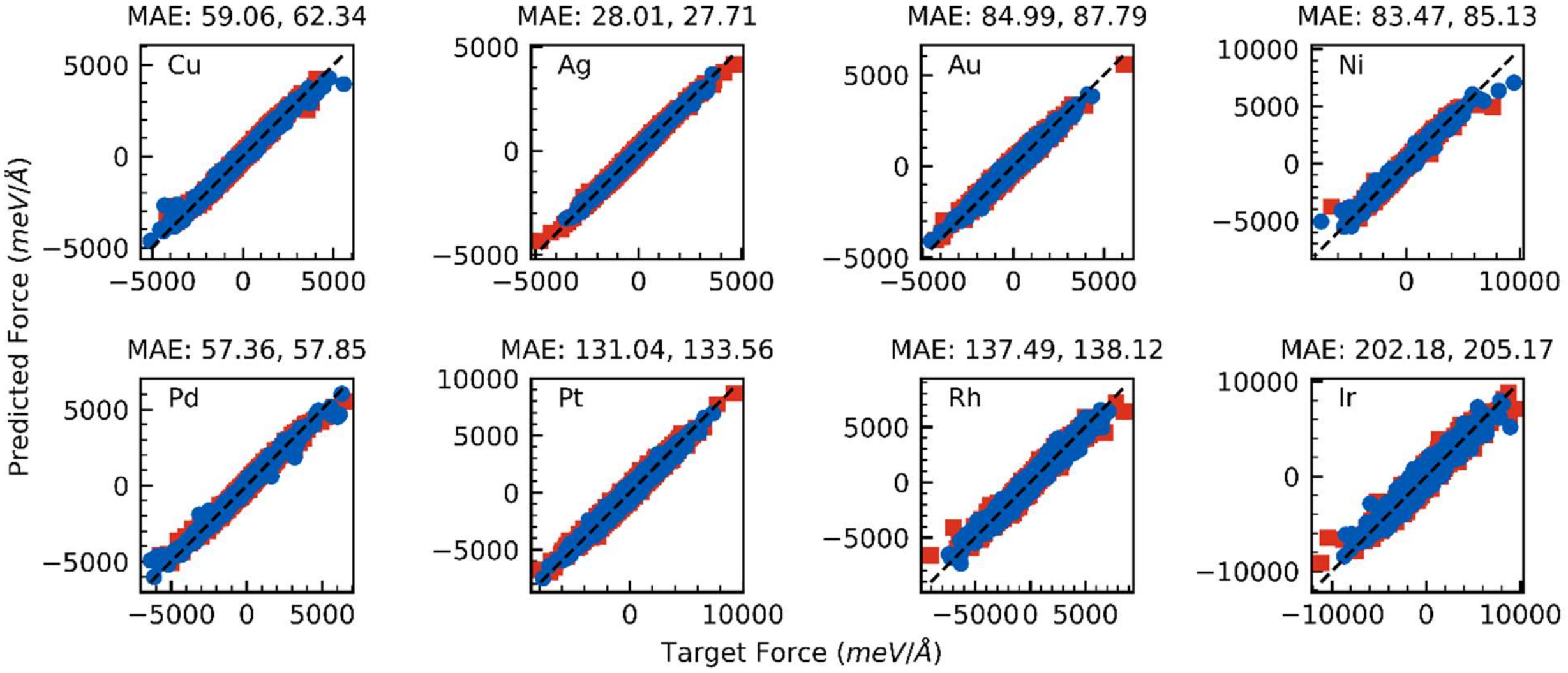

*SI-Figure 21. Parity plots of models discovered with GP1-c showing performance on forces. The values on the title of each subplot are the MAE on training and validation, respectively. Validation values are blue circles, and training values are red squares.*

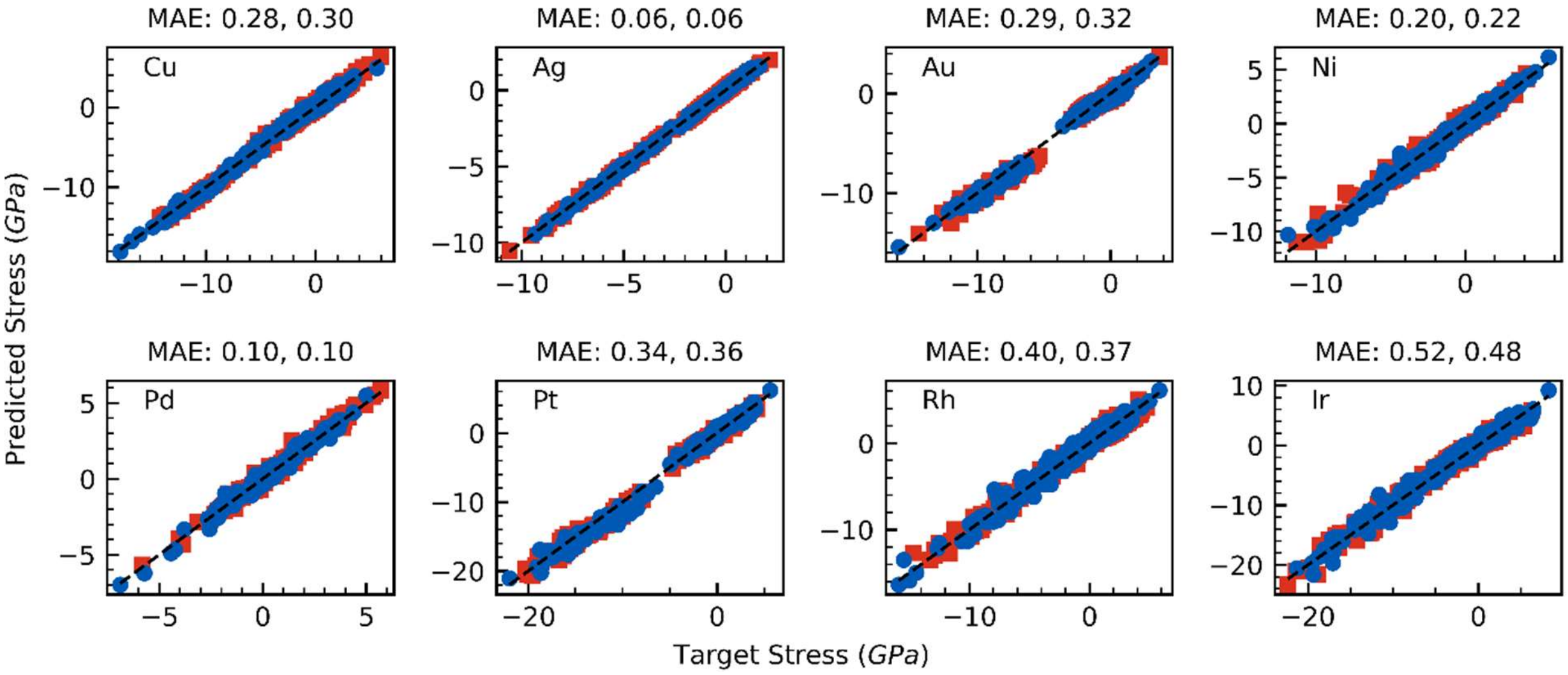

*SI-Figure 22. Parity plots of models discovered with GP1-c showing performance on stresses. The values on the title of each subplot are the MAE on training and validation, respectively. Validation values are blue circles, and training values are red squares.*

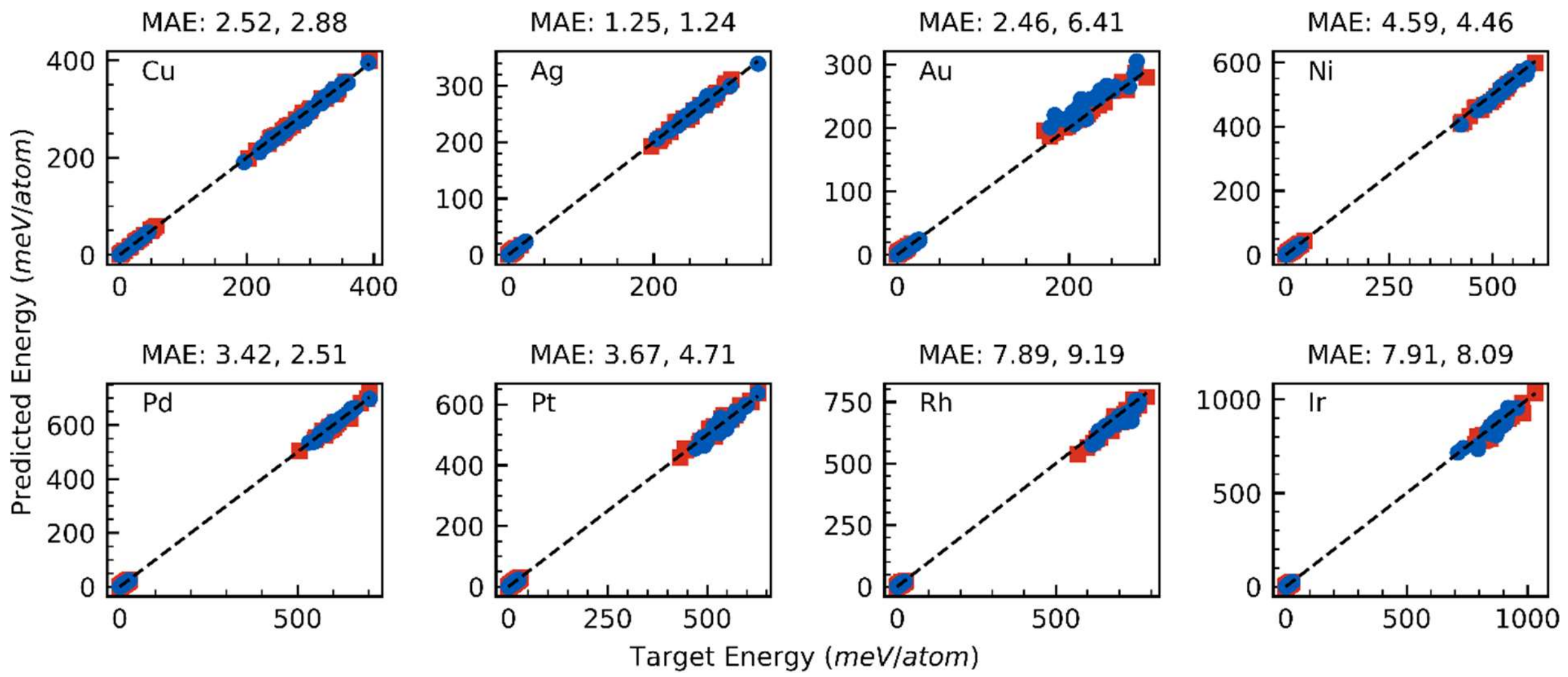

*SI-Figure 23. Parity plots of models discovered with GP2-c showing performance on energies. The values on the title of each subplot are the MAE on training and validation, respectively. Validation values are blue circles, and training values are red squares.*

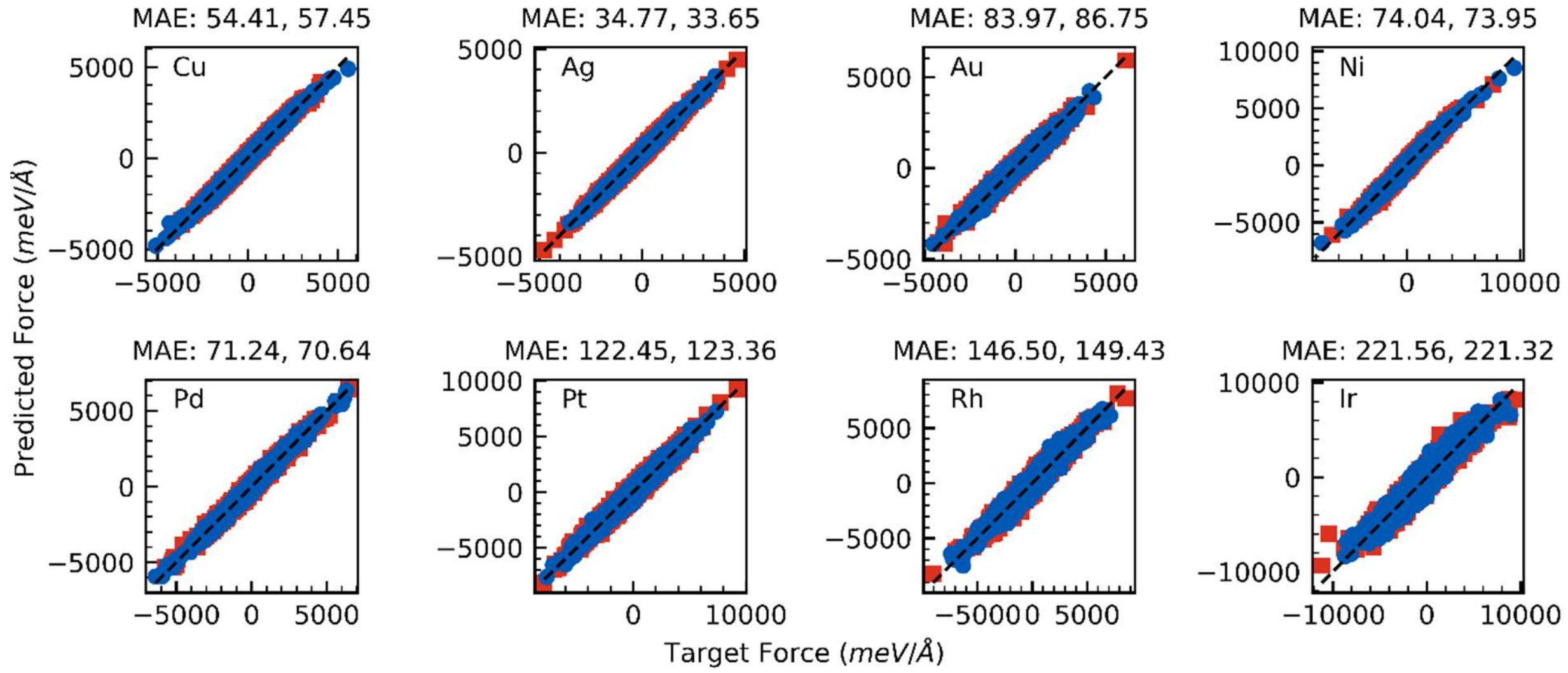

*SI-Figure 24. Parity plots of models discovered with GP2-c showing performance on forces. The values on the title of each subplot are the MAE on training and validation, respectively. Validation values are blue circles, and training values are red squares.*

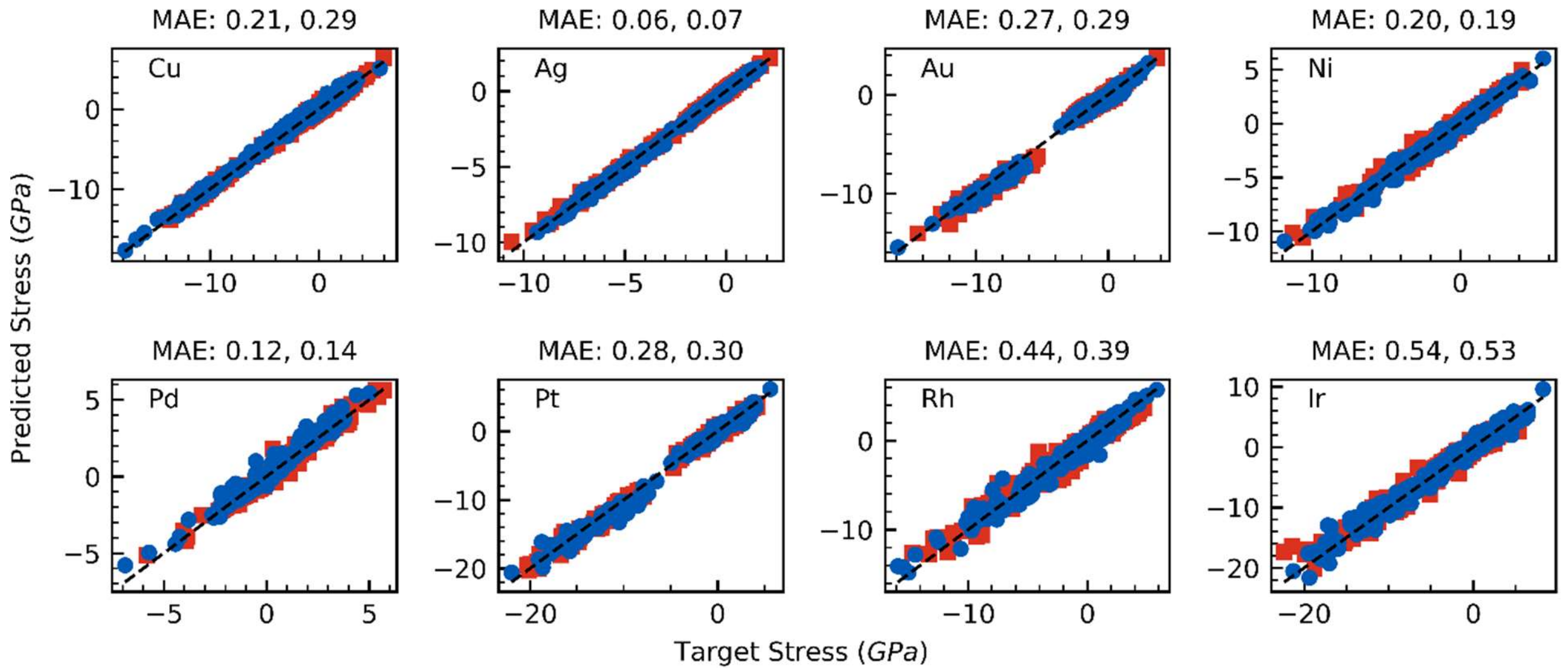

*SI-Figure 25. Parity plots of models discovered with GP2-c showing performance on stresses. The values on the title of each subplot are the MAE on training and validation, respectively. Validation values are blue circles, and training values are red squares.*

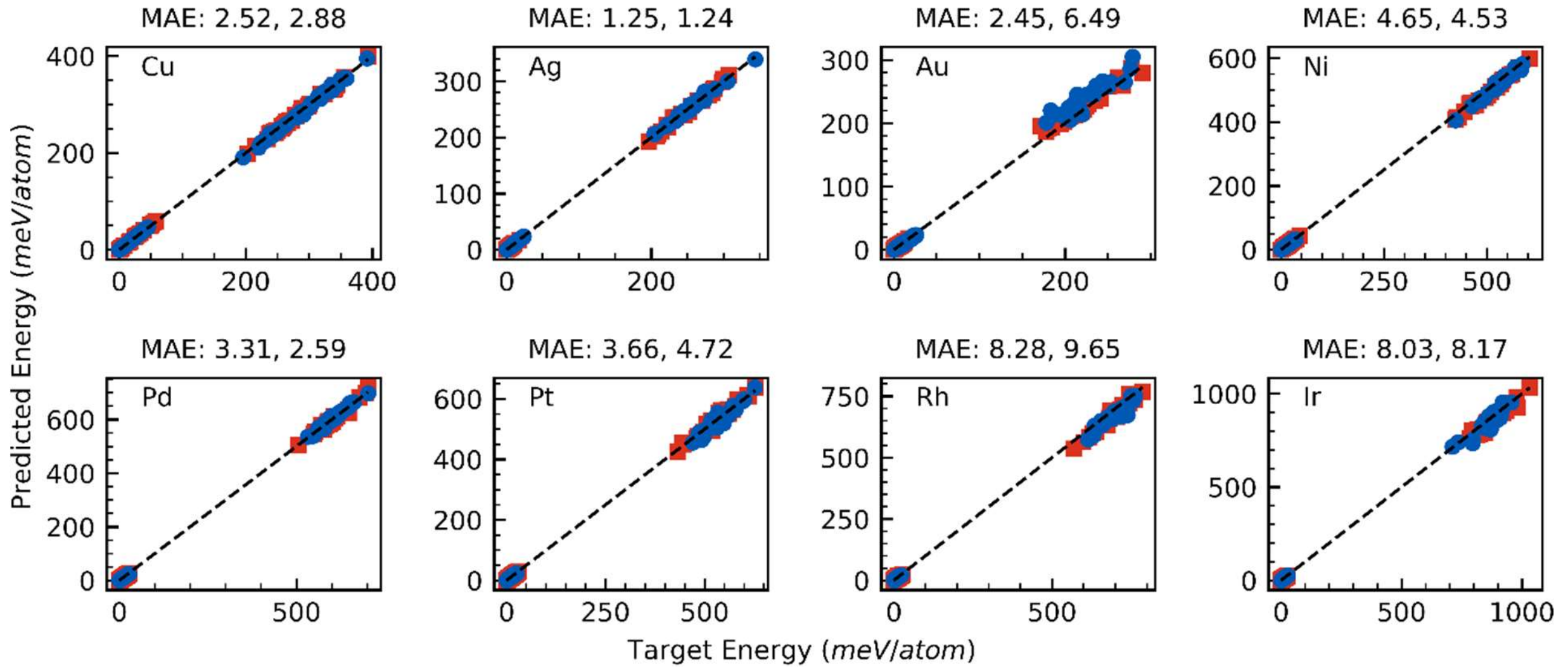

*SI-Figure 26. Parity plots of models discovered with GP3-c showing performance on energies. The values on the title of each subplot are the MAE on training and validation, respectively. Validation values are blue circles, and training values are red squares.*

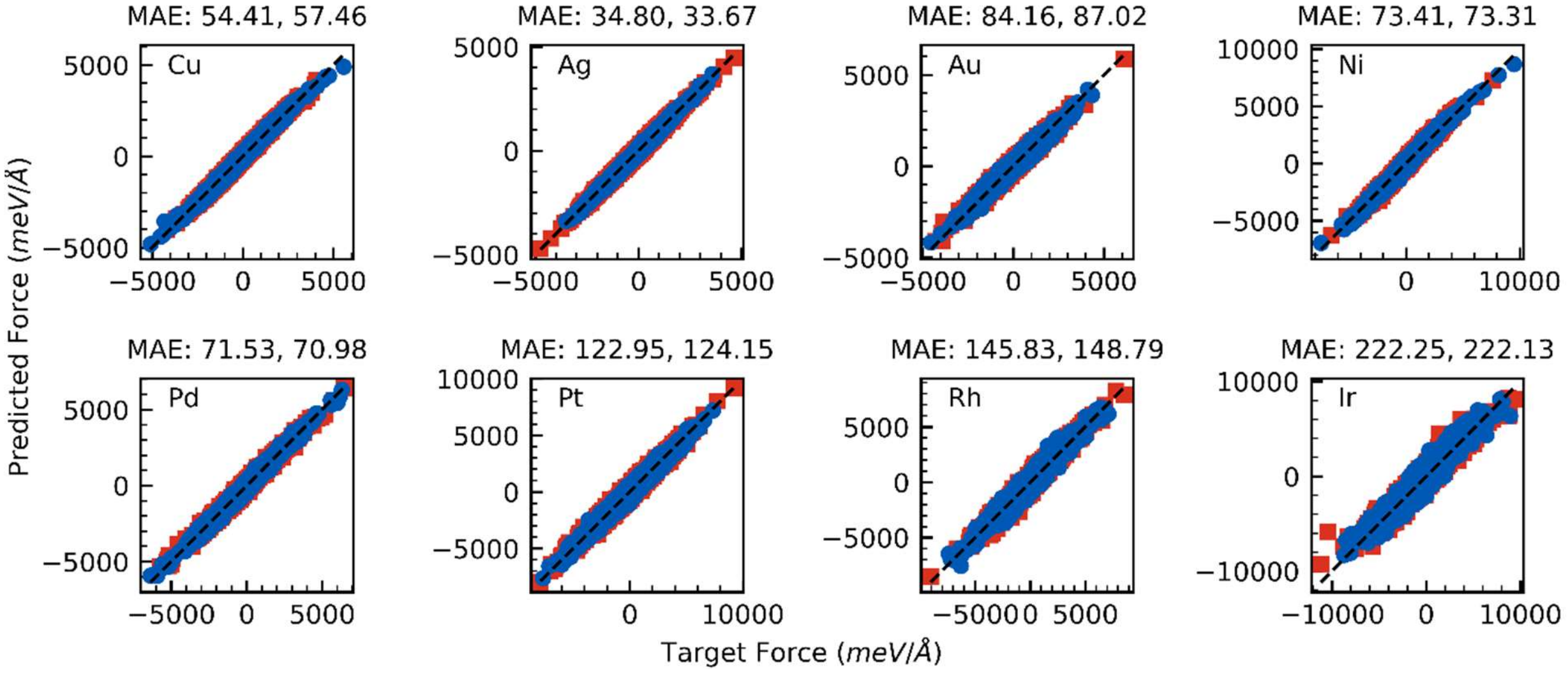

*SI-Figure 27. Parity plots of models discovered with GP3-c showing performance on forces. The values on the title of each subplot are the MAE on training and validation, respectively. Validation values are blue circles, and training values are red squares.*

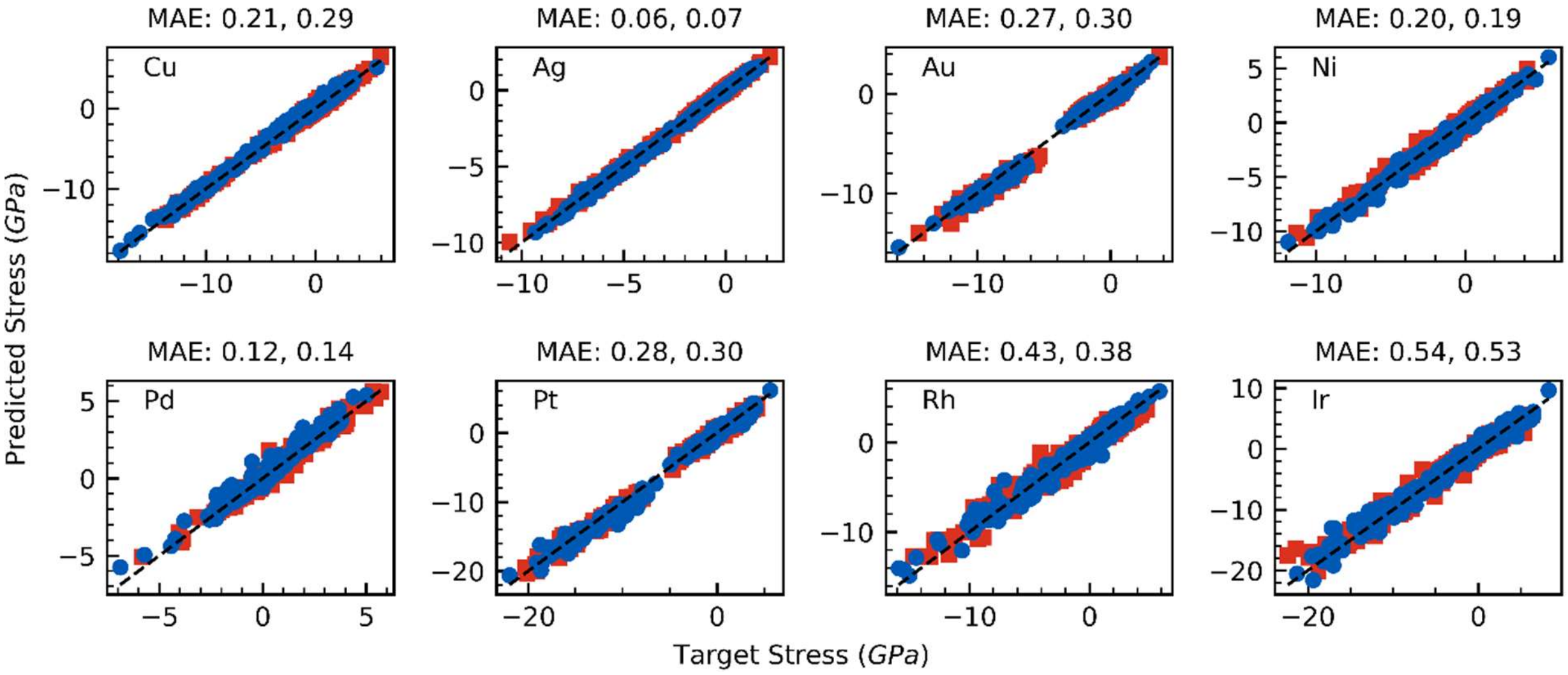

*SI-Figure 28. Parity plots of models discovered with GP3-c showing performance on stresses. The values on the title of each subplot are the MAE on training and validation, respectively. Validation values are blue circles, and training values are red squares.*

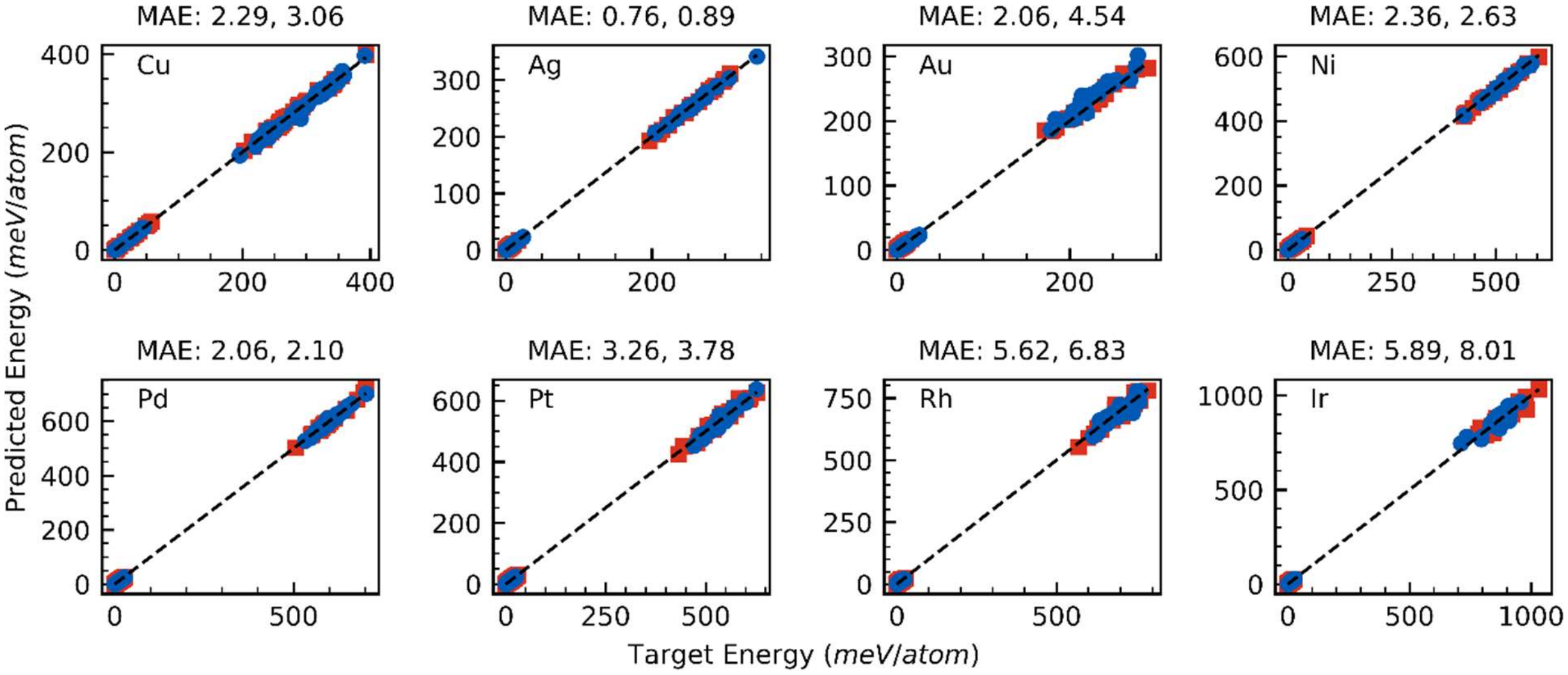

*SI-Figure 29. Parity plots of models discovered with LB-c showing performance on energies. The values on the title of each subplot are the MAE on training and validation, respectively. Validation values are blue circles, and training values are red squares.*

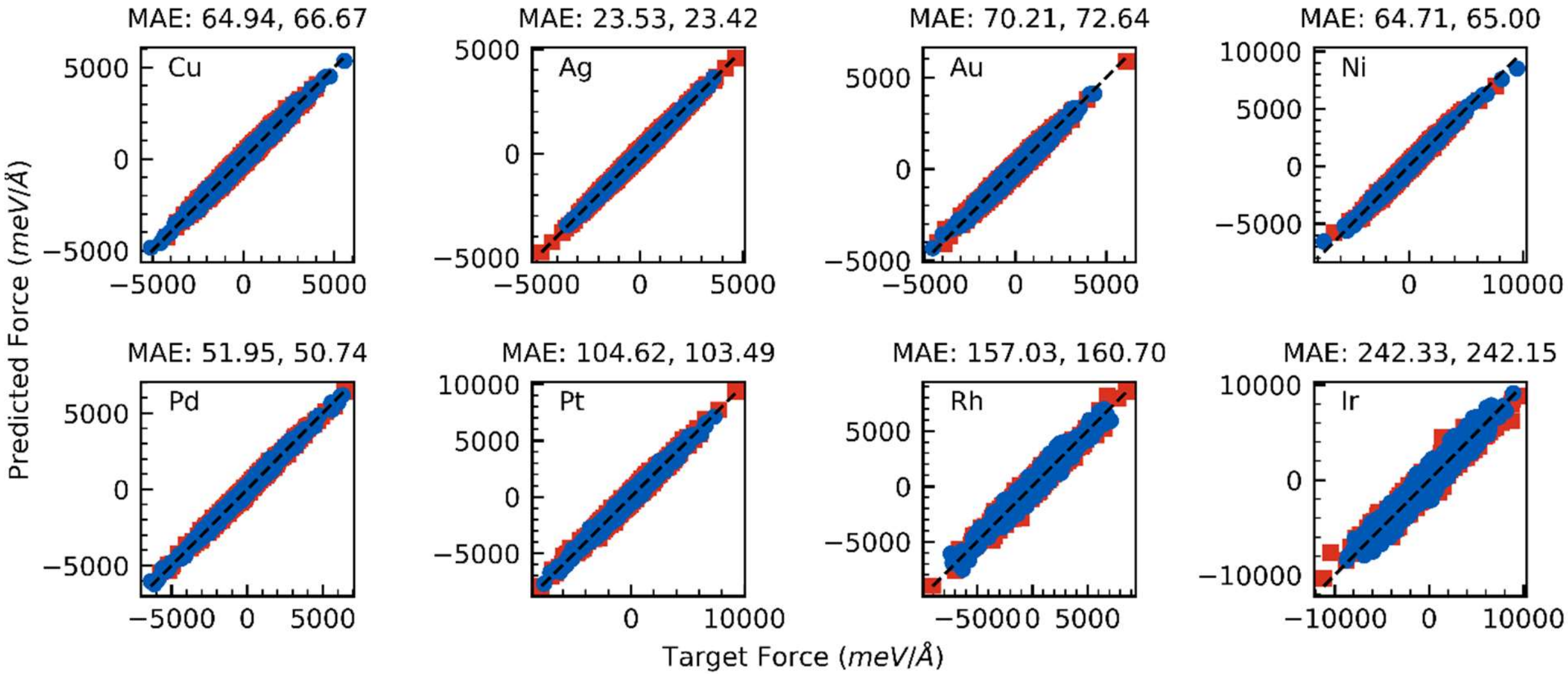

*SI-Figure 30. Parity plots of models discovered with LB-c showing performance on forces. The values on the title of each subplot are the MAE on training and validation, respectively. Validation values are blue circles, and training values are red squares.*

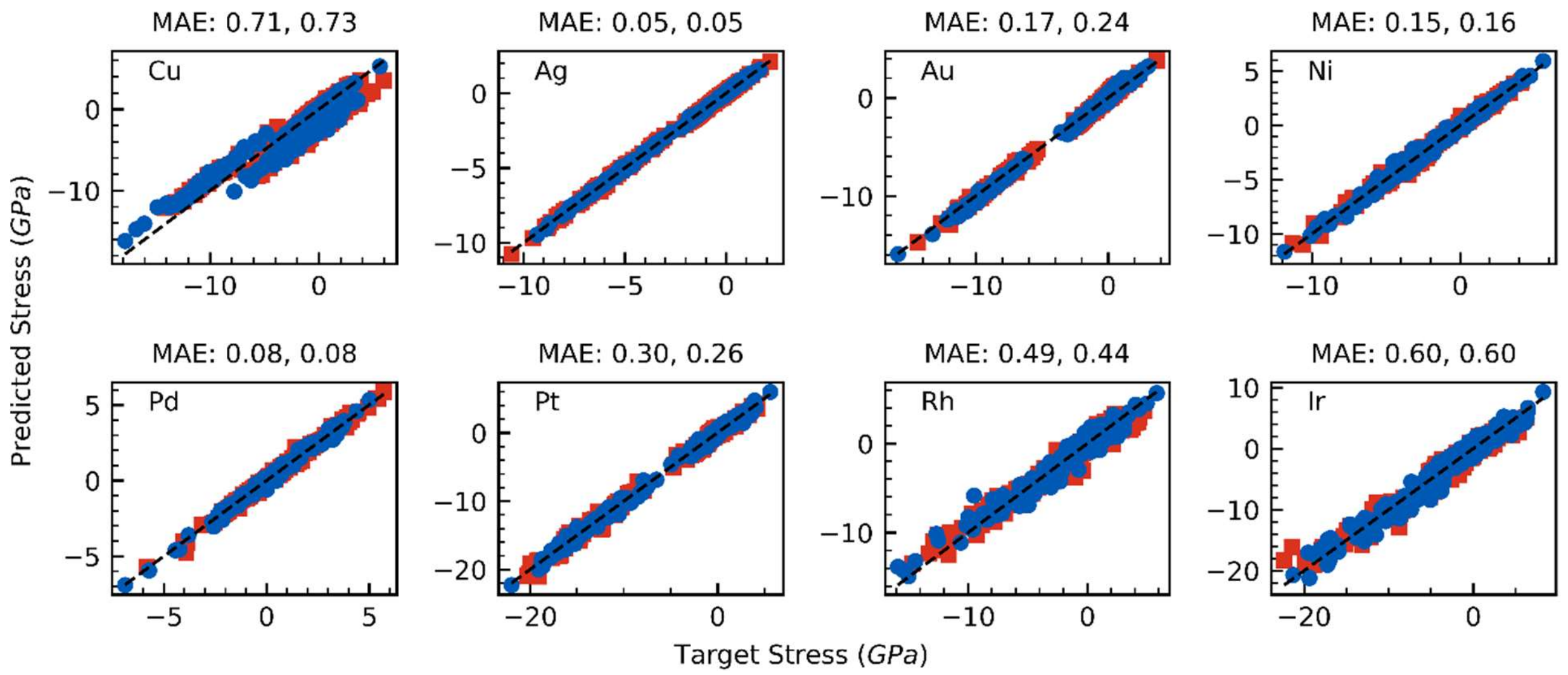

*SI-Figure 31. Parity plots of models discovered with LB-c showing performance on stresses. The values on the title of each subplot are the MAE on training and validation, respectively. Validation values are blue circles, and training values are red squares.*

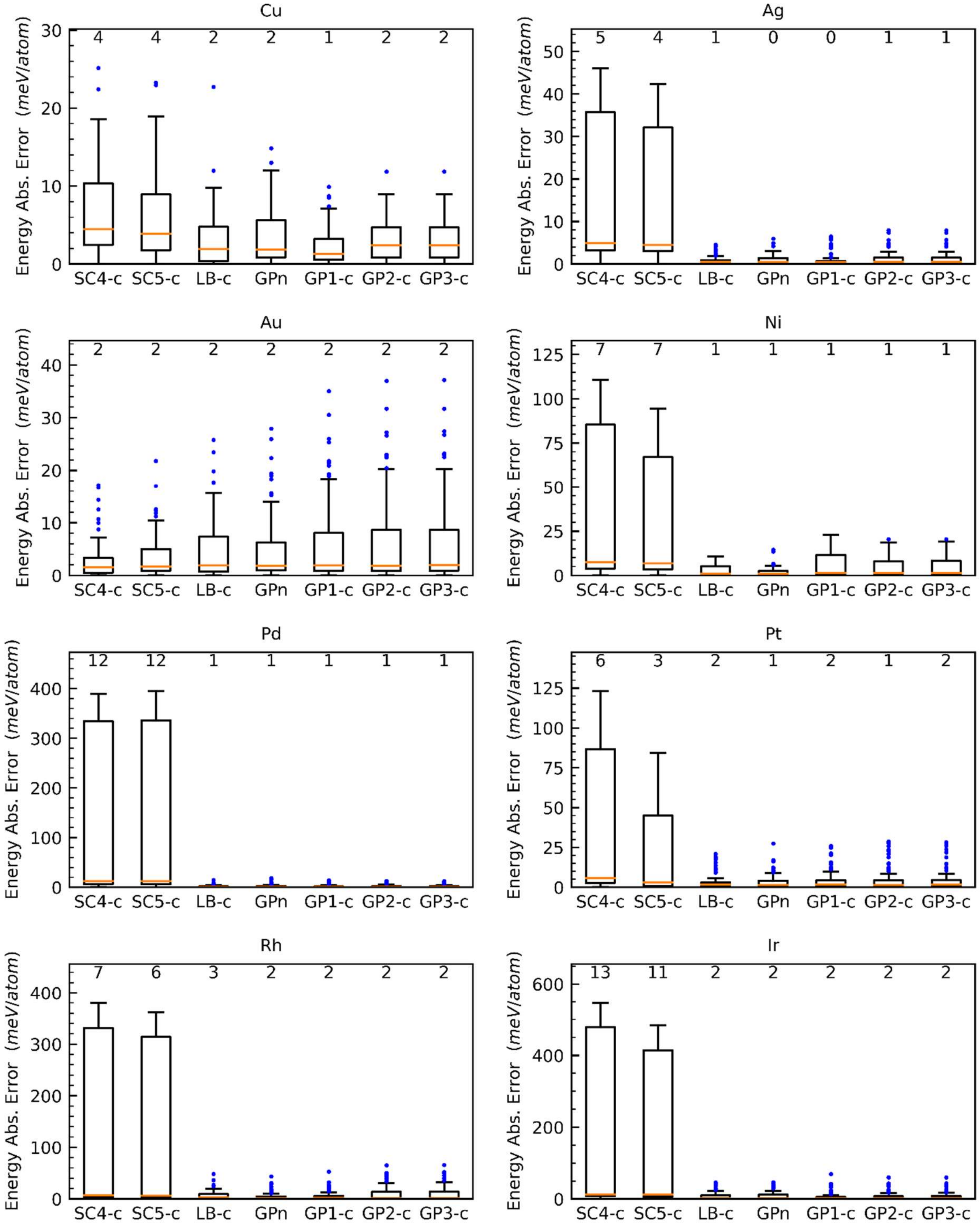

*SI-Figure 32. Boxplot of absolute errors on energies of models for each element*

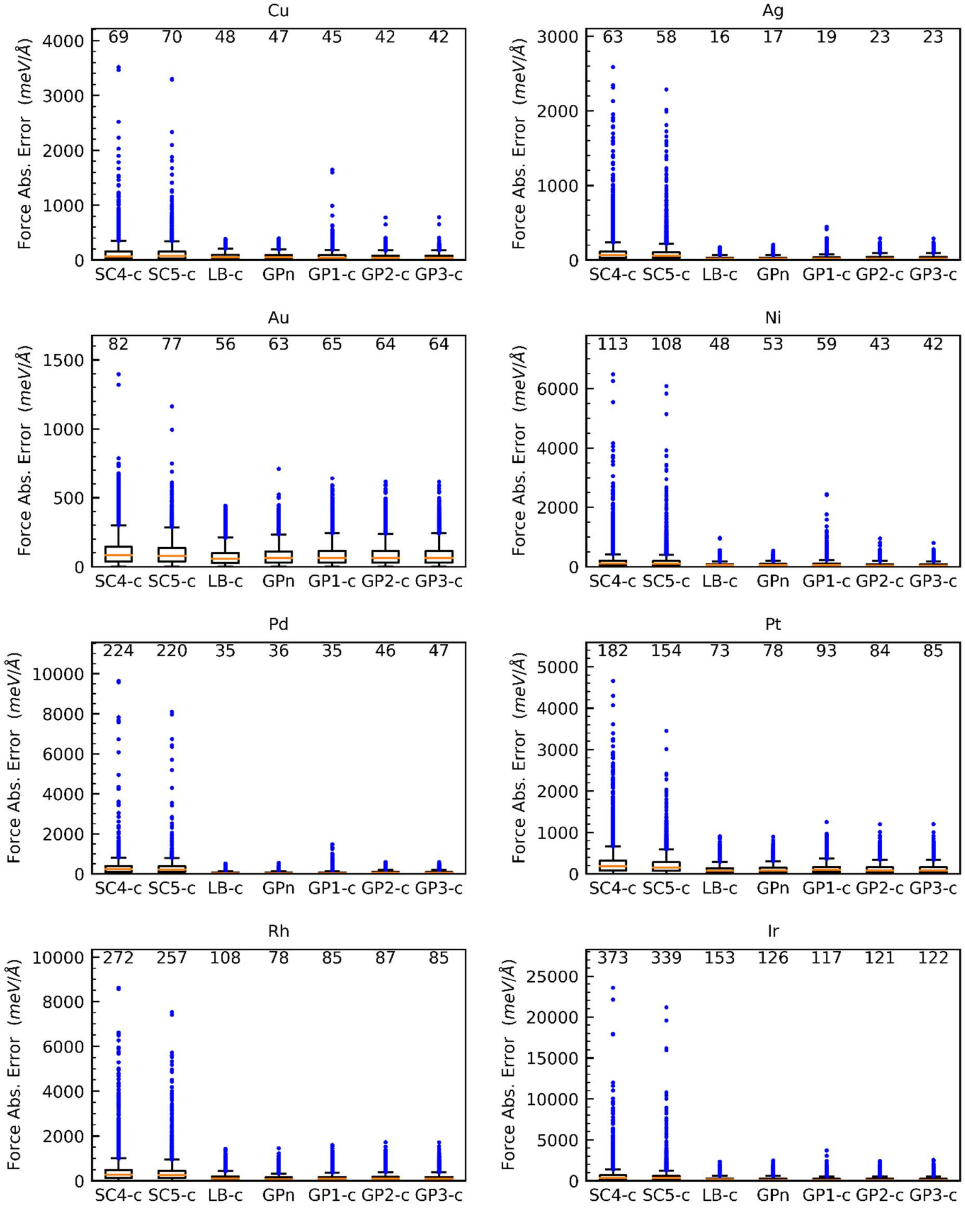

*SI-Figure 33. Boxplot of absolute errors on forces of models for each element*

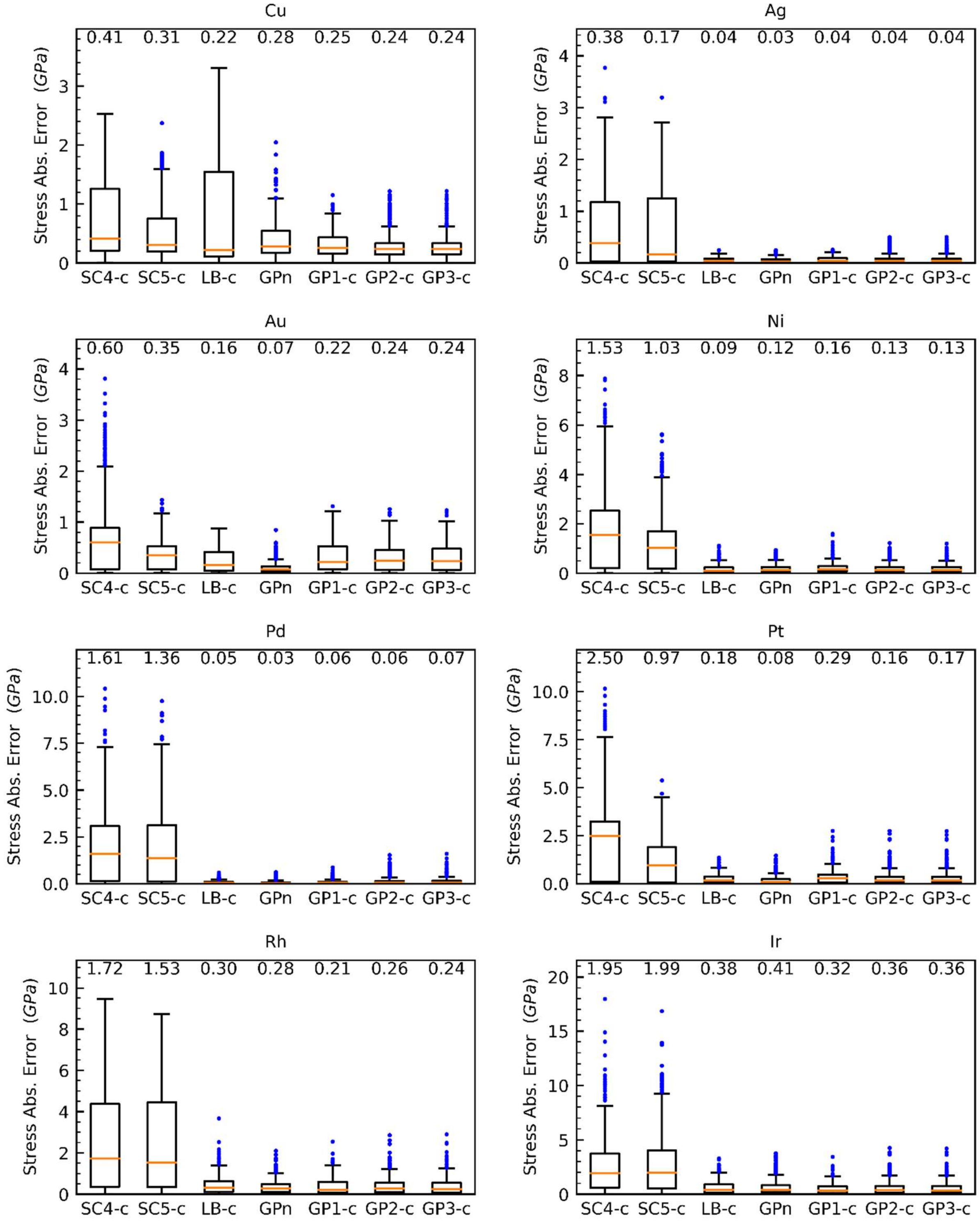

*SI-Figure 34. Boxplot of absolute errors on stresses of models for each element*

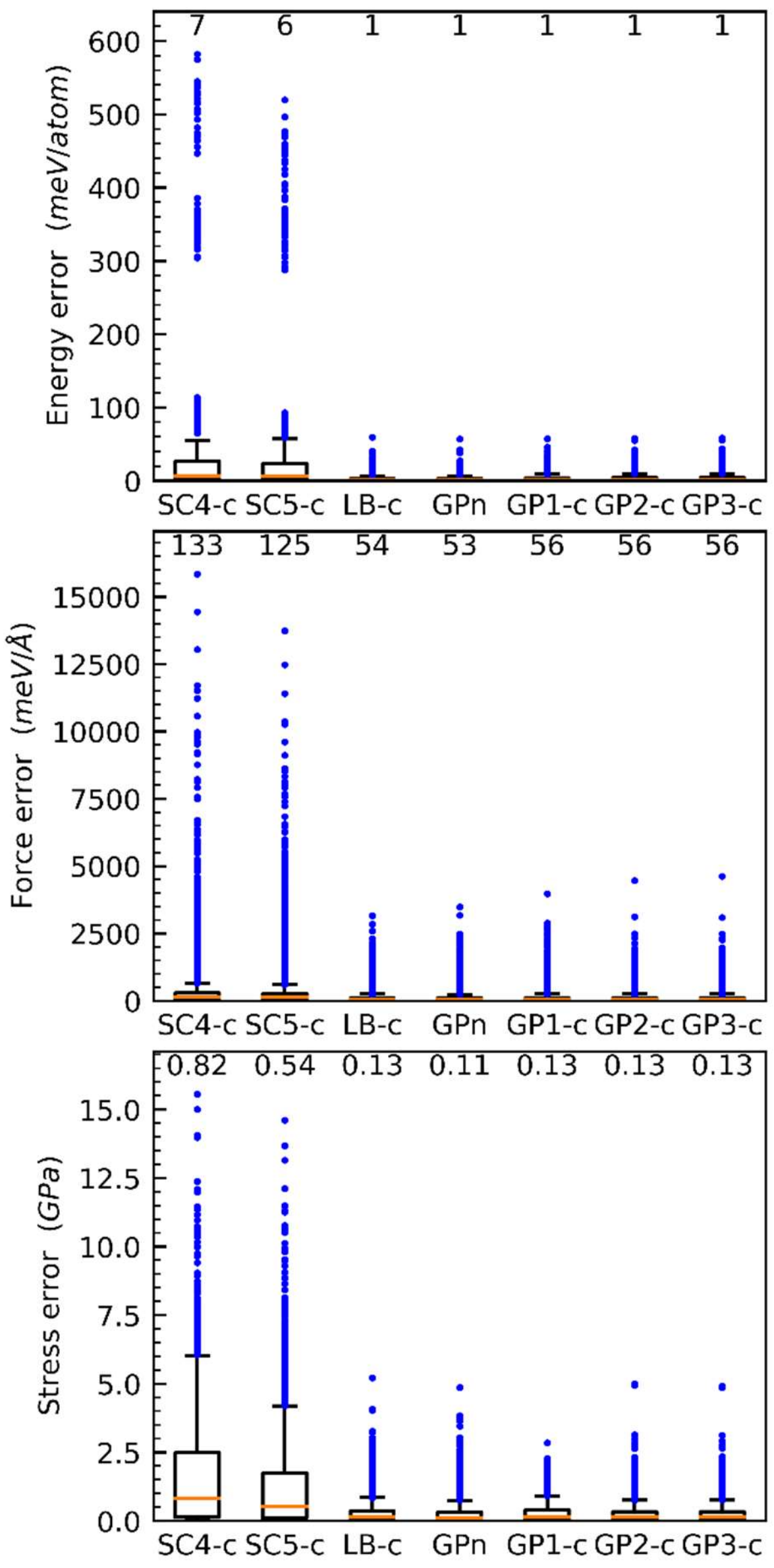

*SI-Figure 35. Boxplot of absolute errors of models across Cu, Ag, Au, Ni, Pd, Pt, Rh, and Ir*

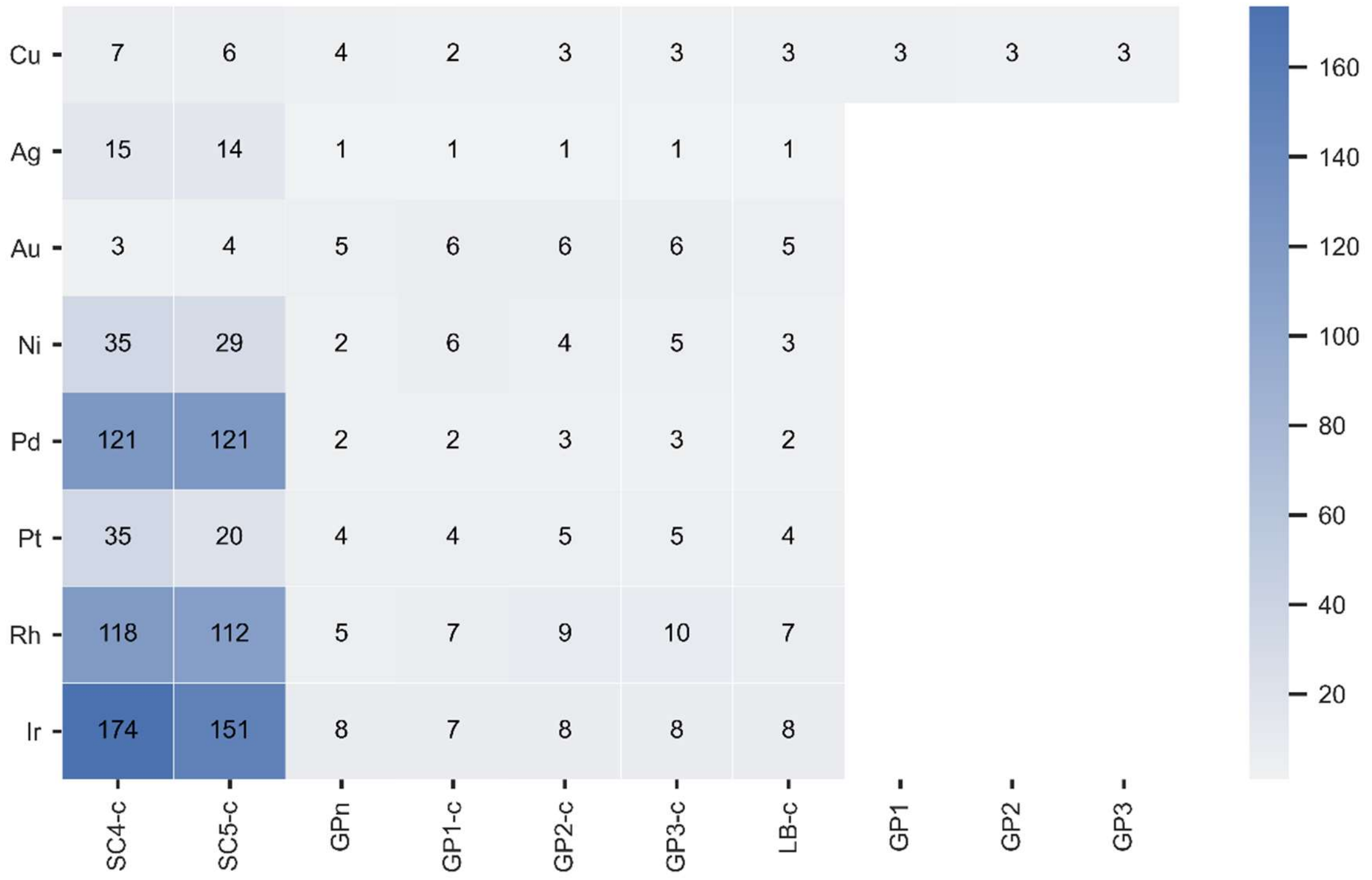

*SI-Figure 36. Energy MAE on validation in meV/atom*

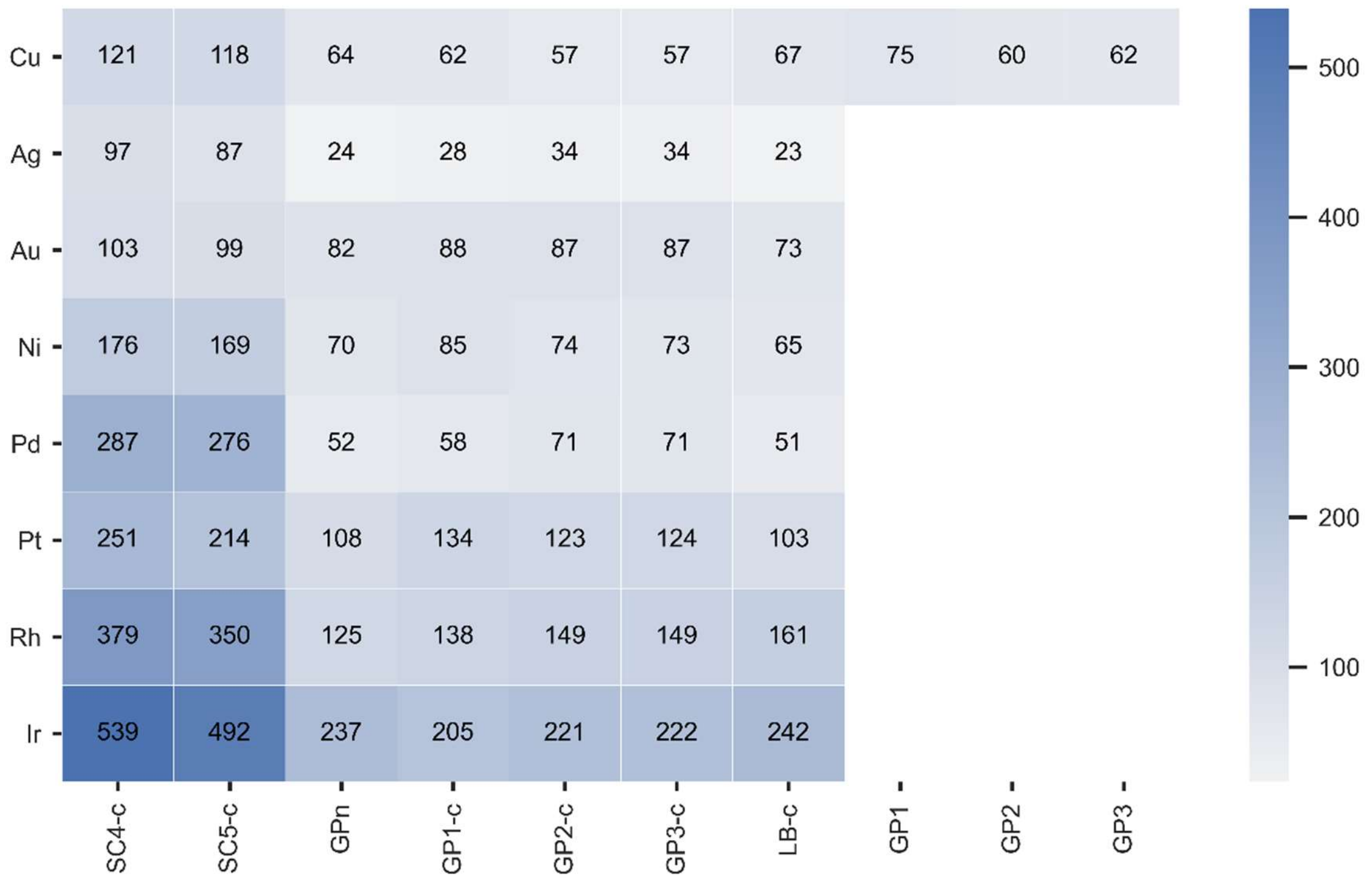

*SI-Figure 37. Force MAE on validation in meV/Å*

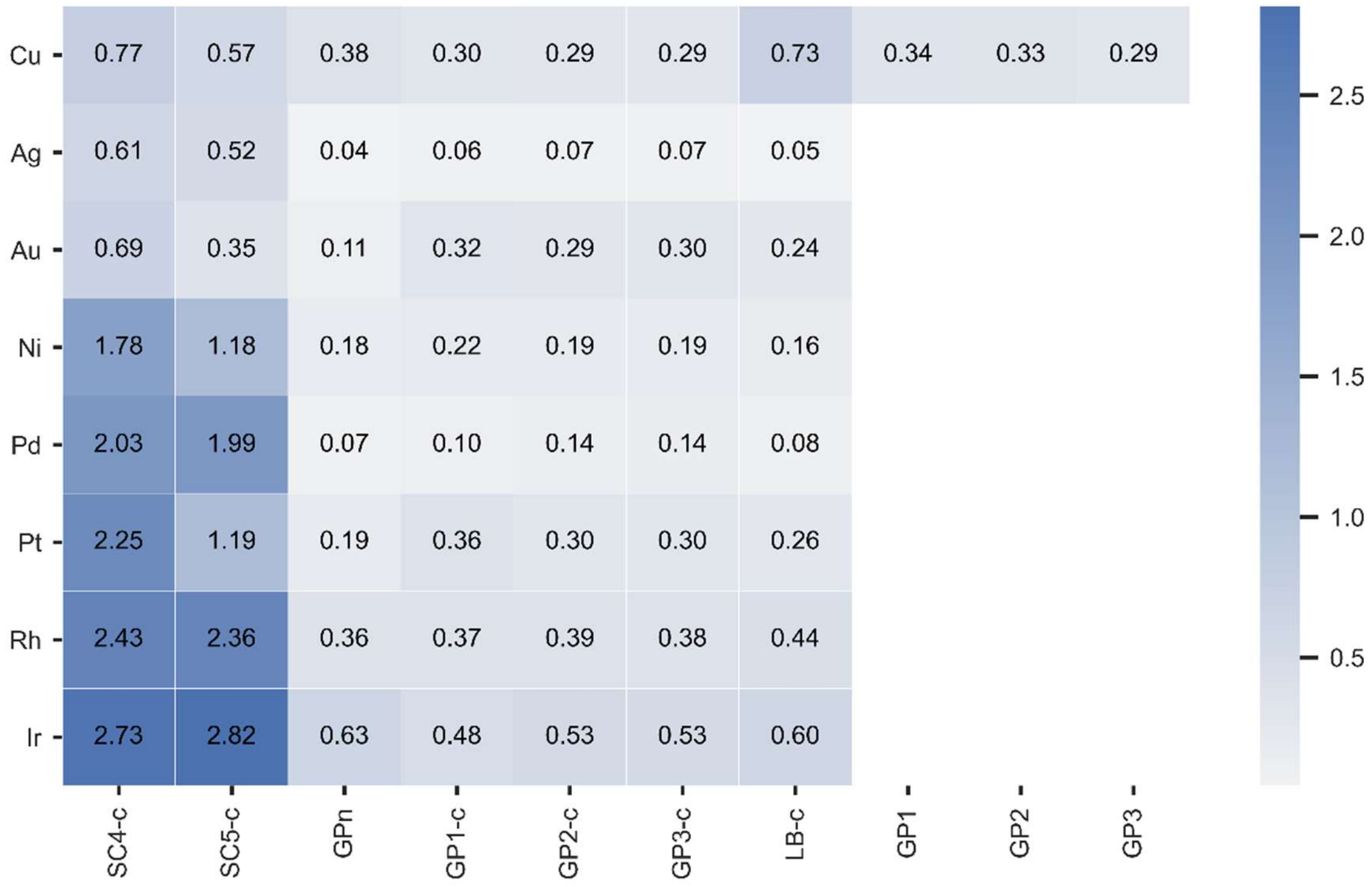

*SI-Figure 38. Stress MAE in GPa*

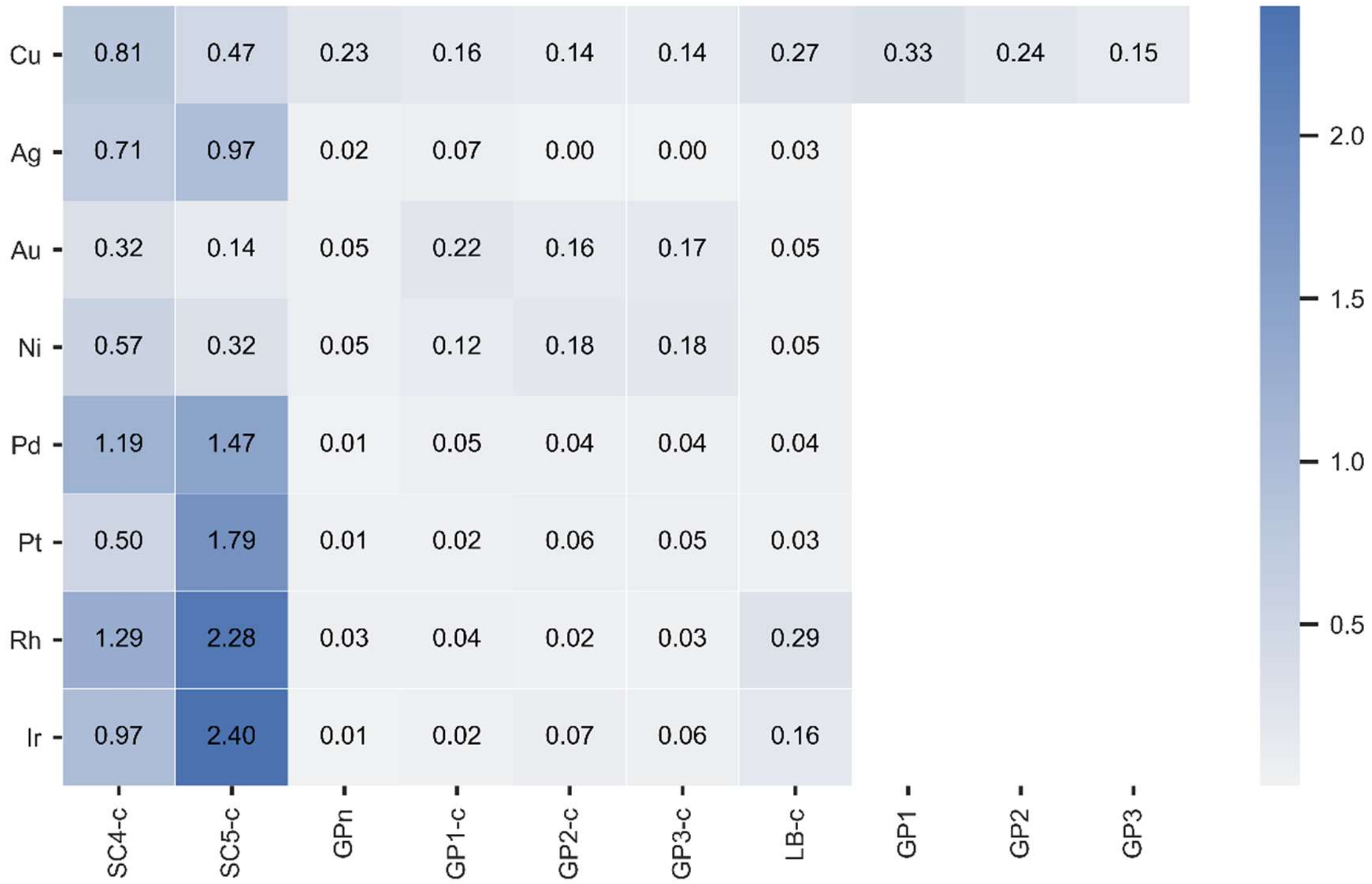

*SI-Figure 39. Absolute percent error on fcc lattice parameter*

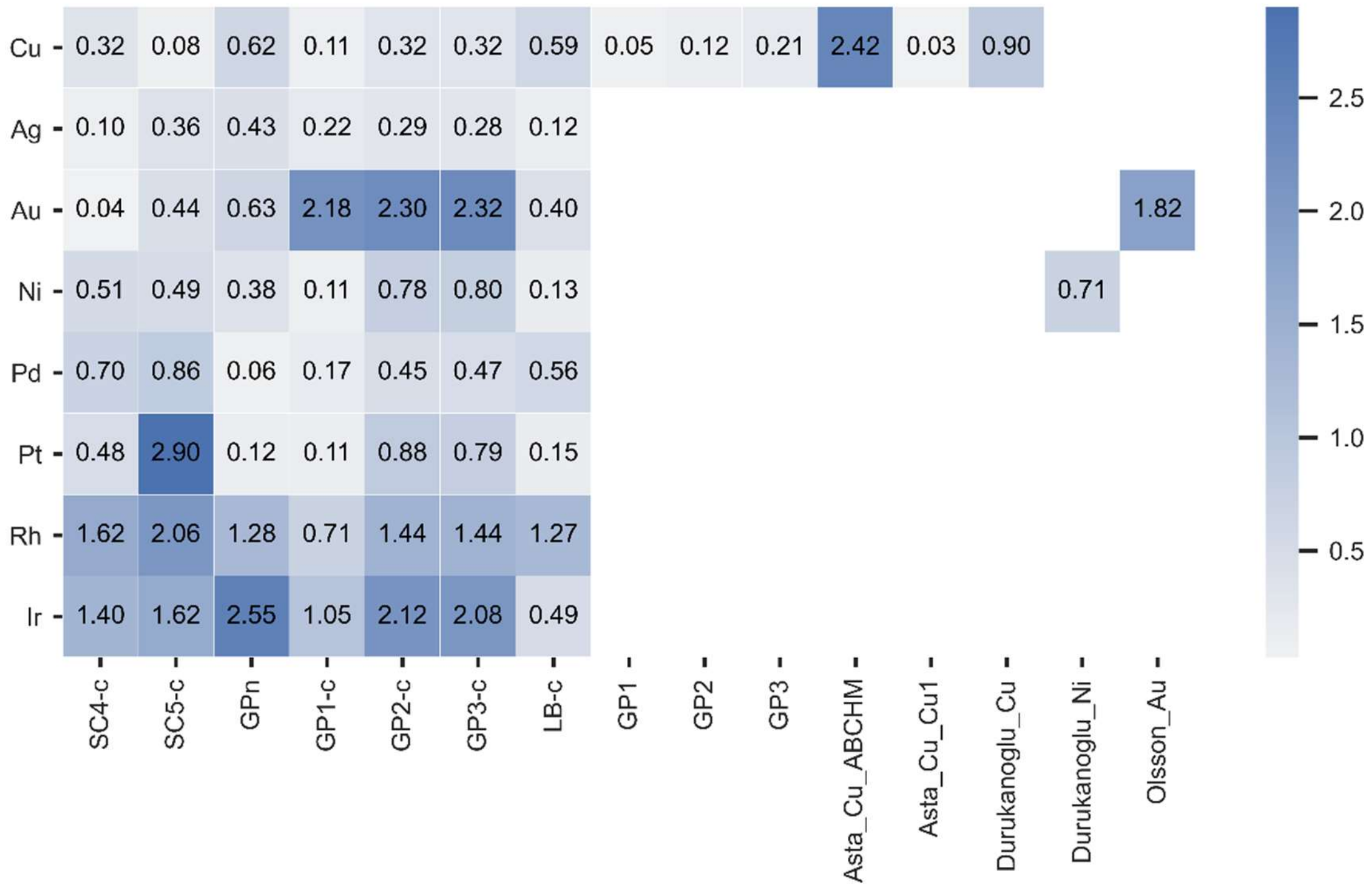

*SI-Figure 40. Absolute percent error on bcc lattice parameter.*

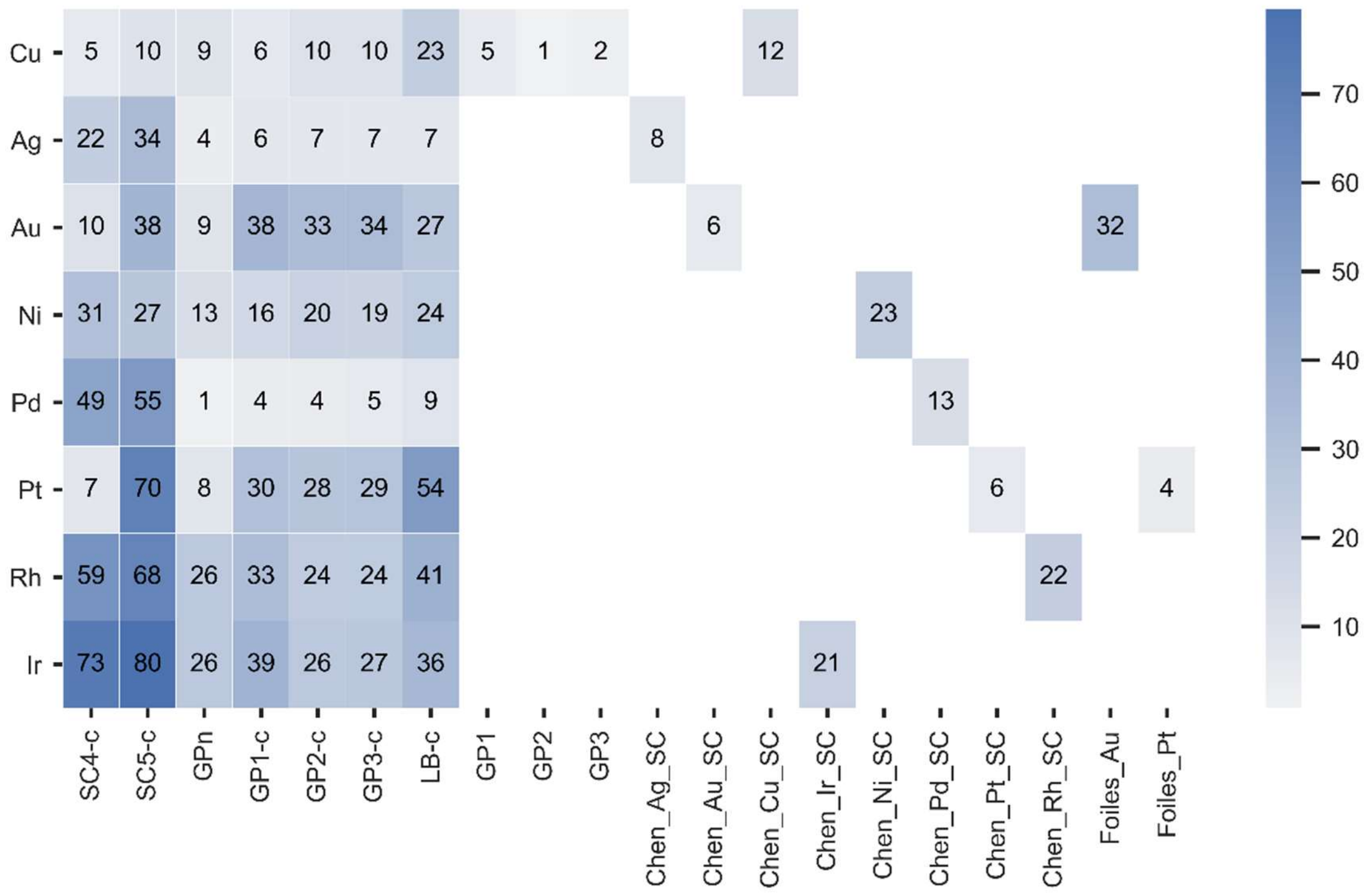

*SI-Figure 41. Mean absolute percent error on the fcc elastic constants C11, C12 and C44.*

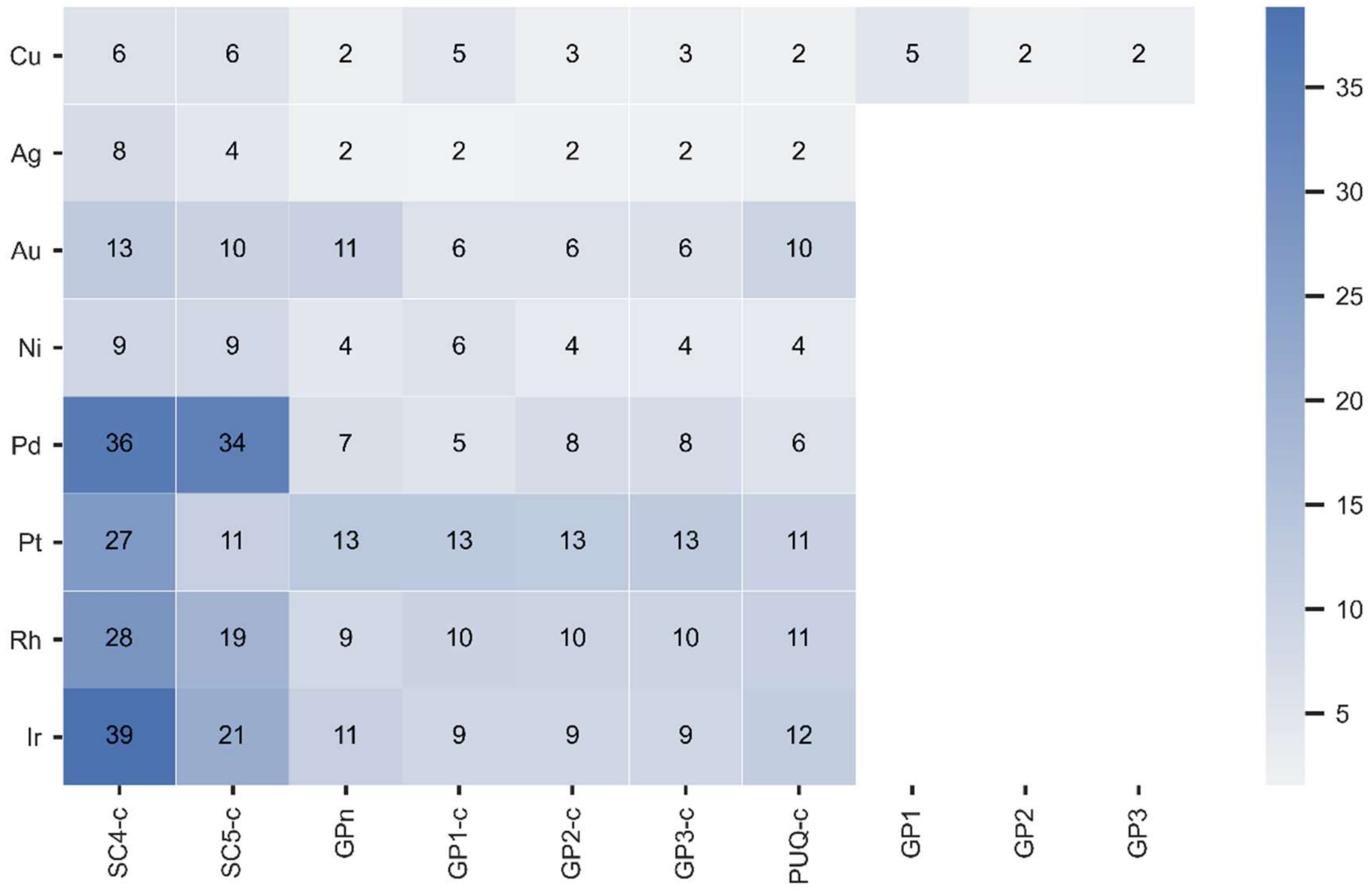

*SI-Figure 42. Mean absolute percent error on phonon frequencies.*

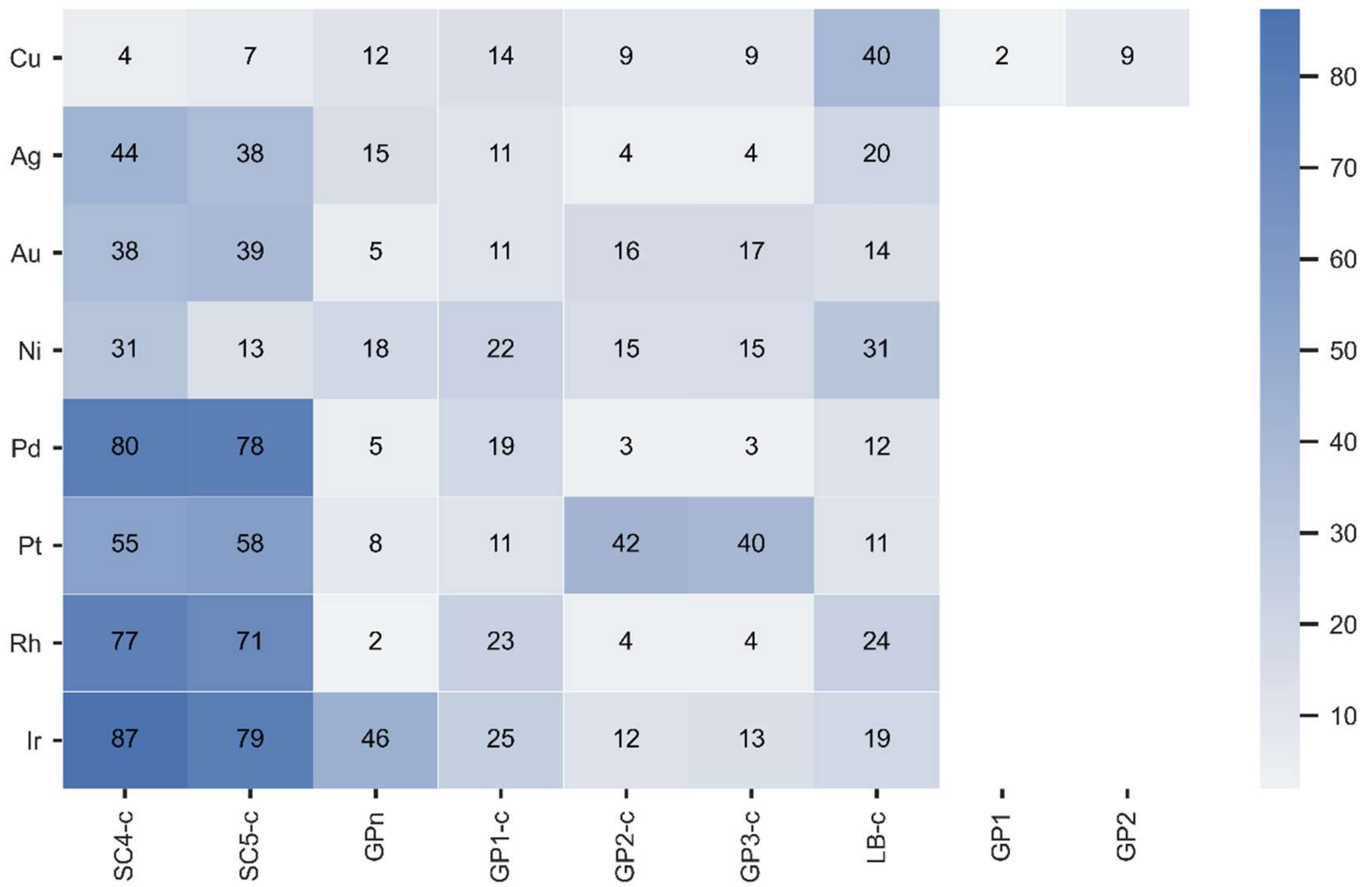

*SI-Figure 43. Mean absolute percent error on 13-low index surface energies.*

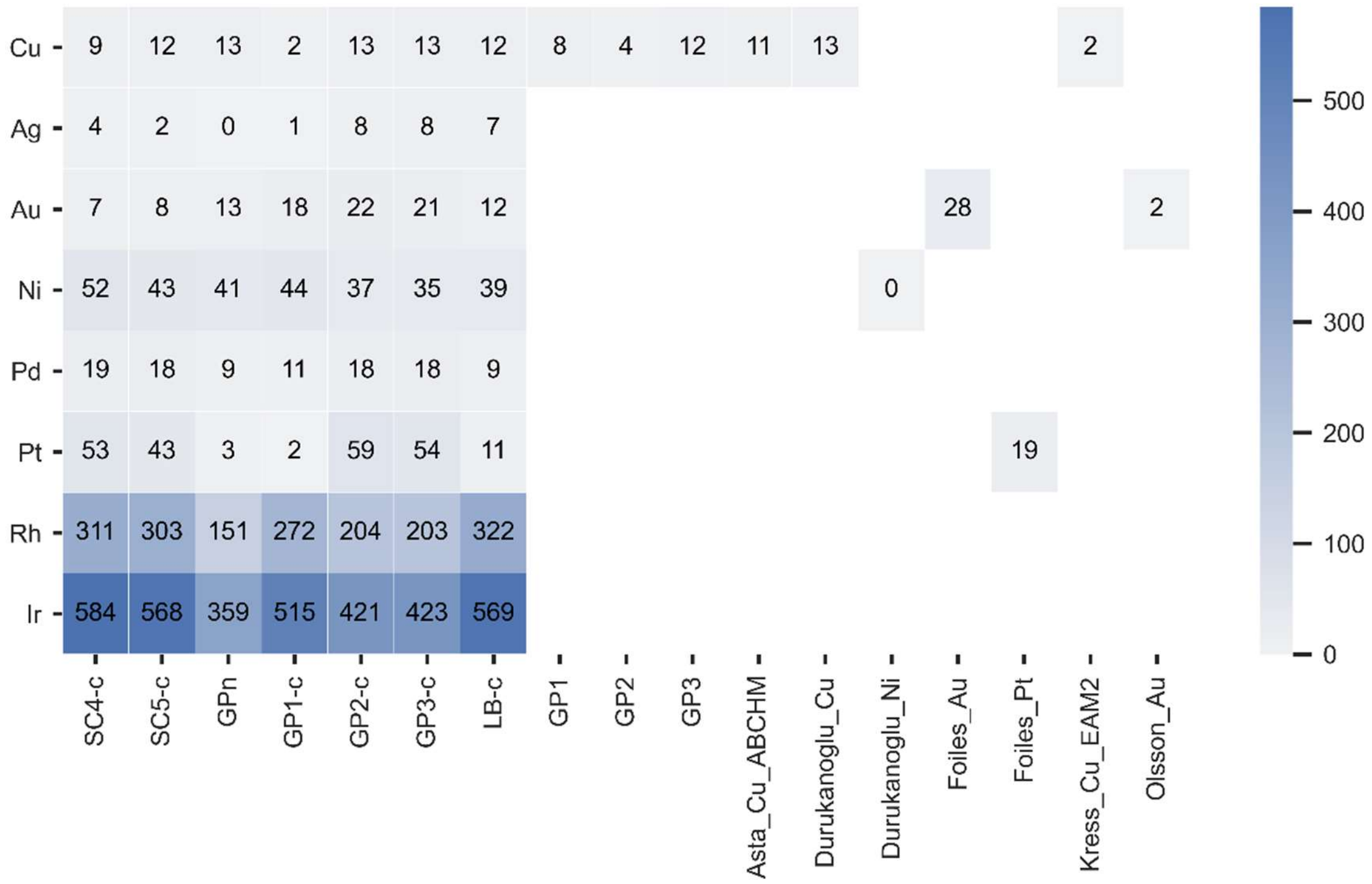

*SI-Figure 44. Absolute error on bcc formation energy in meV*

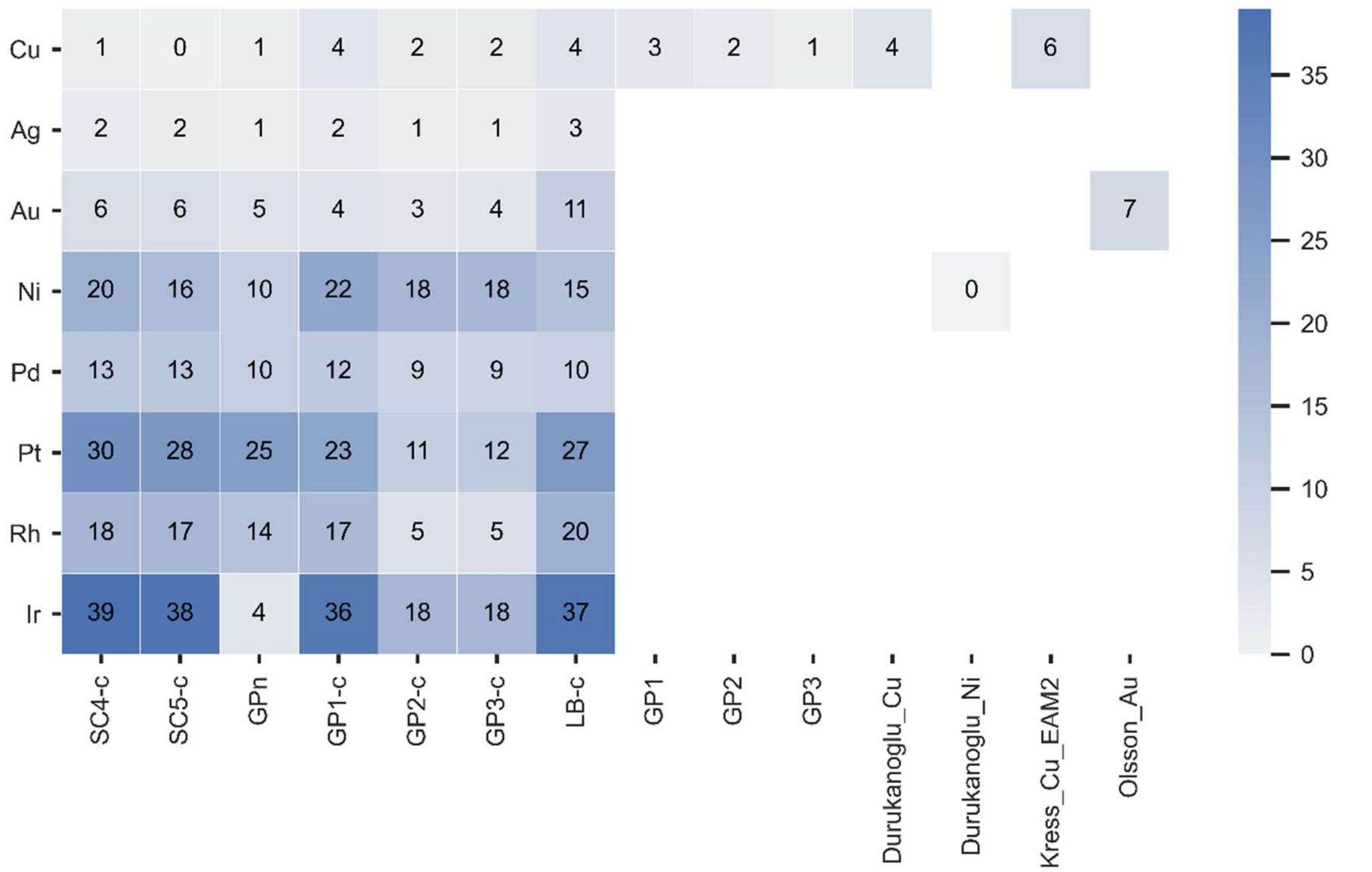

*SI-Figure 45. Absolute error on hcp formation energy in meV*

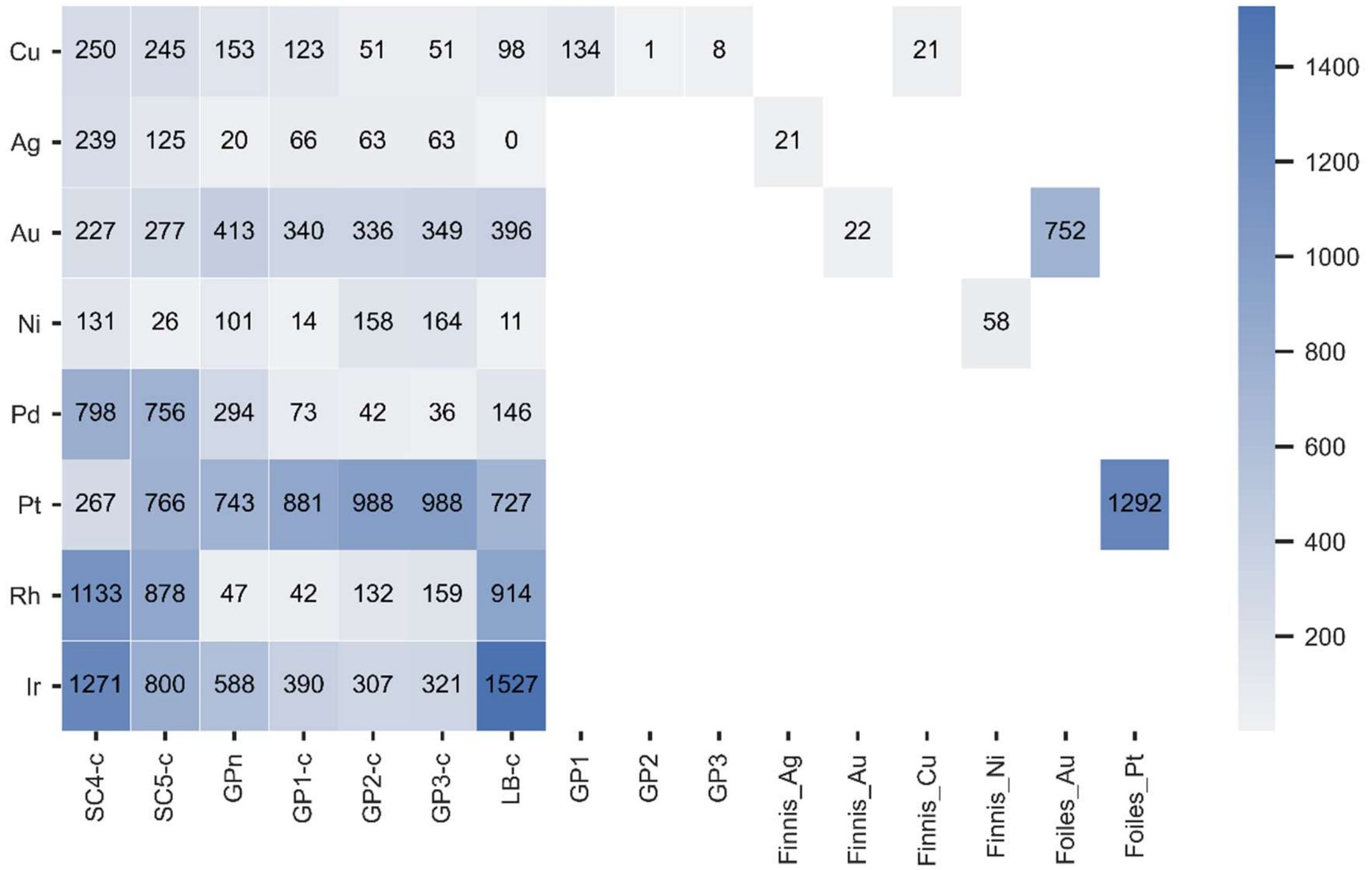

*SI-Figure 46. Absolute error on the vacancy formation energy in meV*

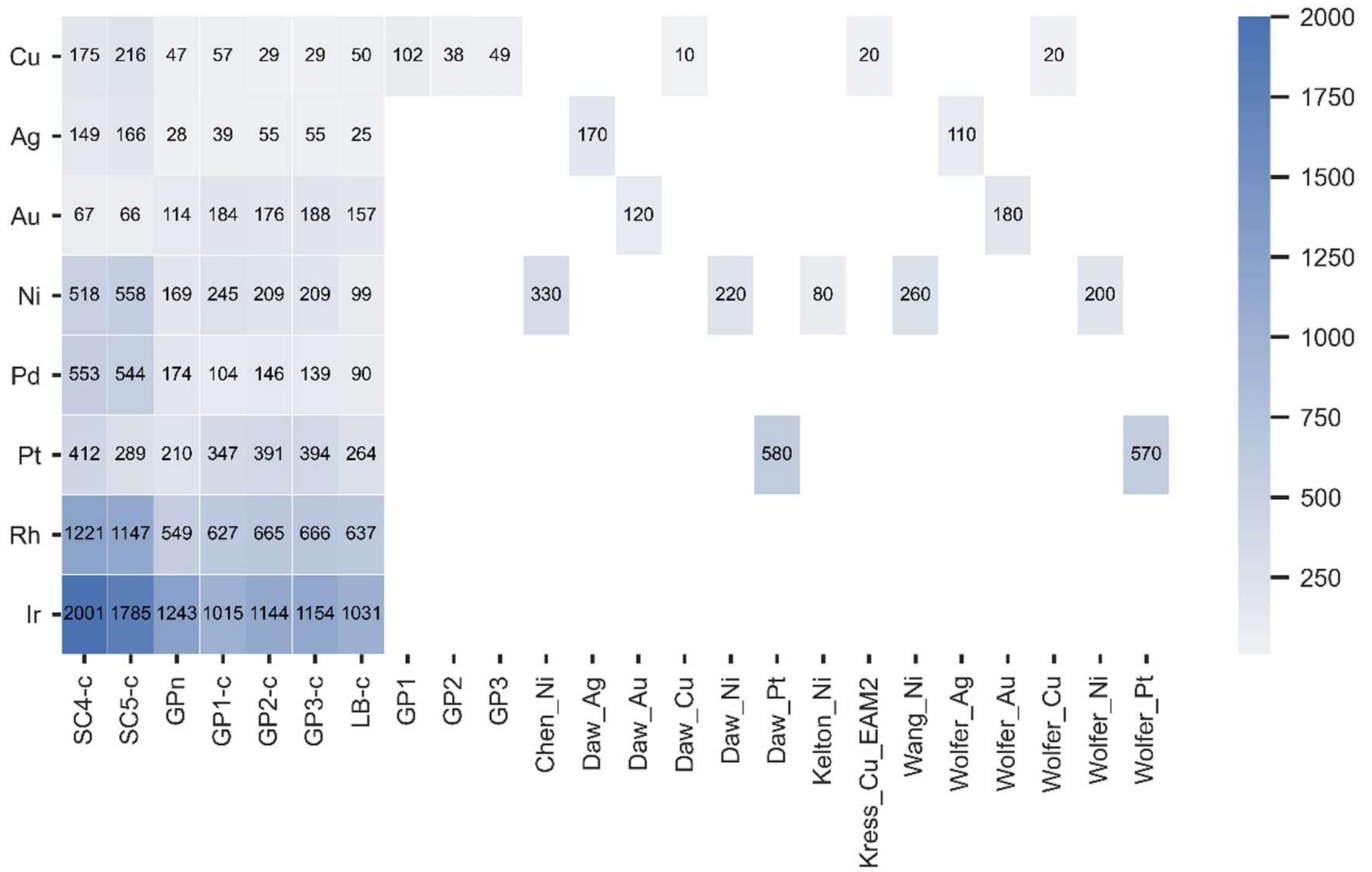

SI-Figure 47. Absolute error on the vacancy migration energy in meV

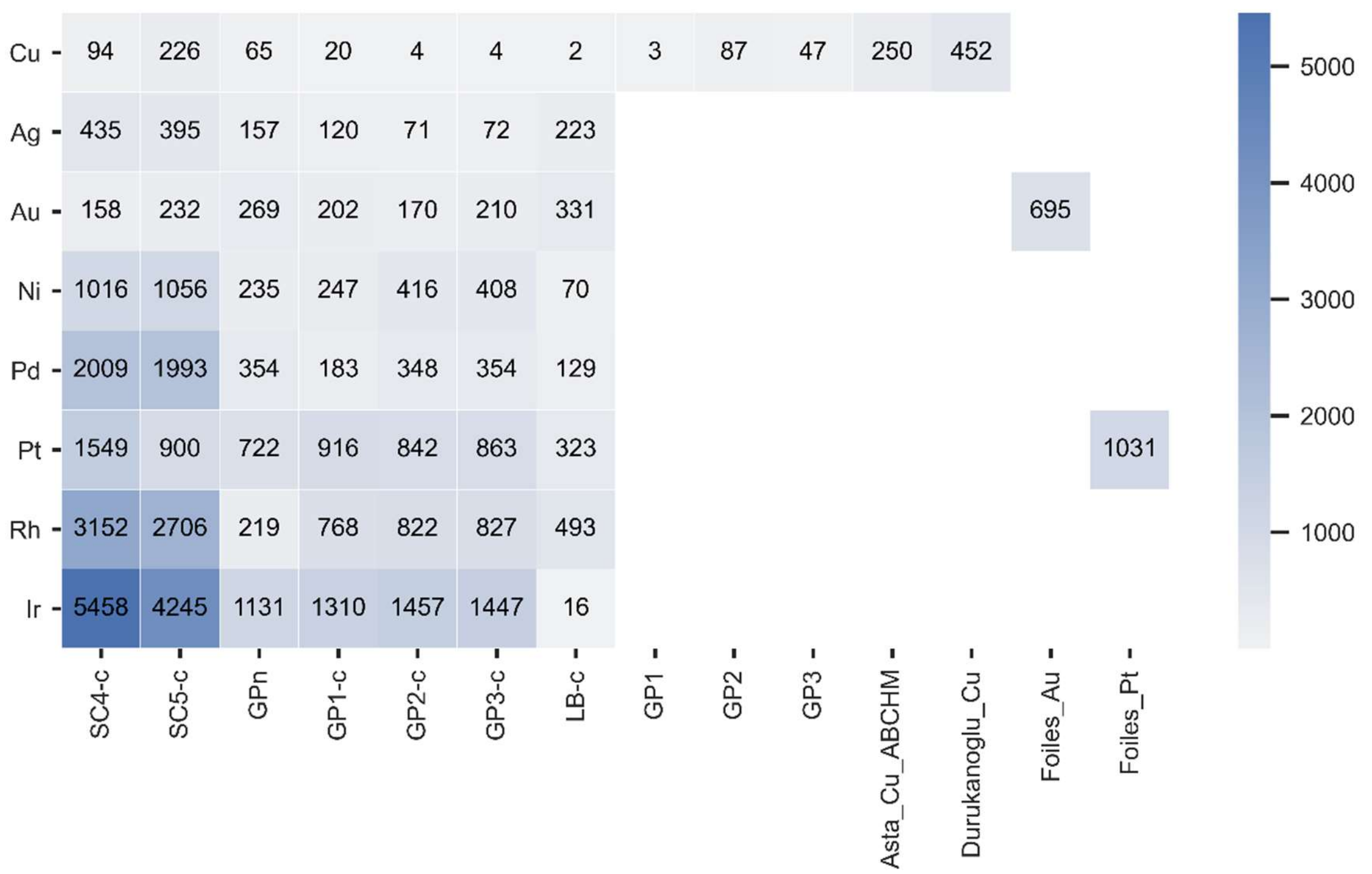

SI-Figure 48. Absolute error on the <100> dumbbell formation energy in meV

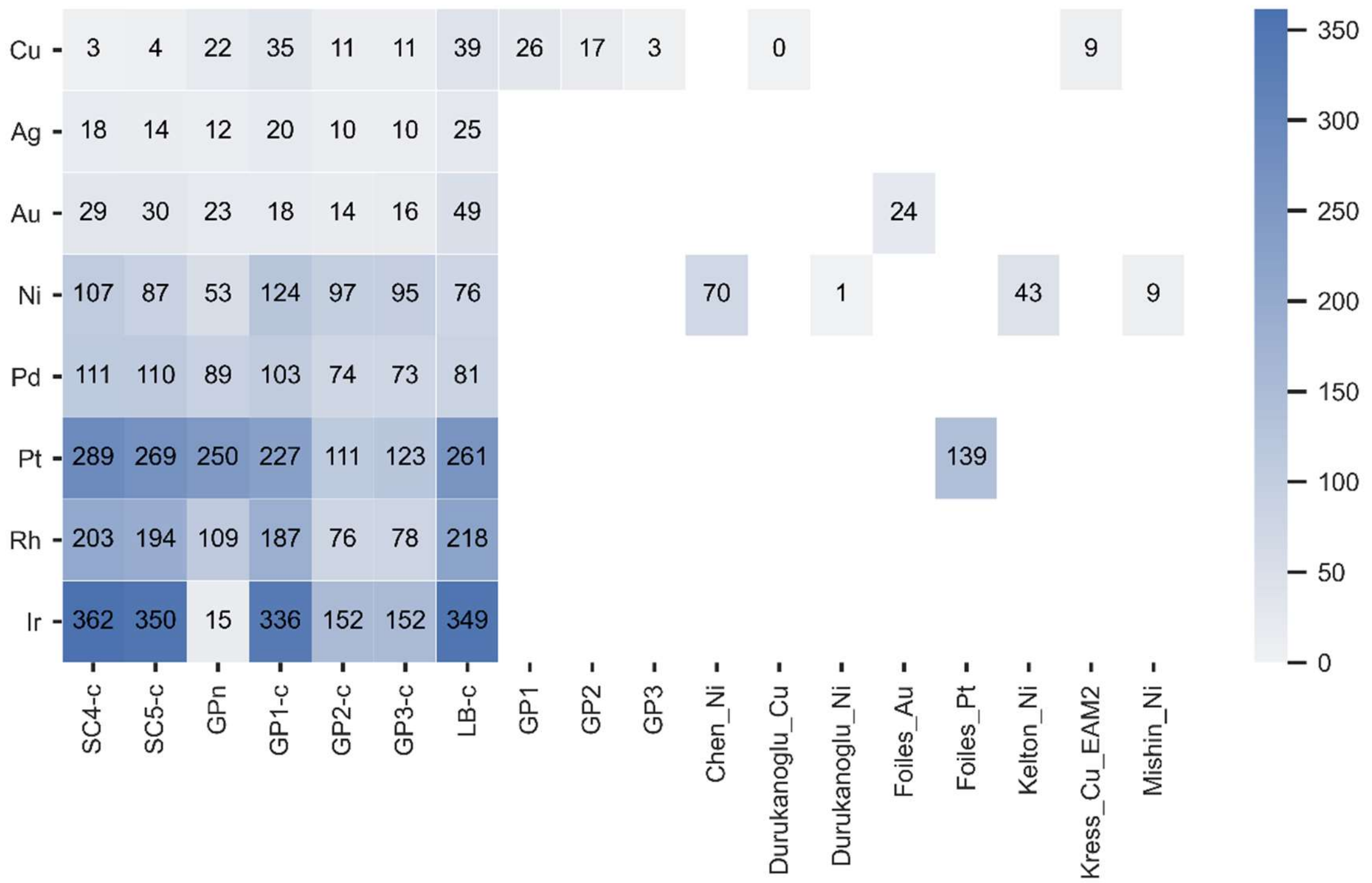

*SI-Figure 49. Absolute error on the intrinsic stacking fault energy in mJ/m²*

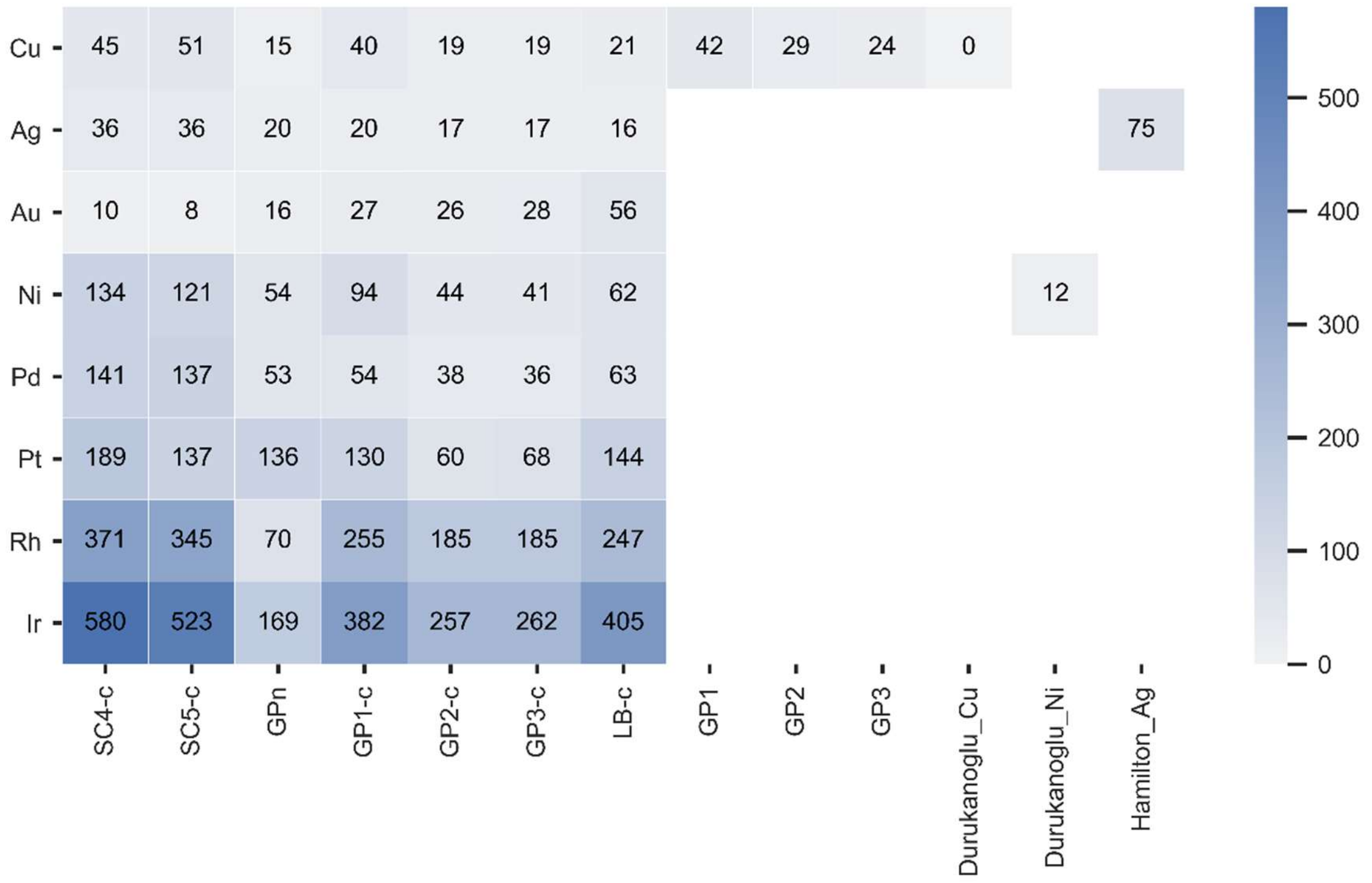

*SI-Figure 50. Absolute error on the unstable stacking fault energy in mJ/m²*

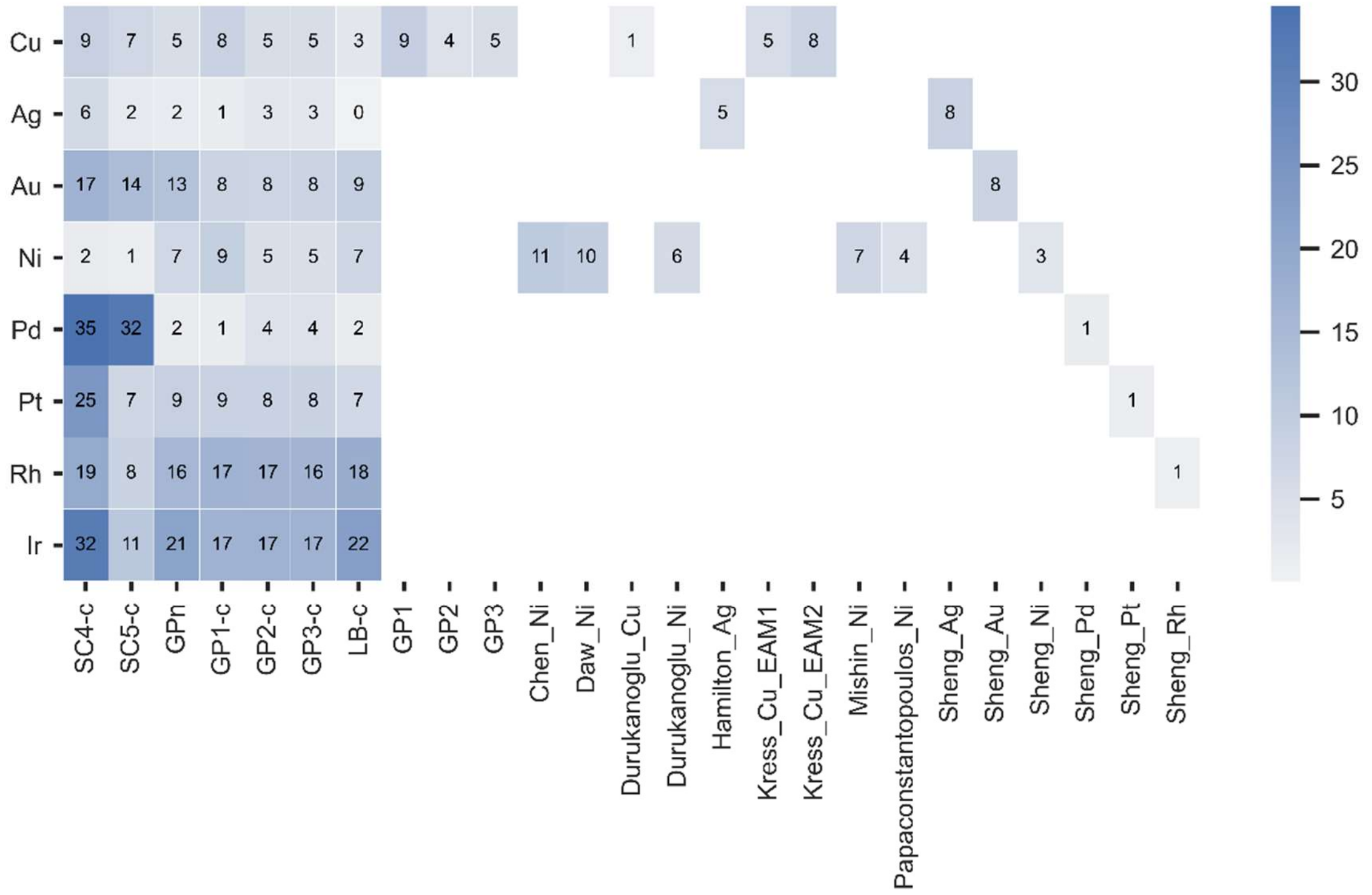

*SI-Figure 51. Absolute percent error on the $\nu_L(K)$ phonon frequency.*

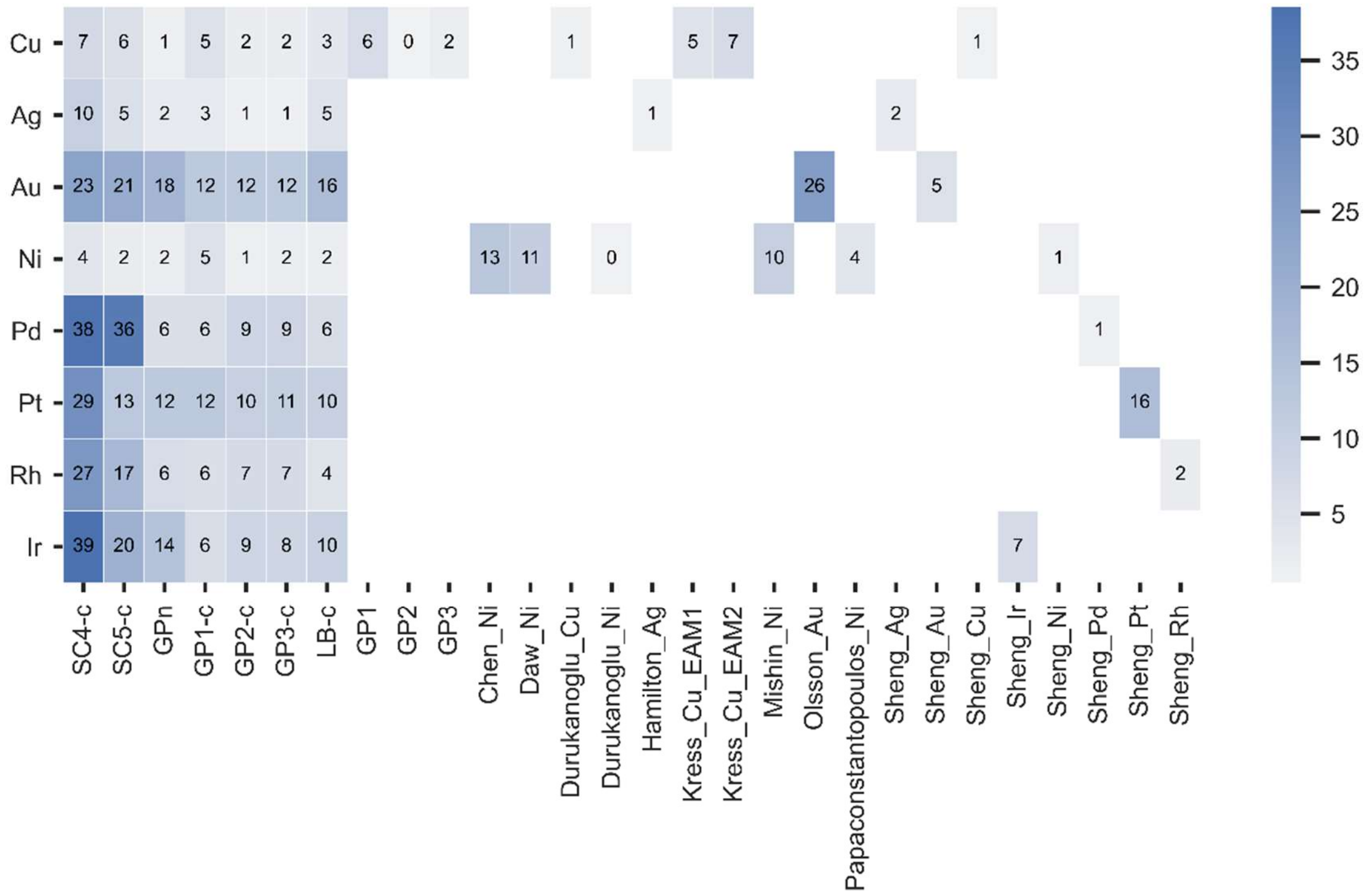

*SI-Figure 52. Absolute percent error on the $\nu_L(L)$ phonon frequency.*

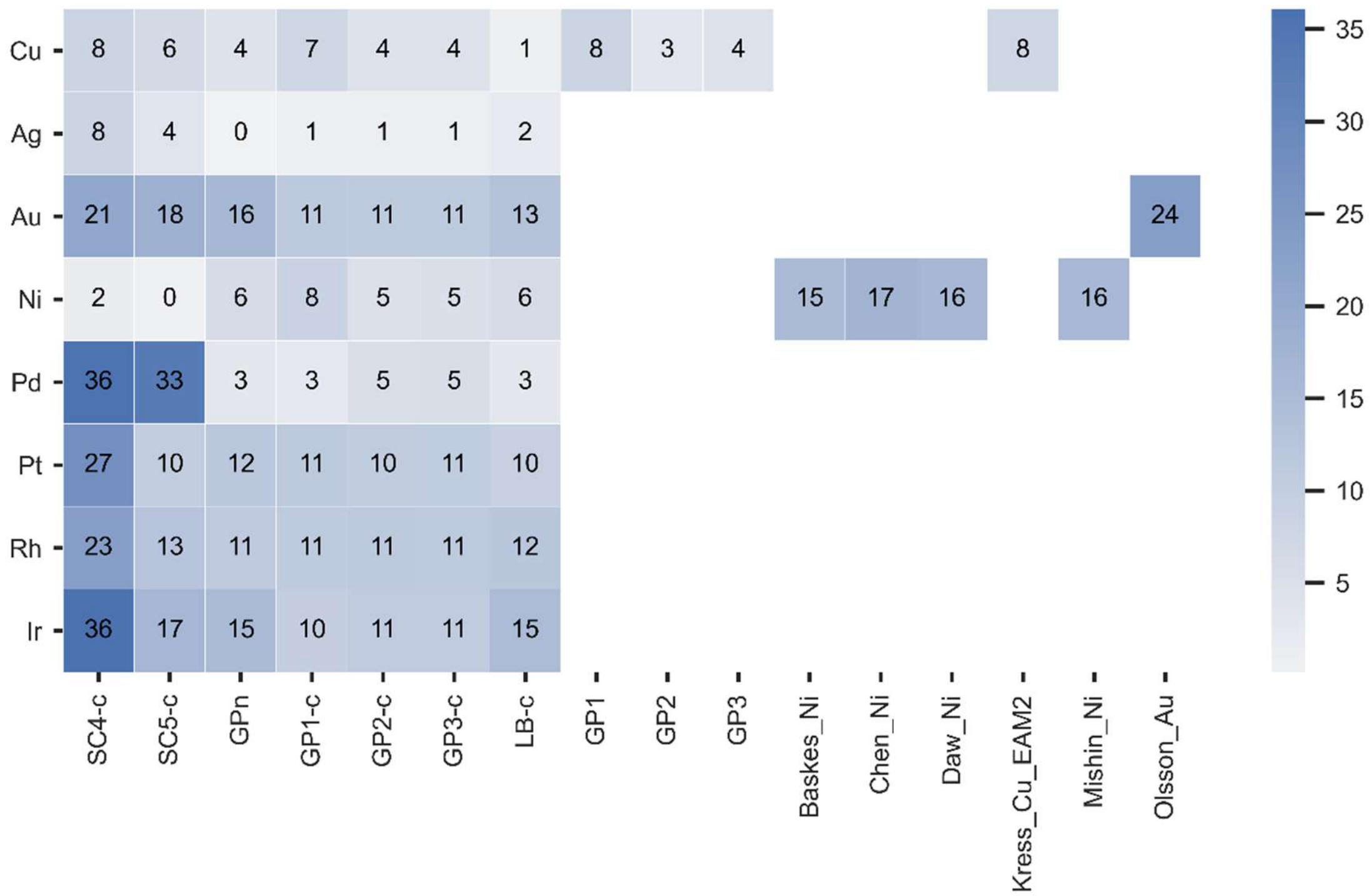

*SI-Figure 53. Absolute percent error on the $\nu_L(X)$ phonon frequency.*

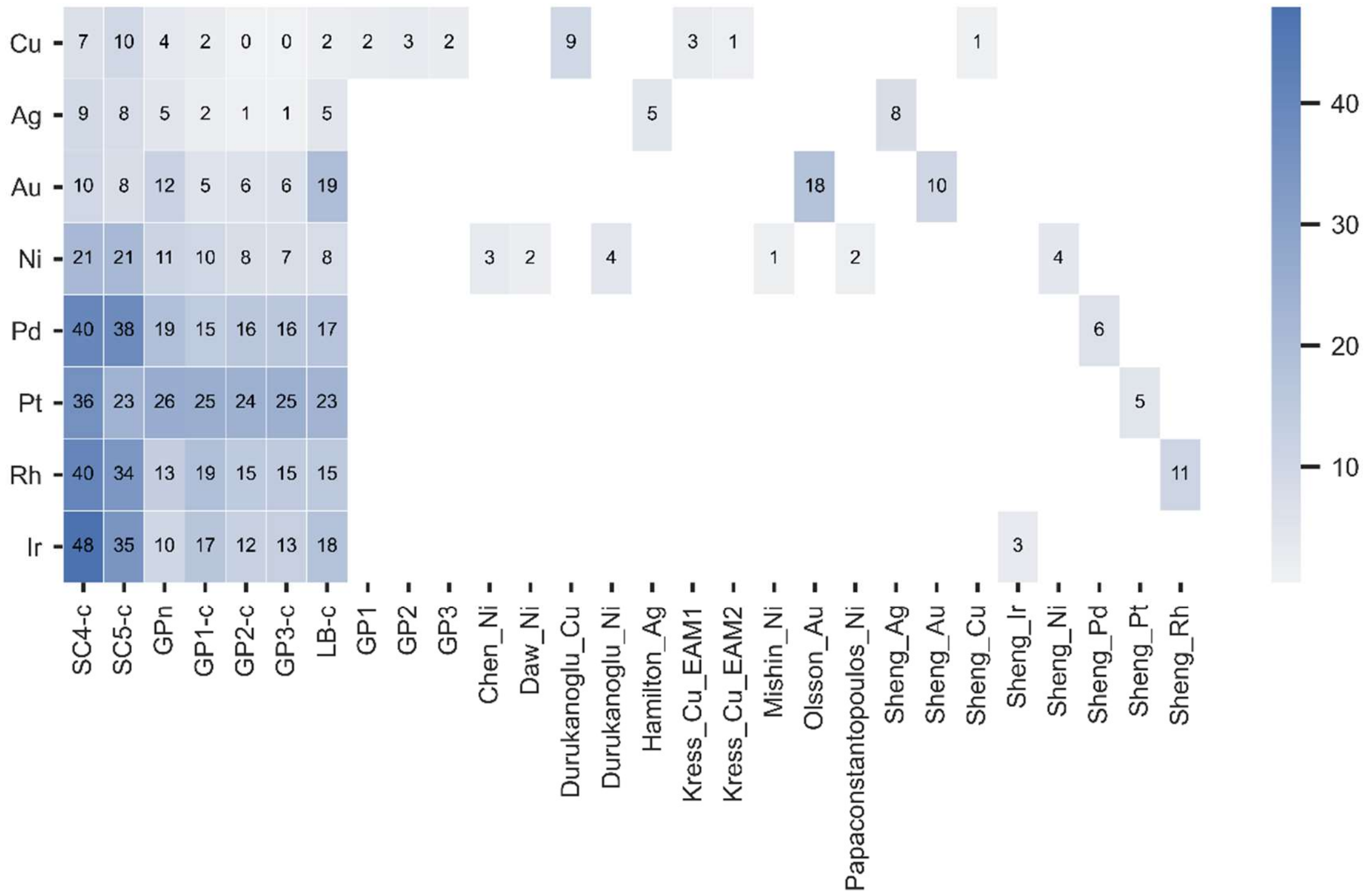

*SI-Figure 54. Absolute percent error on the $\nu_T(L)$ phonon frequency.*

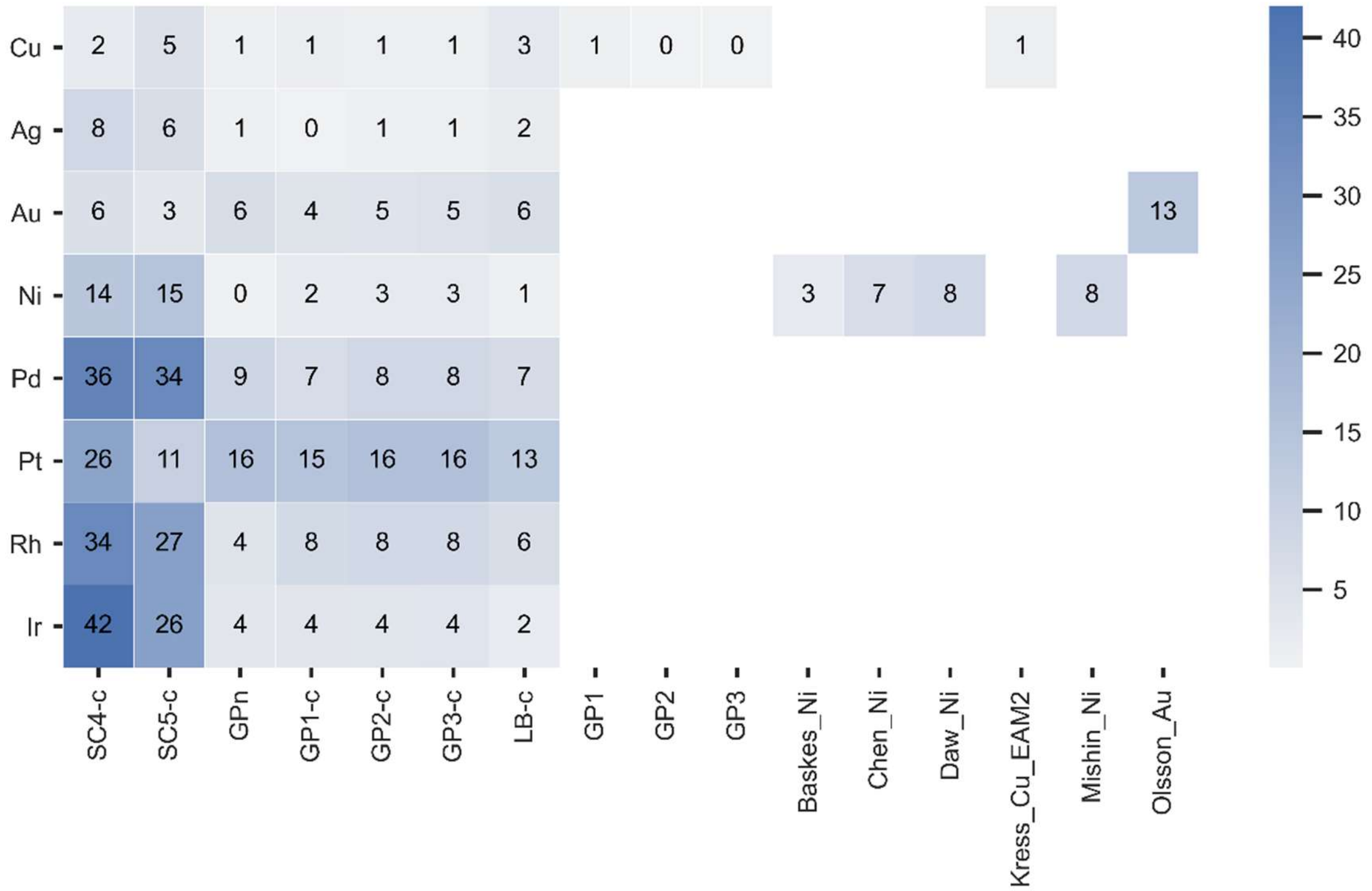

*SI-Figure 55. Absolute percent error on the $\nu_T(X)$ phonon frequency.*

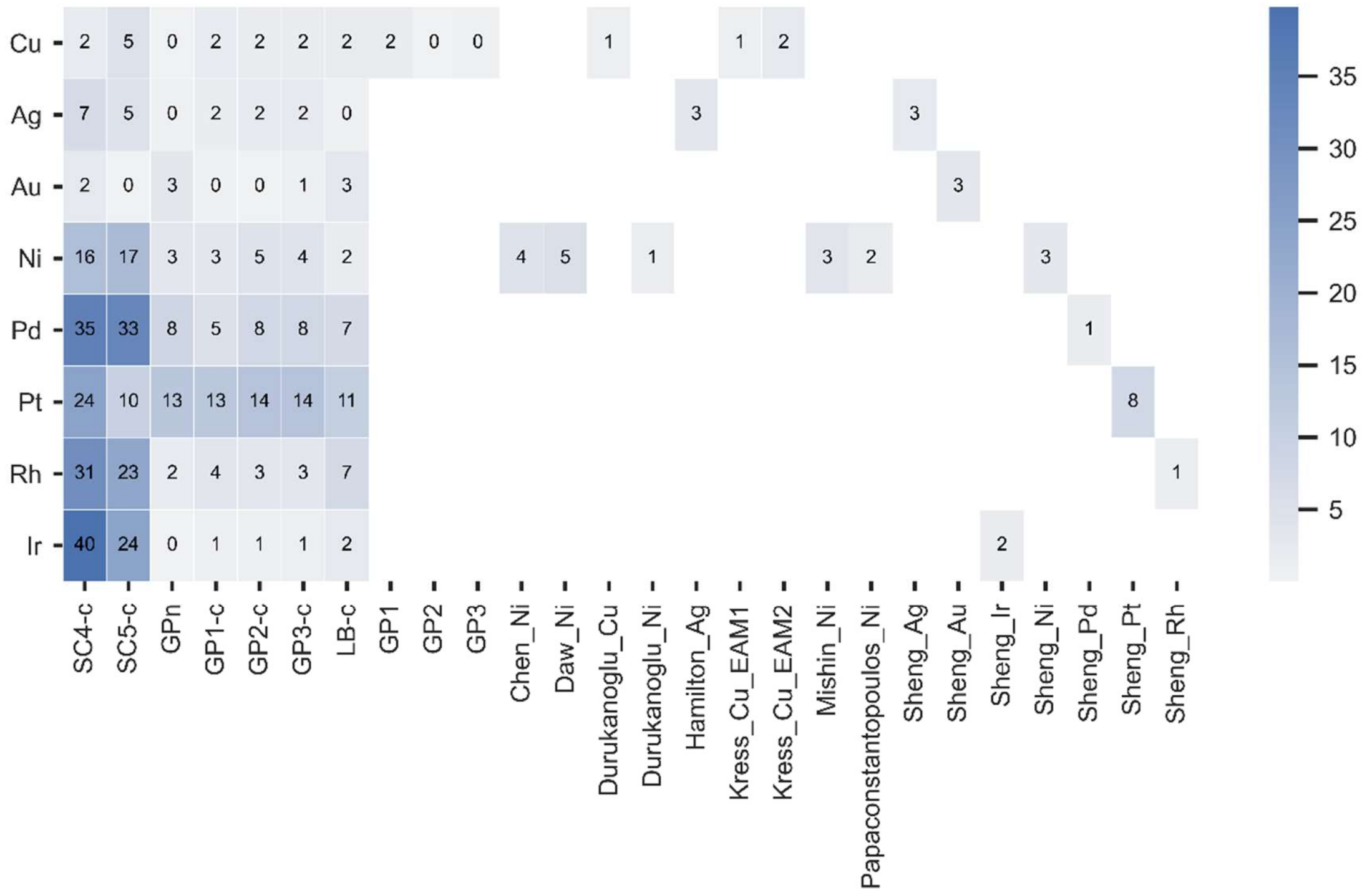

*SI-Figure 56. Absolute percent error on the $\nu_{T1}(K)$ phonon frequency.*

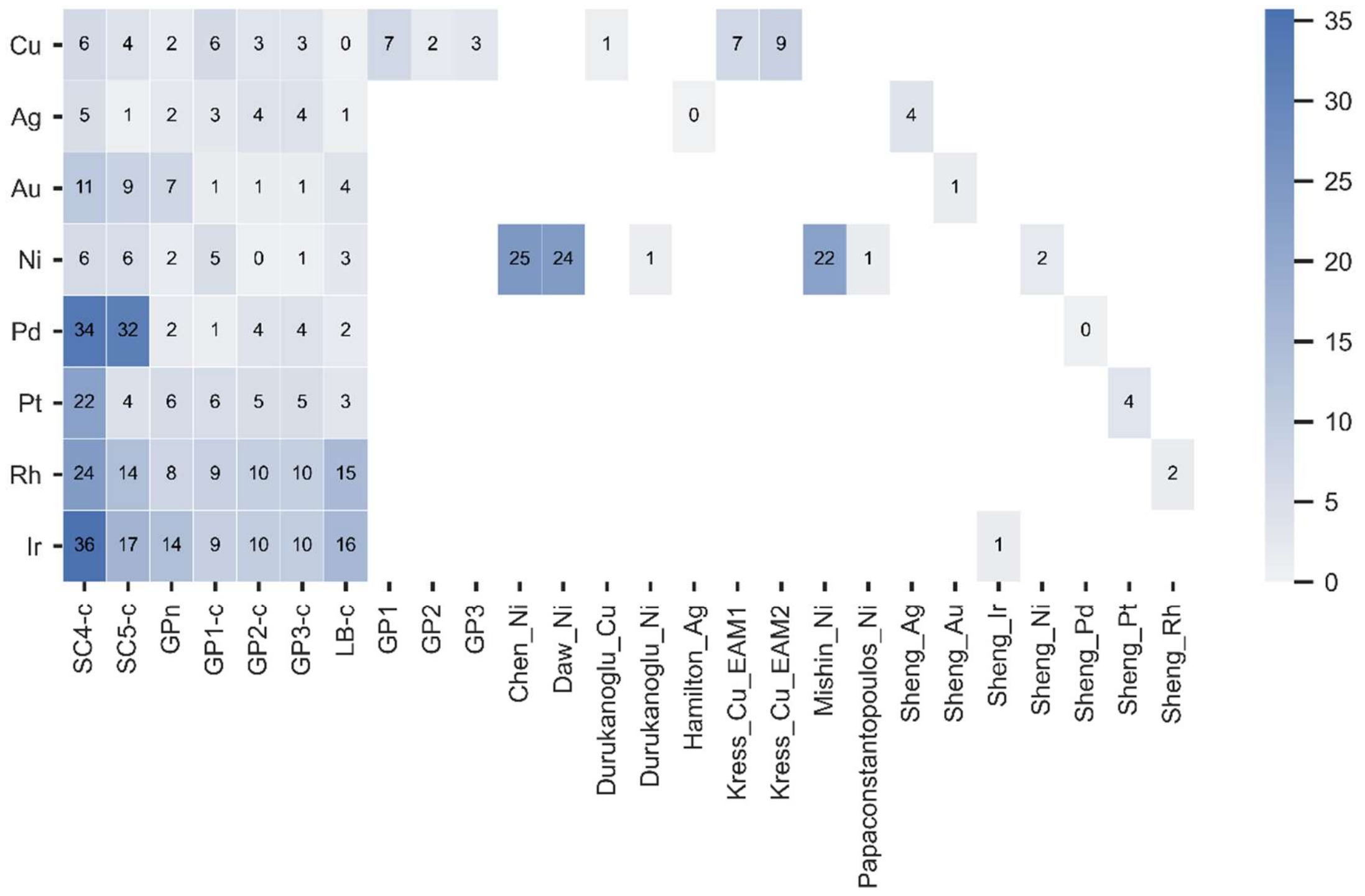

*SI-Figure 57. Absolute percent error on the $\nu_{T2}(K)$ phonon frequency.*

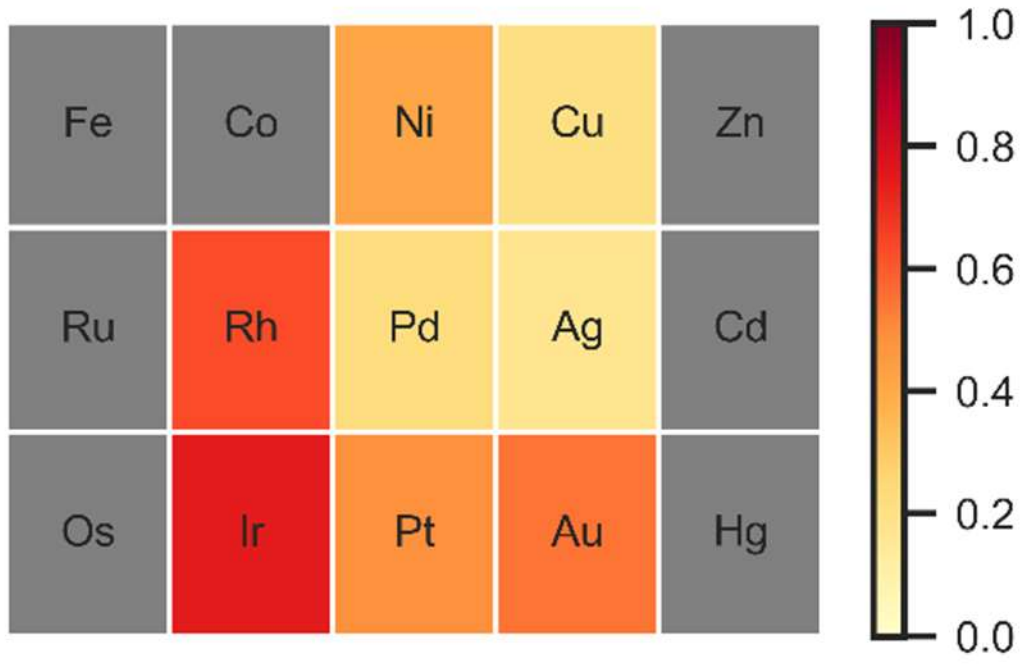

*SI-Figure 58. Average of normalized errors across validation properties for GP1-c models. The validation metrics considered on this plot are in Table 1 (excluding the fitness). The normalization was done using min-max scaling (x-min(x))/(max(x)-min(x))*

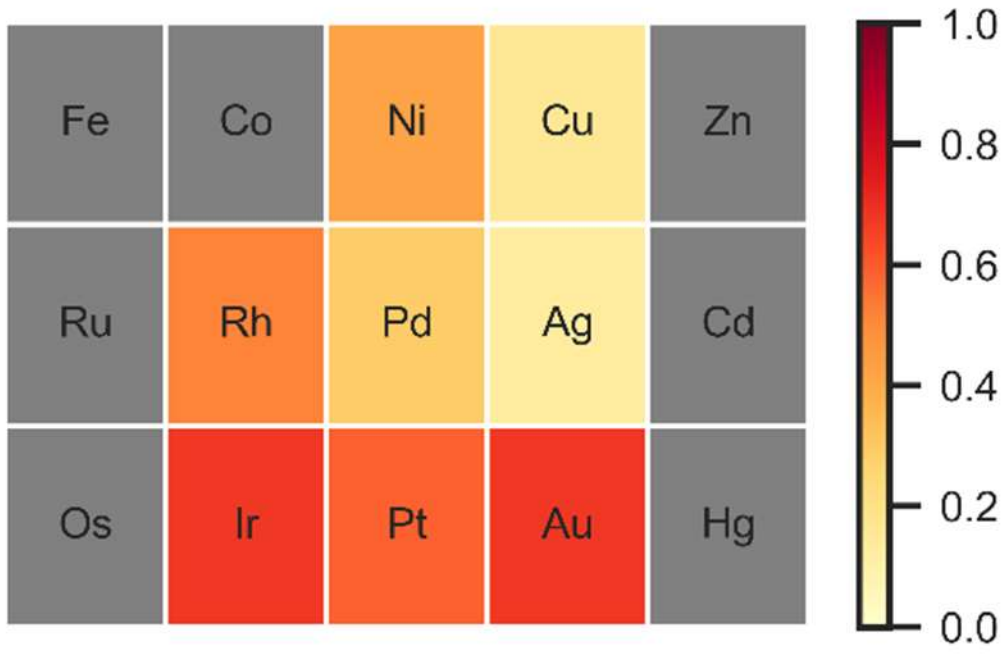

*SI-Figure 59. Average of normalized errors across validation properties for GP3-c models. The validation metrics considered on this plot are in Table 1 (excluding the fitness). The normalization was done using min-max scaling (x-min(x))/(max(x)-min(x))*

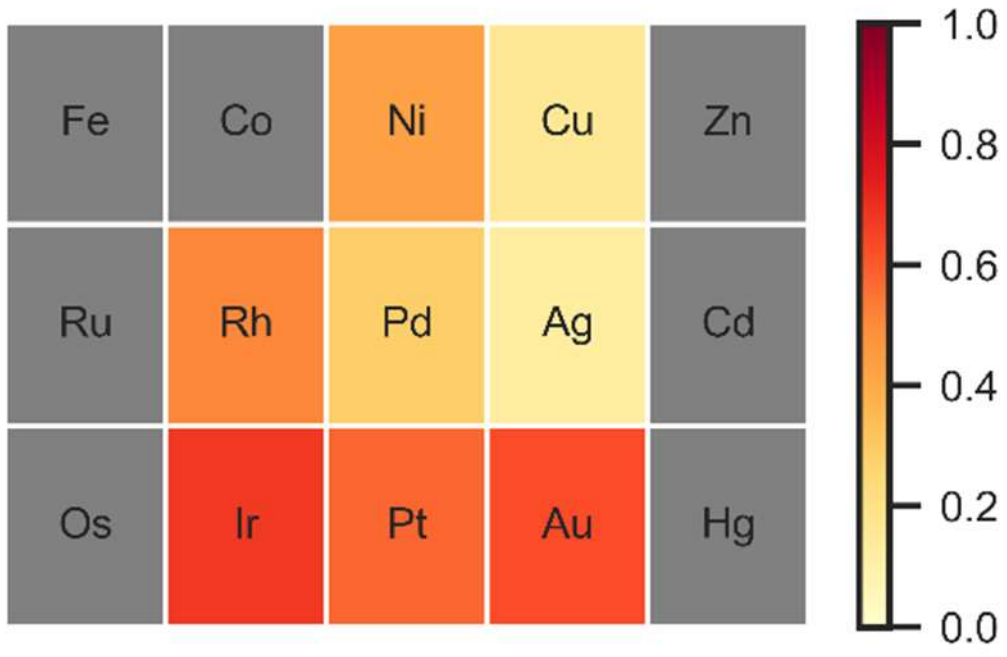

*SI-Figure 60. Average of normalized errors across validation properties for GP2-c models. The validation metrics considered on this plot are in Table 1 (excluding the fitness). The normalization was done using min-max scaling (x-min(x))/(max(x)-min(x))*

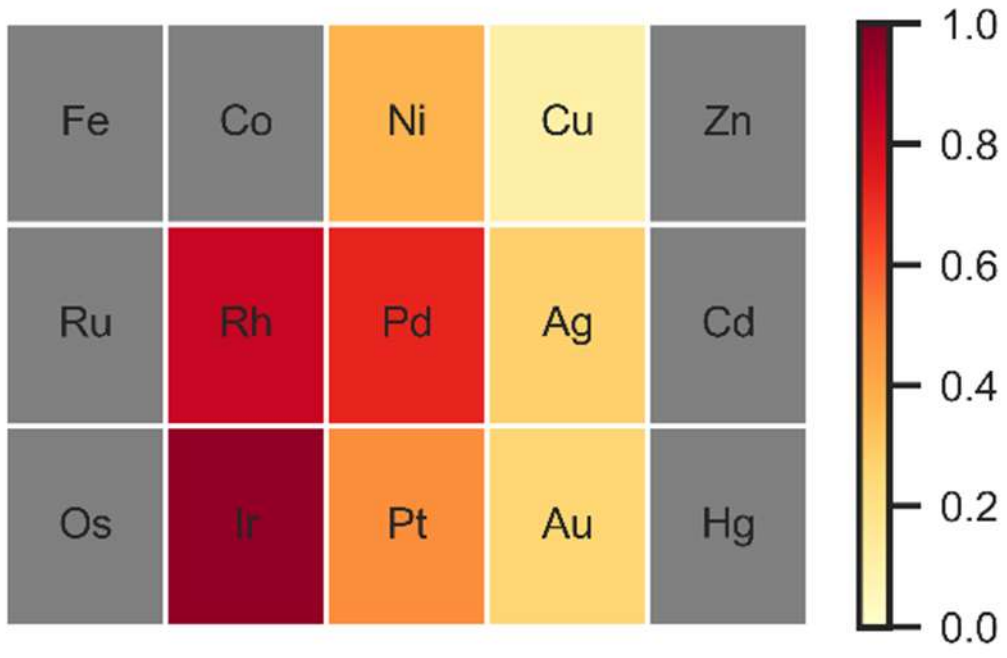

*SI-Figure 61. Average of normalized errors across validation properties for SC4-c models. The validation metrics considered on this plot are in Table 1 (excluding the fitness). The normalization was done using min-max scaling (x-min(x))/(max(x)-min(x))*

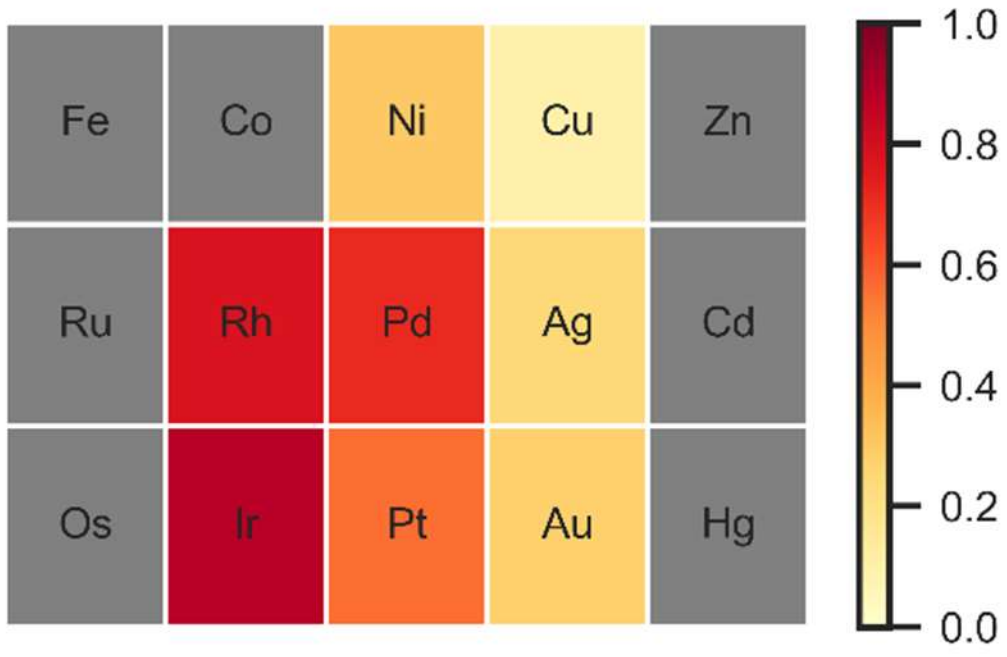

*SI-Figure 62. Average of normalized errors across validation properties for SC5-c models. The validation metrics considered on this plot are in Table 1 (excluding the fitness). The normalization was done using min-max scaling (x-min(x))/(max(x)-min(x))*

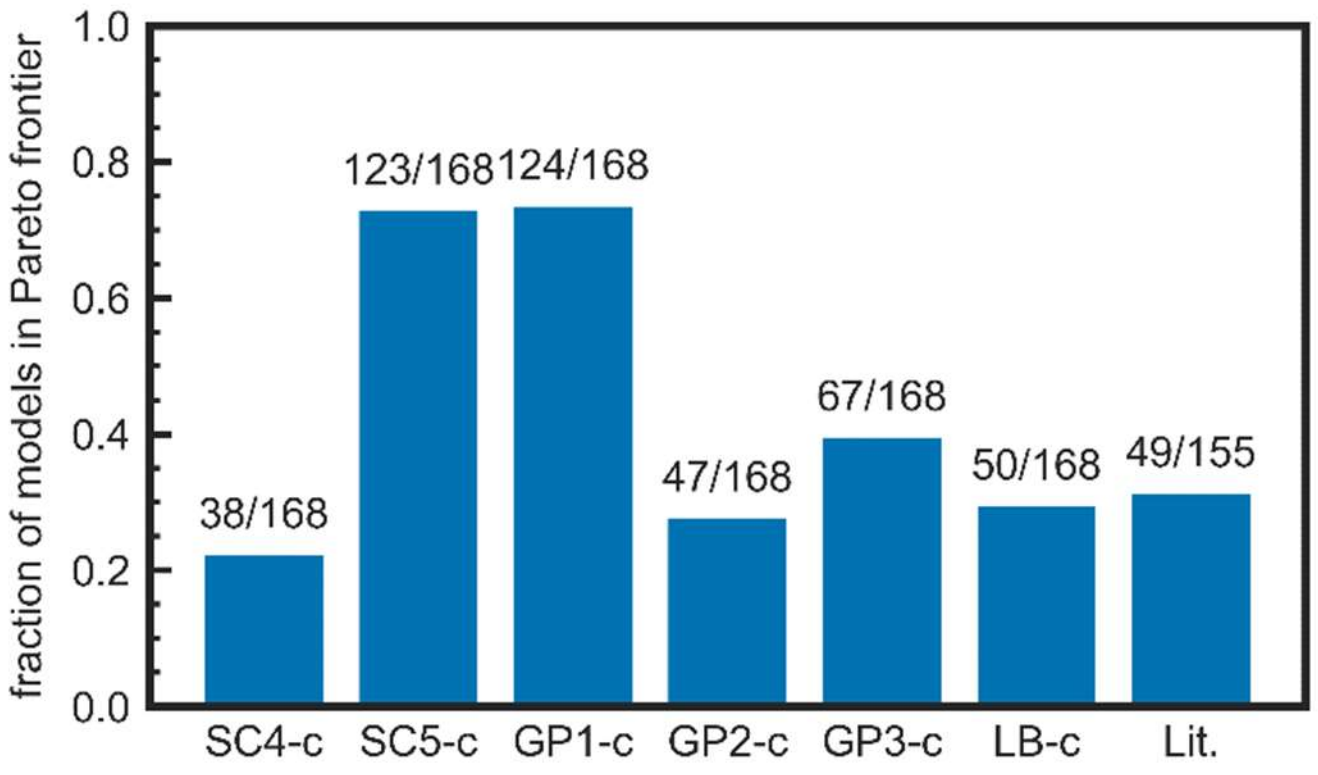

(a)

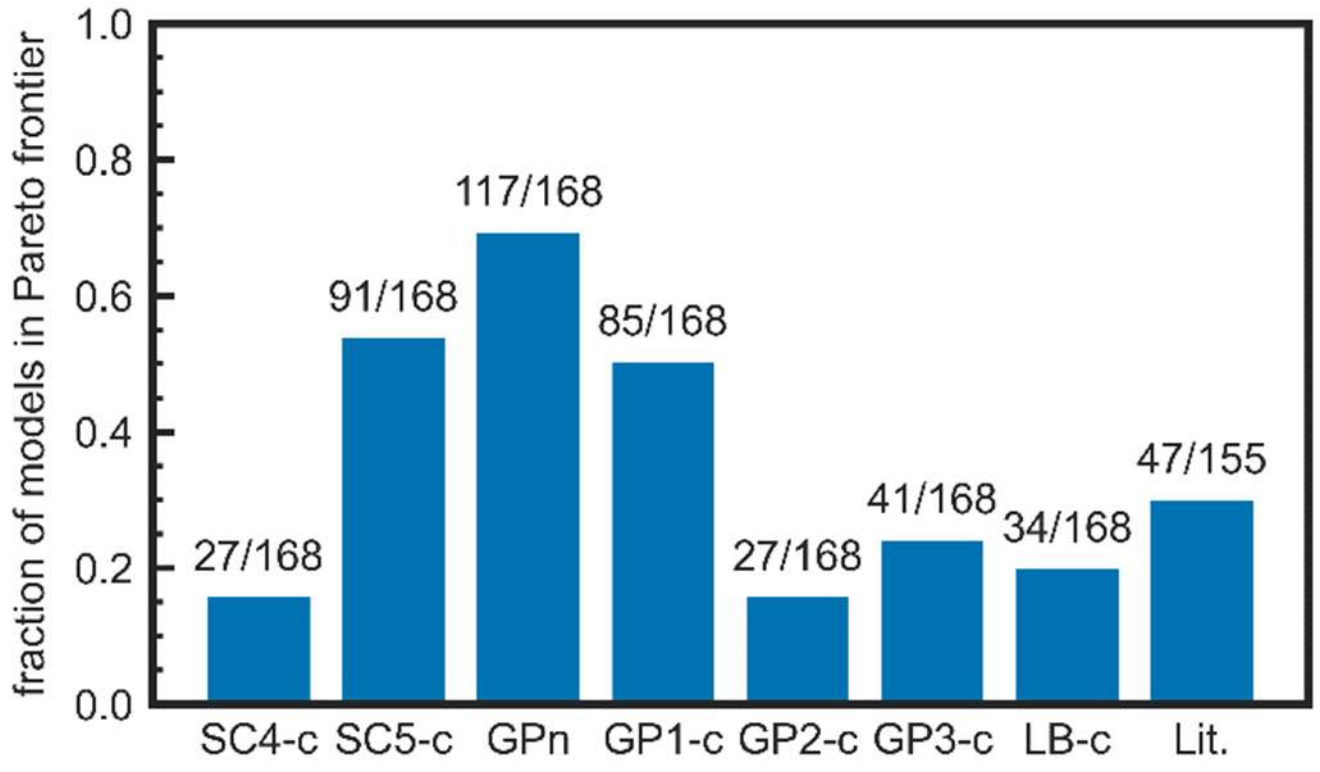

(b)

*SI-Figure 63. Number of times that an EAM-type model belongs to the Pareto frontier divided by the number of times that the model has validation values available across the elements and properties. (a) excluding GPn models to analyze the transferability of GP1-c, GP2-c, and GP3-c. (b) including GPn to analyze how POET can find models with a better complexity-accuracy tradeoff. The metrics considered are the ones shown in Table 1 on the main manuscript (excluding the fitness).*

*SI-Table 2. Cutoff distances*

| Element | Outer cutoff distance, $r_{out}$ (Å) | Inner cutoff distance, $r_{in}$ (Å) |
|---|---|---|
| Cu | 5 | 3 |
| Ag | 5.5 | 3.5 |
| Au | 5.5 | 3.5 |
| Ni | 5 | 3 |
| Pd | 5.5 | 3.5 |
| Pt | 5.5 | 3.5 |
| Rh | 5.25 | 3.25 |
| Ir | 5.4 | 3.4 |

*SI-Table 3. Parameters of GP1 and GP1-c models*

| element | x0 | x1 | x2 | x3 | x4 |
|---|---|---|---|---|---|
| Cu (a) | 10.213032 | 10.213032 | 0.210769 | 0.972441 | 0.328949 |
| Cu | 9.484145768 | 5.085386294 | 0.290921084 | 0.1915873 | 0.200459294 |
| Ag | 9.69475452 | 4.516520485 | 0.000333 | 0.183431686 | 0.235142007 |
| Au | 11.87322967 | 5.399305334 | 0.000638 | 6.614996057 | 0.742988891 |
| Ni | 10.41736297 | 5.551786863 | 0.302376677 | 0.148148977 | 0.157143032 |
| Pd | 10.10090728 | 4.761956156 | 0.306976252 | 0.172988701 | 0.203414297 |
| Pt | 11.70379336 | 5.391638705 | 0.000179 | 5.169953388 | 0.536014872 |
| Rh | 10.29114394 | 4.786653571 | 0.395012273 | 0.04272145 | 0.109993287 |
| Ir | 10.9739431 | 4.880921708 | 0.406723036 | 0.02778075 | 0.087663454 |

Notes: (a) Parameters from [31]. The parameters in GP1 correspond to:

$$E_i = \sum_j (r^{x_0 - x_1 r} - x_2^{\,r}) f(r) + x_3 / \sum_j x_4^{\,r} f(r)$$

*SI-Table 4. Parameters of GP2 and GP2-c models*

| element | x0 | x1 | x2 | x3 | x4 | x5 | x6 | x7 |
|---|---|---|---|---|---|---|---|---|
| Cu (a) | 7.325665 | 3.979468 | 3.935002 | 27.319252 | 11.126676 | 0.034045 | 11.73571 | 2.926828 |
| Cu | 9.708921646 | 2.042161889 | 3.111537267 | 29.79738083 | 12.54239672 | 0.130570106 | 10.32138731 | 2.89972650 |
| Ag | 28.21088746 | 1.300431328 | 2.639679473 | 22.08421391 | 18.51939857 | 0.296911264 | 6.800731775 | 1.82397066 |
| Au | 35.50794814 | 3.371647935 | 3.529615959 | 20.14938691 | 7.260291939 | 7.832280034 | -2.379519739 | -0.0422891 |
| Ni | 5.560674959 | 5.771775656 | 4.807729452 | 35.62554898 | 2.008914012 | 0.000246 | 22.98841482 | 5.2725877 |
| Pd | 9.618326536 | 3.997224815 | 3.352722931 | 41.55663361 | 0.847535341 | 2.213072697 | 4.18967871 | 1.4505722 |
| Pt | 15.41217564 | 4.697571488 | 3.77267856 | 51.27691626 | 4.301949264 | 0.003134623 | 9.684177103 | 1.53515267 |
| Rh | 8.324596442 | 4.582036048 | 3.670327541 | 47.50099338 | 13.49894162 | 2.61E-05 | 25.05670522 | 5.13638985 |
| Ir | 1.665687389 | 9.82561501 | 4.890543442 | 63.83216173 | 6.352562049 | 1.64E-05 | 25.3949301 | 5.14740146 |

Notes: (a) Parameters from [31]. The parameters in GP2 correspond to:

$$E_i = x_0 \sum_j r^{x_1 - x_2 r} f(r) + \left( x_3 - \sum_j \left( x_4 + x_5 r^{x_6 - x_7 r} \right) f(r) \right) / \sum_j f(r)$$

*SI-Table 5. Parameters of GP3 and GP3-c models*

| element | x0 | x1 | x2 | x3 | x4 | x5 | x6 |
|---|---|---|---|---|---|---|---|
| Cu [a] | 7.508311 | 3.979897 | 3.934521 | 28.013689 | 0.031791 | 11.734548 | 2.933153 |
| Cu | 9.653998788 | 2.060534748 | 3.116552313 | 29.79705838 | 0.131622327 | 10.30393221 | 2.896217937 |
| Ag | 26.4855544 | 1.467960106 | 2.679806316 | 22.02488054 | 0.316159588 | 6.676462684 | 1.801372235 |
| Au | 18.85588321 | 4.951473594 | 3.89542606 | 20.0709281 | 3.146357686 | -0.925808305 | 0.003602991 |
| Ni | 7.48071736 | 4.684675392 | 4.471605354 | 35.92501217 | 0.000246 | 23.12325616 | 5.328349011 |
| Pd | 11.05054968 | 3.581177506 | 3.263151364 | 40.97222955 | 0.974045504 | 5.629805775 | 1.686345098 |
| Pt | 9.926992479 | 5.890721459 | 4.067027873 | 50.47696829 | 0.005335237 | 9.087390711 | 1.459929578 |
| Rh | 16.91561527 | 2.441522737 | 3.097197331 | 48.21641282 | 4.12E-05 | 24.67533329 | 5.140494131 |
| Ir | 1.312187465 | 10.27621494 | 4.942567634 | 65.27435096 | 0.000437 | 19.57546211 | 4.188192247 |

(a) Parameters from [31]. The parameters in GP3 correspond to:

$$E_i = x_0 \sum_j r^{x_1 - x_2 r} f(r) + \left( x_3 - x_4 \sum_j r^{x_5 - x_6 r} f(r) \right) / \sum_j f(r)$$

*SI-Table 6. Parameters of SC4-c models*

| element | x0 | x1 | x2 | x3 | x4 |
|---|---|---|---|---|---|
| Cu | 639.4431951 | -9.314032349 | 8.539842097 | -3.939504085 | 0.5 |
| Ag | 27844.05026 | -12.47480779 | 12.68528807 | -5.625307263 | 0.5 |
| Au | 8181.184056 | -9.934009998 | 109.9086417 | -7.831419368 | 0.5 |
| Ni | 659.8848777 | -9.650559919 | 9.100240433 | -4.223886938 | 0.5 |
| Pd | 25084.21852 | -12.89563225 | 38.1980012 | -8.354901196 | 0.5 |
| Pt | 8491.995124 | -10.06919047 | 150.4289378 | -8.108452359 | 0.5 |
| Rh | 22382.02447 | -13.0556093 | 23.92776153 | -7.120089457 | 0.5 |
| Ir | 185595.7209 | -15.40871138 | 32.63989534 | -8.445864459 | 0.5 |

The parameters of SC4-c correspond to: $E_i = \sum_j x_0 r^{x1} f(r) - \left( \sum_j x_2 r^{x_3} f(r) \right)^{x_4}$

*SI-Table 7. Parameters of SC5-c models*

| element | x0 | x1 | x2 | x3 | x4 |
|---|---|---|---|---|---|
| Cu | 631.267176 | -9.406630905 | 38.29001209 | -3.519277139 | 0.064366175 |
| Ag | 27857.11891 | -12.64781229 | 7.159344695 | -4.176935801 | 0.942883625 |
| Au | 8173.588449 | -10.2864238 | 111.0000883 | -7.058621328 | 0.705163433 |
| Ni | 641.8157332 | -9.768708674 | 42.24618404 | -3.296206018 | 0.059023287 |
| Pd | 25086.89713 | -13.11906214 | 38.52709442 | -7.439547099 | 0.670360942 |
| Pt | 8499.317134 | -11.02098225 | 141.8100625 | -5.068173828 | 1.762037756 |
| Rh | 22381.24106 | -13.32746526 | 23.52746732 | -5.130588742 | 1.240077883 |
| Ir | 185597.0729 | -15.64936747 | 29.98698996 | -4.676368063 | 1.887777152 |

The parameters of SC5-c correspond to: $E_i = \sum_j x_0 r^{x1} f(r) - \left( \sum_j x_2 r^{x_3} f(r) \right)^{x_4}$

*SI-Table 8. Parameters of LB-c models*

| element | $D_e$ | $a$ | $r_e$ | $F_0$ | $\gamma$ | $\beta$ | $a_1$ | $\alpha$ | $\varphi$ | $F_1$ |
|---|---|---|---|---|---|---|---|---|---|---|
| Cu | 0.077718 | 2.344814 | 2.572717 | -0.71667 | 3.151194 | 4.714342 | -6.55805 | 2.427183 | 2.413163 | 0.386822 |
| Ag | -6.06466 | 3.595108 | 1.76287 | -13.8068 | 0.693073 | 7.890585 | -17.2178 | 0.853049 | 6.186988 | 8.759516 |
| Au | -13.1 | 4.779872 | 1.714699 | -33.1733 | 0.395984 | 4.326999 | -0.02976 | 7.130727 | -10.4687 | 28.24906 |
| Ni | -10.8456 | 3.646155 | 1.352459 | -7.71595 | 1.395753 | 4.348538 | -0.60062 | 1.932137 | 5.095405 | 1.020529 |
| Pd | -9.89089 | 3.908212 | 1.628892 | -17.686 | 0.342708 | 2.603131 | -0.95136 | 0.297092 | 11.16681 | -3.1042 |
| Pt | -16.3162 | 3.995222 | 1.561789 | -33.162 | 0.590408 | 4.054753 | -0.06768 | 4.624841 | -2.34908 | 20.56575 |
| Rh | 0.2717 | 1.812749 | 2.756533 | -0.11428 | 19.67465 | 3.819038 | -4.54838 | 2.291349 | 2.325303 | -0.04348 |
| Ir | 0.376259 | 1.704919 | 2.841511 | -0.21419 | 14.03302 | 3.755305 | -4.0656 | 2.582749 | 1.540296 | 0.463724 |

The parameters of LB-c correspond to Equations 6, 7, 8, and 9 in the main text.

*SI-Table 9. Temperatures of the DFT molecular dynamics simulations used for generating the training and validation data.*

| Element | Temperature of NVT fcc (K) | Temperature of NVT liquid (K) | Temperature of NPT (K) |
|---|---|---|---|
| Cu | 300 | 1400 | 1400 |
| Ag | 300 | 1535 | 300 |
| Au | 300 | 1637 | 300 |
| Ni | 300 | 2457 | 300 |
| Pd | 300 | 2742 | 300 |
| Pt | 300 | 3070 | 300 |
| Rh | 300 | 3354 | 300 |
| Ir | 300 | 4078.5 | 300 |

*SI-Table 10. Initial parameters of SC models [45]*

| element | x0 | x1 | x2 | x3 | x4 |
|---|---|---|---|---|---|
| Cu | 644.524255 | -9 | 22.97001139 | -6 | 0.5 |
| Ag | 27844.52505 | -12 | 25.11061249 | -6 | 0.5 |
| Au | 8176.059345 | -10 | 121.9754648 | -8 | 0.5 |
| Ni | 651.549205 | -9 | 27.01282713 | -6 | 0.5 |
| Pd | 25086.3566 | -12 | 52.52960227 | -7 | 0.5 |
| Pt | 8496.08965 | -10 | 161.1358259 | -8 | 0.5 |
| Rh | 22379.22732 | -12 | 39.12190388 | -6 | 0.5 |
| Ir | 185600.0961 | -14 | 46.44422713 | -6 | 0.5 |

The parameters of SC correspond to: $E_i = \sum_j x_0 r^{x1} f(r) - \left( \sum_j x_2 r^{x_3} f(r) \right)^{x_4}$

*SI-Table 11. Number of validation datapoints from literature models for each element. For each element and each property, we build a Pareto frontier. We build 21 Pareto frontiers for each element.*

| Element | Number of datapoints from literature models |
|---|---|
| Cu | 37 |
| Ag | 15 |
| Au | 21 |
| Ni | 52 |
| Pd | 6 |
| Pd | 13 |
| Rh | 6 |
| Ir | 5 |

*SI-Table 12. Number of validation properties predicted by models from the literature that have a greater absolute error than the error of a validation property predicted by an interatomic potential model used in this work; labeled N on the table. Note: there are a total of 155 validation properties predicted by models from the literature.*

| Model | N |
|---|---|
| SC4-c | 52 |
| SC5-c | 56 |
| GPn | 82 |
| GP1-c | 72 |
| GP2-c | 79 |
| GP3-c | 80 |
| LB-c | 74 |